\documentclass[british,a4paper,10pt]{article}
\usepackage{babel,amsmath,amssymb,amsthm}
\usepackage{bm}
\usepackage[sans]{dsfont}
\usepackage[mathscr]{euscript}
\usepackage{times}
\usepackage{cite}
\usepackage{mathtools}
\usepackage[all]{xy}
\usepackage{float}
\usepackage{hyperref}
\usepackage{a4wide}
\theoremstyle{plain}
\newtheorem{theorem}{Theorem}
\newtheorem{proposition}[theorem]{Proposition}
\newtheorem{corollary}[theorem]{Corollary}

\theoremstyle{remark}
\newtheorem{remark}{Remark}

\theoremstyle{definition}

\theoremstyle{definition}
\newtheorem{assumption}{Assumption}

\newcommand{\Var}{\operatorname{Var}}
\newcommand{\Cov}{\operatorname{Cov}}
\newcommand{\Tr}{\operatorname{Tr}}
\newcommand{\rmd}{\mathrm{d}}
\newcommand{\rme}{\mathrm{e}}
\newcommand{\rmi}{\mathrm{i}}

\newcommand{\Ebb}{\mathbb{E}}
\newcommand{\Rbb}{\mathbb{R}}
\newcommand{\Cbb}{\mathbb{C}}

\newcommand{\rms}{\mathrm{s}}
\newcommand{\fs}{f_\rms}
\newcommand{\fsT}{f_{\rms, T}}
\newcommand{\openone}{\mathds{1}}
\newcommand{\ind}{\mathtt{1}}

\newcommand{\norm}[1]{\left\Vert#1\right\Vert}
\newcommand{\abs}[1]{\left\vert#1\right\vert}

 \newcommand{\Ecal}{\mathcal{E}}

\newcommand{\Lcal}{\mathcal{L}}
 
 \newcommand{\Ncal}{\mathcal{N}}
\newcommand{\Qcal}{\mathcal{Q}}

\title{The eight-port homodyne detector: the effect of imperfections on quantum random number generation and
on detection of quadratures}
\author{Alberto Barchielli\thanks{{} \ also: Istituto Nazionale di Alta Matematica (INDAM-GNAMPA), and  Politecnico di Milano, Dipartimento di Matematica \ (email: alberto.barchielli@polimi.it )}
\\ Istituto Nazionale di Fisica Nucleare (INFN), Sezione di Milano,
\\
and Alberto Santamato\thanks{{} email: asantamato@cnit.it } \\ Photonics
Networks and Technologies Lab  -- \\ Consorzio Nazionale Interuniversitario per le Telecomunicazioni, Pisa, Italy}

\begin{document}
\maketitle

\begin{abstract}
The eight-port homodyne detector is an optical circuit designed to perform the monitoring of two quadratures of an optical field, the signal. By using quantum Bose fields and quantum stochastic calculus, we give a complete quantum description of this apparatus, when used as quadrature detector in continuous time. We can treat either the travelling waves in the optical circuit, either the observables involved in the detection part: two couples of photodiodes, postprocessing of the output currents\ldots The analysis includes imperfections, such as not perfectly balanced beam splitters, detector efficiency, electronic noise, phase and intensity noise in the laser acting as local oscillator; this last noise is modeled by using mixtures of field coherent states as statistical operator of the laser component. Due to the monitoring in continuous time, the output is a stochastic process and its full probability distribution is obtained. When the output process is sampled at discrete times, the quantum description can be reduced to discrete mode operators, but at the price of having random operators, which contain also the noise of the local oscillator. Consequently, the local oscillator noise has a very different effect on the detection results with respect to an additive noise, such as the noise in the electronic components.
As an application, the problem of secure random number generation is considered, based on the local oscillator shot noise. The rate of random bits that can be generated is quantified by the min-entropy; the possibility of classical and quantum side information is taken into account by suitable conditional min-entropies. The final rate depends on which parts of the apparatus are considered to be secure and on which ones are considered to be exposed to the intervention of an intruder. In some experimentally realistic situations, the entropy losses are computed, depending on the values of the parameters quantifying the imperfections.
\end{abstract}

\maketitle

\tableofcontents

\section{Introduction}\label{sec:intro}

Sources of true random numbers, those that are not generated from algorithms but from stochastic processes, are of great interest especially in areas such as cryptography, simulation and secure communication. Particularly interesting are those random number generators which rely on stochastic processes of quantum origin; these are fundamentally unpredictable due to the aleatory nature of the measurement outcome of quantum states. A single photon travelling through a balanced beam splitter and detected at the two outputs with single photon detectors \cite{Stefanov00} is a paradigmatic example of quantum random number generator (QRNG); the randomness is quantified by the entropy of the superposition of being in one output or the other. The entropy in this case is maximal for a perfectly balanced beam splitter since the probability distribution of the two outcomes is a uniform distribution of the two values (say 0 and 1).
Quite recently QRNGs that do not require single photons, but based on continuous variable (CV) measurements, have been proved to be very efficient and technologically less demanding \cite{Extractor,Haw+15,Vill17,+AS+18,Smith2019,Thewes2019,APFPS20,Huang+20,Lupo+21}; these devices perform homodyne detection and  exploit the entropy of the stochastic process underpinning the shot noise generated from balanced receivers when the signal input is the vacuum state. Improvements of this approach have also been demonstrated \cite{Vill18}; the simultaneous measurement of complementary quadratures of the vacuum state with double homodyne detection generates not only genuine  but also secure random numbers. In fact,  following this technique it is possible to counteract the influence of an adversary that is trying to exploit untrusted elements of the system to steer the statistics of the random numbers towards his own benefits.

The eight-port homodyne detector \cite{KiuL08b,FOP05,LPS10,Leo10,Dmo19} is a combination of an optical circuit and photodetectors, devised to ``measure'' two quadratures of a quantum ``signal'', possibly two complementary quadratures. The quantum treatment of both single and double homodyne detection is usually done by using discrete bosonic modes \cite{LPS10,Leo10,FOP05,KiuL08a,KiuL08b,BKS14,Vill17,Vill18}; thanks to the use of continuous Bose fields we allow for a more general  description in terms of travelling waves and of detection in continuous time  \cite{RCCBS95,Bar86,ZolG97,Bar06,ZGVit15}.
In this work we want to focus our attention on the detailed analysis of the eight-port apparatus, described in Fig.\ \ref{fig:optcir}: it
is composed by four beam splitters  and four photodiodes, one at each output of the beam splitters 3 and 4.
The inputs are the \emph{signal} (port 1) and the \emph{local oscillator} (LO -- port 3); the beam splitter 1 mixes the signal with a vacuum input, while beam splitter 2 mixes the LO with another vacuum input. We include imperfections and noises in the analysis of this circuit and   we  show that it is still possible to use it as a detector apparatus. We also produce formulae useful to calibrate non-ideal experimental implementations of QRNG modules.

The purpose of the device is to measure two field quadratures of the input signal and this is realized when
the LO  power is much larger than the signal power (i.e.\ in the limit of infinitely strong local oscillator).
We are interested in the distribution of the two output currents generated by the difference of the pairs of photocurrents produced by each couple of  photodiodes.
In the limit of strong LO, as we shall see, it is possible to demonstrate that the apparatus is indeed monitoring in continuous time two quadratures; when the signal is in the vacuum state, the variance of the each quadrature is reduced to pure shot noise (up to imperfections to be discussed), which is an ideal source of entropy.  This noise is the source of randomness that is used to produce random bits. The distribution of the measured photocurrent is approximately Gaussian centered around a zero mean with a variance dependent on the power of the local oscillator.
Continuous time observations of real signals are involved in the description; however, when this apparatus is used for QRNG, the signals are sampled at discrete times and the CV outputs are discretized in their turn. This is due to the natural discrete sampling that an oscilloscope or an ADC (analog to digital converter) operates on the input signals. The discrete distribution obtained this way can be used to produce the actual uniform distribution of independent samples needed for true random number generators. The transformation from the output statistics (which is close to a Gaussian distribution) to a uniform distribution
can be done using established techniques described for example in \cite{Extractor}.

Independently from the application as detector or as random number generator, the system we are considering consists of linear optical elements with travelling light waves,  constantly monitored by the four photodiodes. To give a full quantum description of this system
we shall use quantum Bose fields and \emph{quantum stochastic calculus} (QSC) \cite{HudP84,GarC85,Parthas92}. An important aim of this work is also to use this concrete application to show how to construct a consistent quantum theory of optical circuits \cite{ASth,BG21} and \emph{photo-detection in continuous time} \cite{Bar06,Bar90,Bar91}, even when imbalanced beam splitters, detector response functions, laser and electronic noises have to be taken into account. Here, quantum and classical noises find a consistent description, based on general quantum measurement theory \cite{KiuL08a,WisM10,Hol82,Bar86,BG12,BarG13}. The whole approach presented here can be generalized also to  other optical circuits, such as the ones presented in \cite{FOP05,Raym03}, or to problems of quantum communication, such as \emph{quantum key distribution} (QKD) \cite{APFPS20,Qin2018,YMC+11}.

\subsection{Plan of the paper}

In Sec.\ \ref{sec:ocpd} we discuss the quantum treatment of the eight-port detector. We introduce the necessary quantum fields and the unitary operators representing the beam splitters and other linear components in the circuit. The measurement stage, the photodetectors, is represented by projection valued measures (pvm) and positive operator valued measures (POVM) in continuous time.
We allow for not balanced beam splitters.
The circuits also includes a variable phase shifter between the beam splitters 2 and 3 and another one between the beam splitters 2 and 4.
At one input of beam splitter 2 there is a laser  (the local oscillator, LO) which we consider of very strong power with respect to a possible signal in the other inputs. The laser state is a mixture of field coherent states; phase fluctuations and intensity noise are included in the formulation of such a state. The action of the photo-detectors is described by the pvm  associated with the photo-counting operators; this counting process is smoothed in time by the photodiodes and the resulting photocurrents are subtracted in order to implement the balanced homodyne detection.
More precisely we work with the characteristic operators (Fourier transform) \cite{Bar86,ZolG97,Bar06} of the counting pvm in continuous time and of the POVM describing the subtracted photocurrents. The theoretical results are rooted in QSC, which is essential in the whole construction. In the main text we present the relevant results and give a physical interpretation of the formulae, while some mathematical details are left to the Appendices \ref{app:fields}, \ref{Phi:counts}, \ref{sec:PhiX}.

The limit of strong LO is done in Sec.\ \ref{sec:Qj}. Here we have to prove the existence of this limit for the full probability distribution of the scaled photocurrents; again the mathematical proofs are left to Appendix \ref{app:strLO}.
For a suitable choice of the two phases $\psi_j$, the two subtracted currents become proportional  to two complementary quadrature operators meaning that the circuit considered in the strong intensity limit  is indeed measuring simultaneously the two quadratures. In the ideal case, this result shows that this circuit is a physical implementation of the theoretical POVM introduced in \cite{Bar86} for time continuous measurements; however, now we take into account also the effects of imbalanced beam splitters, inefficiencies of the photodiodes and LO noises. We also discuss how the increase of phase noises can degrade the homodyne detection into heterodyne detection in a continuous way.

The outputs of the apparatus of Fig.\ \ref{fig:optcir} are often sampled at discrete times, either when it is used for QRNG or as a detection apparatus for the signal quadratures. The previous general results are particularized to this situation in Sec.\ \ref{dsampling}. Moments and probability distributions of the involved observables are obtained; again mathematical computations are left to Appendices \ref{app:dsproof} and \ref{app:probdens}. In this situation it is possible to express the various physical quantities through discrete mode operators, but, thanks to  our general approach in continuous time, one sees that these mode operators are random, because their structure is based on the function describing the coherent state of the LO and this function is random due to the intensity noise and phase fluctuations.

In Sec.\ \ref{sec:hmin} we apply the previous results to the problem of QRNG. Now the signal is in the vacuum state and the outputs are sampled at discrete times; moreover, the CV outputs are naturally discretized by the ADC apparatus.
Following a common approach presented in other works we use the classical min-entropy and the conditional min-entropy \cite{Extractor,KRS09,Vill18,+AS+18} to evaluate the rate of secure bits that can be extracted form the system under practical and non ideal situations (not perfect balancing of the beam splitters, some inefficiency of the detectors, presence of laser intensity and phase noises).
Our work generalizes and enlarges the analysis presented in \cite{APFPS20} and \cite{Vill18} since we allow for an imperfect realization of the double homodyne detection scheme, not only for additive electronic noise. By choosing some realistic values for the free parameters we give also examples where the number of secure bits per sample can be computed.
The problem of QRNG is not only that the extracted bits must be truly random, but also that they must be unknown to a possible adversary.
There are subtleties to be taken into consideration depending on what it is assumed to be ``secure'' and ``trusted'', with respect to what is ``untrusted'', due to the possibility that an eavesdropper takes advantage of this untrusted part of the system to gain knowledge on the generated data or to simply corrupt the generation process \cite{Qin2018,Smith2019,Thewes2019}. In a first approach we consider untrusted only the classical noise: the laser noise and the
the electronic noise due to  the amplification chain after the photodiodes; the apparatus and, in particular, the input ports are trusted. Then, we consider also the case in which the signal port is not secure and the so called \emph{quantum side information} has to be taken into account. Via the introduction of a conditional min-entropy, this classification influences  the quantification of the secure rate of random bits that can be extracted from the system.
In the examples we show also how  the generation rate of ``secure bits'' is influenced by the imperfections in the circuit, as imbalance of the beam splitters and not unit efficiency of the photodiodes. In this section we discuss also the different roles of the LO fluctuations and of the electronic noise, which is a purely additive noise.

As a particular case, the circuit of Fig.\ \ref{fig:optcir} includes also the case of single homodyning. In Sec.\ \ref{sec:singlehom} we explain how to obtain this case and we discuss its application to the QRNG problem; again quantitative examples are given.
Some final comments are in Sec.\ \ref{sec:conclusion}.

\section{Optical circuit and photon detection}\label{sec:ocpd}

\begin{center}
\begin{figure}[H]
\includegraphics[scale=0.8]{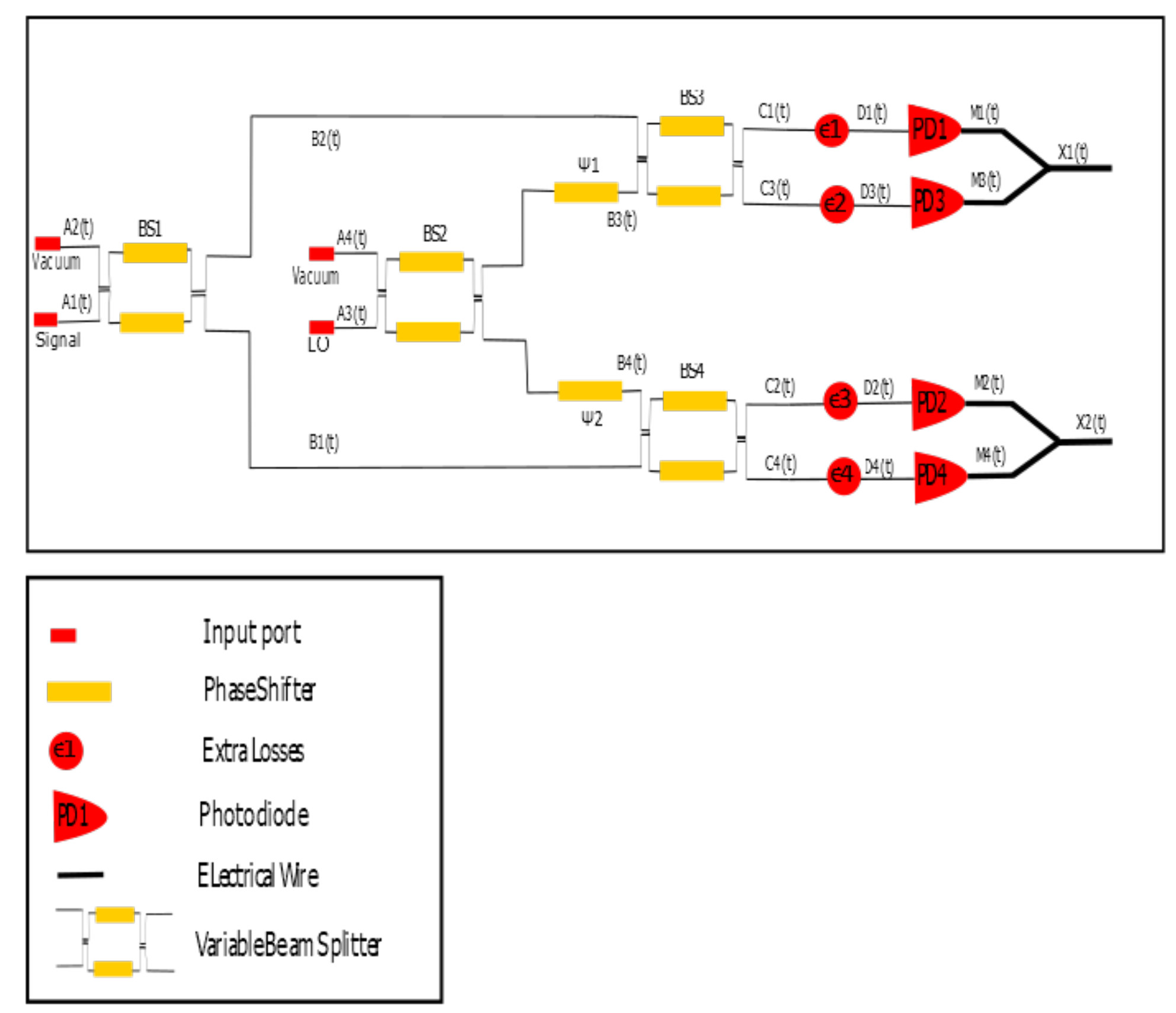}
\caption{\label{fig:optcir} The eight-port optical circuit for a double homodyne detection. \ \ Input fields: $A_j(t)$.  \ \ Transformed fields: \ stage 1 $B_j(t)$, \ stage 2 $C_j(t)$. \ \ 
Detected fields: $D_j(t)$. \quad  Detector outputs: $M_j(t)$. \ \ Balanced outputs: $X_j(t)$.}
\end{figure}
\end{center}

To give a quantum description of optical circuits, such as the one of Figure \ref{fig:optcir}, and of the photon detection scheme, we need to introduce suitable quantum fields representing the travelling waves inside the circuit and some notions of QSC, needed to handle the equations used in the description of the detection stage \cite{Bar86,Parthas92,ZolG97,Bar06,ZGVit15}. Let us stress that the original motivation for the introduction of QSC was related to quantum open system theory \cite{HudP84,GarC85,Parthas92,ZolG97}.

We introduce $d$ Bose fields $a_j(t)$ satisfying the canonical commutation rules (CCRs)
\begin{equation}\label{CCR}
[a_i(s), a_j(t)]=0, \qquad [a_i(s), a_j^\dagger(t)]=\delta_{ij}\delta(t-s).
\end{equation}
In quantum field theory, the CCRs admit inequivalent representations. In order to have the existence of the vacuum and of the coherent states we fix the Fock representation and denote by $\Gamma=\Gamma_1\otimes \Gamma_2\otimes \cdots$ the \emph{symmetric Fock space} (see \eqref{Focksp} in Appendix \ref{app:fields}), the Hilbert space where the fields  $a_j(t)$ act. The \emph{coherent vectors} $e_j(f)\in \Gamma_j$, with
$f\in L^2(\Rbb)$, are defined by \eqref{expV} and satisfy
\begin{equation}\label{ae(f)}
a_j(t)e_j(f)=f(t)e_j(f);
\end{equation}
this equation  shows that $e_j(f)$ is indeed a coherent vector for the annihilation operators.
Note that $e_j(0)$ represents the vacuum state for the field component $a_j(t)$.

To develop the theory of quantum stochastic differential equations, the integral version of the $a_j$-fields is needed, together with the integral of quadratic expressions preserving the number of quanta:
\begin{equation}\label{fdensity}
A_j(t)=\int_0^t a_j(s)\rmd s, \qquad \Lambda_{ij}^A(t)=\int_0^t a^\dagger_i(s)a_j(s)\rmd s.
\end{equation}
The operators $\Lambda_{ij}^A(t)$ were named \emph{gauge process} \cite{HudP84}; note that $\Lambda_{jj}^A(t)$ is the \emph{number process} for the field $j$ (indeed it has the integer numbers as eigenvalues). By \eqref{ae(f)} we get
$\langle e_j(f)|\Lambda_{jj}(t)|e_j(f)\rangle =\int_0^t\abs{f(s)}^2\rmd t$ and this quantity represents the mean number of photons in the time interval $(0,t)$; then, $\abs{f(t)}^2$ is the instantaneous mean number of photons per unit of time.

The approximations involved in the use of these fields to represent the electromagnetic field are justified when the interactions and the refraction indexes are little dependent on frequency in a large band around a principal frequency $\omega_0$; this is the so called ``broadband, quasi-monochromatic approximation''. Moreover, to use these fields in an optical circuit means to use a ``quantum travelling wave formulation'', an approximation which is in some way opposed to the use a single mode or a few discrete modes. By using these fields and QSC it is also possible to develop a theory of direct, homodyne, heterodyne detection in continuous time \cite{SYM79pII,YS80pIII,ZolG97,ASth,BG21,Bar06,Bar91,Bar90,WisM10,YMC+11}.

By suitable unitary transformations, which preserve the canonical commutation relations, it is possible to represent the optical linear devices which compose an optical circuit \cite{ASth,BG21}. In particular we shall need the transformation generated by a beam splitter of transmissivity $\eta\in [0,1]$: two input fields
$A_1(t)$ and $A_2(t)$ are mixed together and transformed in the fields $B_1(t)$ e $B_2(t)$, given by
\begin{equation}\label{AtoB1}
B_1(t)=\sqrt{\eta}\,A_1(t)+\rmi \sqrt{1-\eta}\, A_2(t),
\qquad
B_2(t)=\rmi\sqrt{1-\eta}\, A_1(t)+ \sqrt{\eta}\, A_2(t).
\end{equation}

In the optical circuit that we shall analyse, the polarization does not play any role; when it is not so, polarization can be taken into account by doubling the fields and also linear devices depending on the polarization can be introduced \cite{ASth}.

\subsection{Optical circuit}\label{sec:optcirc}
The circuit under study is drawn in Figure \ref{fig:optcir}. It presents 4 input ports, into which the input fields $A_j(t)$ enter, and 4 output ports, from which the 4 output fields $D_j(t)$ leave the circuit; the output fields are detected by 4 photodiodes.

The field $A_1(t)$ carries the \emph{input signal}, and it is split in two fields by mixing it at the beam splitter BS1 with the field $A_2(t)$ in the vacuum state. The field $A_3(t)$ carries the laser light playing the role of \emph{local oscillator} (LO), and it is split in two fields by mixing it at the beam splitter BS2 with the field $A_4(t)$ in the vacuum state. As discussed in Sec.\ \ref{sec:efficiency}, to model the light losses in the circuit and the efficiency of the detectors we add before the output ports 4 fictitious beam splitters and 4 auxiliary input fields $A_{j+}(t)$ in the vacuum state. So, to analyze the circuit of Figure \ref{fig:optcir} we use $d=8$ field components.

This circuit is suggested as a realization of a double homodyne detection in \cite[Sec.\ 5.6.1]{FOP05}, \cite[Sec.\ 4.5.1]{WisM10},  \cite[Fig.\ 5]{BKS14},  for instance.  Here we want to perform a fully quantum analysis of this circuit, taking into account
noise sources and imperfections (e.g.\ laser noise and unbalancing of the beam splitters), as suggested in  \cite{APFPS20} for the case of a single homodyne apparatus.

\subsubsection{Beam splitters and  phase shifters}

The circuit is composed by four beam splitters BSj of transmissivity $\eta_j$, $j=1,\ldots,4$ and by two tunable phase shifters.  All the optical paths from the inputs to the outputs are assumed to be equal; this allows to neglect the delays in passing from a beam splitter to the other. By suitable chosen phase shifts in the fields, also imperfections in the optical paths could be taken into account in our formalism.

At the beam splitter BS1 the signal field $A_1(t)$ is mixed with the vacuum field $A_2(t)$ and produces the output fields $B_1(t)$ and $B_2(t)$, which turn out to be given by
\begin{subequations}\label{AtoB}
\begin{equation}
B_1(t)=\sqrt{\eta_1}\,A_1(t)+\rmi \sqrt{1-\eta_1}\, A_2(t),
\qquad
B_2(t)=\rmi\sqrt{1-\eta_1}\, A_1(t)+ \sqrt{\eta_1}\, A_2(t).
\end{equation}
At the beam splitter BS2 the LO field $A_3(t)$ is mixed with the vacuum field $A_4(t)$ and, after the two tunable phase shifts $\psi_j$,
it gives rise to the output fields $B_3(t)$ and $B_4(t)$, given by
\begin{equation}
B_3(t)=\rme^{\rmi \psi_1}\left[\sqrt{\eta_2}\,A_3(t)+\rmi   \sqrt{1-\eta_2}\,A_4(t)\right],
\qquad
B_4(t)=\rme^{\rmi \psi_2} \left[\rmi   \sqrt{1-\eta_2}\,A_3(t)+ \sqrt{\eta_2}\,A_4(t)\right].
\end{equation}
\end{subequations}
It will be useful to have a notation for the difference of the tunable phases:
\begin{equation}\label{phij}
\phi=\psi_2-\psi_1.
\end{equation}

Then, the fields $B_1(t)$ and $B_3(t)$ are mixed together at the beam splitter BS3, giving rise to the fields $C_1(t)$ and $C_3(t)$,
while the fields $B_2(t)$ and $B_4(t)$ are mixed at BS4 and give rise to $C_2(t)$ e $C_4(t)$. The output fields turn out to be given by
\begin{subequations}\label{BtoC}
\begin{gather}
C_1(t)=\sqrt{\eta_3}\,B_1(t)+\rmi \sqrt{1-\eta_3}\, B_3(t),
\qquad
C_3(t)=\rmi \sqrt{1-\eta_3}\, B_1(t)+ \sqrt{\eta_3}\, B_3(t),
\\
C_2(t)=\sqrt{\eta_4}\, B_2(t)+\rmi \sqrt{1-\eta_4}\, B_4(t),
\qquad
C_4(t)=\rmi \sqrt{1-\eta_4}\,B_2+\sqrt{\eta_4}\, B_4(t).
\end{gather}
\end{subequations}

\subsubsection{Losses and efficiency of the detectors}\label{sec:efficiency}
To model the field losses in an optical path and/or in a photodetector and to maintain the CCRs \eqref{CCR} for the fields, it is usual to insert a virtual beam splitter of transmissivity less than one \cite{APFPS20}. The beam enters a first input port, while the vacuum enters the second input port. The attenuated beam comes out from the transmission output port, while the lost light comes out from the other port.

In our case we add four input fields $A_{j+}(t)$ in the vacuum state and four beam splitters of transmissivity $\epsilon_j\in (0,1]$; at the 8 outputs we have the observed field $D_j(t)$,
which reach the photodetectors, and the lost fields  $D_{j+}(t)$,
$j=1,\ldots,4$. The transformation of the field operators is
\begin{equation}\label{CtoD}
D_j(t)=\sqrt{\epsilon_j}\,C_j(t)+\rmi \sqrt{1-\epsilon_j}\, A_{j+}(t),
\qquad
D_{j+}(t)=\rmi\sqrt{1-\epsilon_j}\, C_j(t)+ \sqrt{\epsilon_j}\, A_{j+}(t).
\end{equation}
Let us stress that preserving the CCRs \eqref{CCR} is needed to have a consistent quantum description; this means that the attenuation of the optical signal goes together an increase of the noise due to the new vacuum inputs.

\begin{remark}\label{effreg} As suggested in \cite{APFPS20}, the efficiency coefficients $\epsilon_j$
can be considered partially tunable; indeed, by inserting a variable optical attenuator in series before each output port we can diminish the efficiency coefficient. So, if $\epsilon_j^{\rm max}\in (0,1]$ is the coefficient value due to light losses and inefficiency of the detector,  the effective coefficient $\epsilon_j$ is (roughly) tunable in the interval $\left(0,\epsilon_j^{\rm max}\right]$.
\end{remark}

By combining Eqs.\ \eqref{AtoB}, \eqref{BtoC}, \eqref{CtoD} we can express the output fields $D_j(t)$ in terms of the input fields $A_j(t)$ and of the auxiliary fields $A_{j+}(t)$; the total transformation for the field densities is given in Eqs.\ \eqref{fieldsdj}.
From these densities one can also express, in terms of the input fields $a_j(t),\,a_{j+}(t)$, the components
$\Lambda_{ij}^D(t)=\int_0^t d_i^\dagger(s)d_j(s)\,\rmd s$
of the gauge process; the number operators $\Lambda_{jj}^D(t)$ will be used in Sec.\ \ref{sec:counting} in modeling the photodetectors.

\subsection{The field state}\label{sec:sysst}

As already noticed, the fields $A_2(t),\, A_4(t),\, A_{j+}(t)$ are in the vacuum state. Then, if we denote by $\rho_{13}^T$ the reduced state of signal and LO (in the time interval $(0,T)$), the total state of the field is
\begin{equation}\label{rhoT}
\rho^T= \rho_{13}^T\otimes \rho^\bot, \qquad
\rho^\bot = |e_2(0)\rangle\langle e_2(0)|\otimes |e_4(0)\rangle\langle e_4(0)|
\otimes {\prod_{j=1}^4 }^\otimes|e_{j+}(0)\rangle\langle e_{j+}(0)|.
\end{equation}

To represent the laser light of the LO we use the mixture of coherent states
\begin{equation}
\rho_{3}^{T}=\Ebb_f[|e_3(f_T)\rangle \langle e_3(f_T)|], \quad f_T(t)= f(t)\ind_{(0,T)}(t),
\end{equation}
where $f(t)$ is a complex stochastic process; by the comments after Eq.\ \eqref{fdensity}, $\abs{f(t)}^2$ is the instantaneous laser intensity. As for a mono\-chromatic wave, a typical trajectory of this process is not in $L^2(\Rbb)$, as required to define a field coherent vector. However, it can be assumed to be locally square integrable and by multiplying it by the indicator function $\ind_{(0,T)}(t)$ we get $f_T\in L^2(\Rbb)$. The time $T$ is taken to be large and sent to $+\infty$ in the final physical formulae.

\begin{remark}\label{rem:P}
For the means and other moments done under the law $P_f$ of the process $f$ and of the other correlated processes we use the notation $\Ebb_f$, $\Var_f$, $\Cov_f$\ldots On the other side,
for means and moments with respect to the probability law $P$ obtained from the total field state and the various positive operator valued measures (POVM) representing the quantum observables we shall use the notation $\Ebb_P$, $\Var_P$, $\Cov_P$\ldots
\end{remark}

\begin{remark}\label{rem:phloc} In some applications of homodyne detection, when squeezing can be relevant, a good phase coherence between signal and LO must be maintained in time, the LO must be \emph{phase-locked} to the signal \cite{RCCBS95}. To guarantee this, the same laser is used to produce the LO and to stimulate the quantum system of interest (an atom \cite{BG12,BarG13}, a quantum oscillator \cite{BG21}, a non linear medium \cite{AZVit16}, \ldots). This means that signal and LO can be correlated; to model this situation, we take as signal/LO state
\begin{equation}\label{rho13}
\rho_{13}^T= \Ebb_f\left[\rho_{13}^{f,T}\right],\quad \rho_{13}^{f,T}= \rho^{f,T}_{1} \otimes |e_3(f_T)\rangle\langle e_3(f_T)|,
\end{equation}
where $\rho^{f,T}_{1}$ is a random statistical operator for the signal field, depending on the process $f$ or correlated in some way with it.
\end{remark}

We already introduced the reduced random signal/LO state $\rho_{13}^{f,T}$; it is useful to introduce also the random total state and some simplified notations
\begin{equation}\label{qexpect}
\rho^T_f=\rho_{13}^{f,T}\otimes \rho^\bot, \qquad \langle \bullet \rangle_T^f=\Tr\left\{\bullet \rho^T_f\right\},\qquad
\langle \bullet \rangle_T=\Tr\left\{\bullet \rho^T\right\}=\Ebb_f\left[ \langle \bullet \rangle_T^f\right].
\end{equation}

\subsubsection{Laser models} \label{sec:LOstate}
The process $f(t)$ plays the role of reference wave when the whole apparatus as used as a detector for the signal light.
To represent a laser of nominal frequency $\omega_0$ we can take $f(t)=\lambda\rme^{-\rmi \omega_0 t}\times \cdots$, where the dots stay for other contributions representing the main noises affecting the laser light. The first important noise is the \emph{phase noise} and a good model for this is the \emph{phase-diffusion model of a laser} \cite[Sec.\ 11.4.1]{ScuZ97}, \cite[Sec.\ 2.7.3]{Botta08}: a term proportional to a Wiener process is added to the phase $\omega_0 t$, whose effect is to give a finite coherence length. Another important noise is the one generated by the fluctuations of the laser intensity, often represented by the \emph{relative intensity noise} (RIN) \cite{Botta08,Mesc07,APFPS20}:
\begin{equation}\label{nRIN}
n_{\rm RIN}(t)=\frac{\abs{f(t)}^2-\Ebb_f\left[\abs{f(t)}^2\right]}{\Ebb_f\left[\abs{f(t)}^2\right]}.
\end{equation}
As explained in \cite[Sec.\ 2.8]{Botta08}, lasers are constructed in such a way that they do not have other peaks in a large frequency band around their carrier frequency and the proposed noise models should be compliant with these requirements.

We collect all these features in a single model for the laser, a modification of the phase diffusion model:
\begin{equation}\label{LOf+}
f(t)= \lambda\rme^{-\rmi \omega_0 t-\rmi \sqrt{2\gamma_0}\, W(t)}\,u(t), \qquad \lambda=\abs\lambda \rme^{\rmi\theta}, \qquad
\theta\in \Rbb, \qquad \omega_0>0, \qquad \gamma_0>0,
\end{equation}
where $W(t)$ is a standard Wiener process.
Moreover, we take $u(t)$ to be a real Gaussian process, independent of $W(t)$, such that
\begin{equation}\label{ucorr}
\Ebb_f[u(t)]=w, \qquad \Cov_f[u(t),u(s)]=v(t-s), \qquad \Ebb_f\left[u(t)^2\right]\equiv w^2+v(0)=1 .
\end{equation}
The processes $W(t)$, $u(t)$ and the random variable $\theta$ are taken to be independent. Computations of some moments of $f(t)$ are given in Appendix \ref{sec:LOprop}.
By the comments below Eq.\ \eqref{fdensity}, $f(t)$ has the dimensions of a time to $-1/2$. By the last requirement in \eqref{ucorr}, $w$ and $u(t)$ are taken to be pure numbers, so that $\lambda $ has the dimensions of a time to $-1/2$.

By \eqref{LOf+} and \eqref{ucorr}, the mean intensity of the laser is constant in time:
\begin{equation}\label{Eabsf2}
\Ebb_f\left[\abs{f(t)}^2\right]=\abs\lambda^2;
\end{equation}
$\abs\lambda^2$ is the mean number of photons per unit of time. By \eqref{Efff3}, the RIN correlations turn out to have the expression
\begin{equation}\label{RINcov}
\Cov_f\left[n_{\rm RIN}(t),n_{\rm RIN}(s)\right]=\Ebb_f\left[n_{\rm RIN}(t)n_{\rm RIN}(s)\right]
=2v(t-s)^2+4w^2v(t-s).
\end{equation}

The correlation function $v(t)$, defined in \eqref{ucorr}, turns out to be even, $v(-t)=v(t)\in \Rbb$, and positive definite; so, we have $v(0)\geq 0$ and
\begin{equation}\label{LOf3}
v(t) =\frac 1 {2\pi}\int_{-\infty}^{+\infty}\rme^{\rmi\nu t} \tilde v(\nu)\,\rmd \nu, \quad
\tilde v(\nu)=\tilde v(-\nu)\geq 0;
\end{equation}
we assume also $\tilde v(\nu)\in L^1(\Rbb)$.

An important feature of the laser model is the intensity spectrum of the process $f(t)$, defined by
\begin{equation}\label{LOspectrum}
\Pi_f(\mu)=\lim_{T\to +\infty}\Ebb_f\biggl[\abs{\frac 1 {\sqrt T}\int_0^T \rme^{\rmi \mu t}f(t)\rmd t}^2\biggr];
\end{equation}
note that $\Pi_f(\mu)$ turns out to be a pure number. By using \eqref{LOf+}, \eqref{LOf3}, \eqref{Efff2}, we get
\begin{equation}
\Pi_f(\mu)=\frac{2\abs\lambda^2w^2 \gamma_0}{\gamma_0^{\,2} + \left(\mu-\omega_0\right)^2}+\int_\Rbb \rmd \nu\,\frac {\abs\lambda^2\tilde v(\nu) \gamma_0/\pi}{\gamma_0^{\,2}+ \left(\mu-\omega_0-\nu\right)^2}\,.
\end{equation}
If needed, the laser parameters $\gamma_0$, $\omega_0$, $\tilde v(\nu)$ could be estimated in a calibration stage, for instance by measuring the intensity spectrum.

Similarly, we can introduce the spectrum of the intensity fluctuations; from \eqref{RINcov} and \eqref{LOf3}, by straightforward computations we get
\begin{equation*}
\Pi_{\rm RIN}(\mu)=\lim_{T\to+\infty}\frac 1 T\int_0^T\rmd t\int_0^T\rmd s\, \rme^{\rmi\mu\left(t-s\right)}
\Ebb_f\left[n_{\rm RIN}(t)n_{\rm RIN}(s)\right] = 4w^2 \tilde v(\mu)+ \int_\Rbb \frac {\tilde v(\nu)\tilde v(\mu-\nu)} \pi\,\rmd \nu.
\end{equation*}
To agree with the request that the intensity fluctuations do not introduce
new peaks in the laser spectrum we need $\tilde v(\nu) $ to be sufficiently flat.

\subsection{The photodetectors}\label{sec:counting}
Let us consider now the monitoring of the photon flux intensity at the four output ports; we start by considering direct detection (in continuous time) of the photons \cite{ZolG97,WisM10,SYM79pII,YS80pIII,Bar06,Bar91}.

When the detectors at the output ports are perfect photocounters, we can say that we are measuring the quantum observables represented by the number operators
\begin{equation}\label{numop}
\hat N_j(t)= \Lambda^D_{jj}(t)=\int_0^t d_j^\dagger(s)d_j(s)\,\rmd s, \qquad 0\leq t\leq T, \qquad j=1,\ldots,4.
\end{equation}

\begin{remark}\label{rem:compatible}
The essential point is that, by the CCRs satisfied by the fields $d_j(t)$, these are compatible observables: the set
$\{\hat N_j(t),\ 0\leq t\leq T, \ j=1,2,3,4\}$ is a family of commuting self-adjoint operators to which a \emph{projection valued measure} (pvm) is associated. The commutativity also for different times is the key point which allows for a quantum theory of measurements in continuous time \cite{Bar86,Bar06}. All the observables we shall introduce in the following will be functions of the number operators, and, so, also these other observables will be represented by commuting operators.
\end{remark}
The family of number operators is a continuity of operators (with respect to the index ``time''); so, the associated pvm is a measure on an infinite-dimensional value space. Apart from this, by the usual rules of a quantum theory, this measure and the system state (introduced in Sec.\ \ref{sec:sysst}) give rise to a probability measure $P$ for the observed counts $N_j(t)$.
To handle this implicitly defined pvm, it is useful to introduce its Fourier transform, the \emph{characteristic operator} \cite{Bar86,ZolG97,Bar06}, which we discuss in Appendix \ref{Phi:counts}. In any case, as the probabilities are given by the pvm of the commuting number operators, all the moments of the observed counts can be expressed by means the usual quantum expectations; by using the notations introduced in Remark \ref{rem:P} and in Eq.\ \eqref{qexpect}, we have
\begin{equation}\label{Nmoments}
\Ebb_P\left[N_{j_1}(t_1)N_{j_2}(t_2)\cdots\right]=\langle \hat N_{j_1}(t_1)\hat N_{j_2}(t_2)\cdots\rangle_T.
\end{equation}

\begin{remark}  \label{rem:Phi0N} In the case of vanishing signal, as shown in Appendix \ref{sec:0N}, the counting processes $N_j(t)$  are a mixture of Poisson processes with random intensities
\begin{equation}\label{intensities}
J_1^f(t)=\eta_2\left(1-\eta_3\right)\epsilon_1\abs{f(t)}^2, \qquad J_2^f(t)=\left(1-\eta_2\right)\left(1-\eta_4\right)\epsilon_2\abs{f(t)}^2 .
\end{equation}
In Appendix \ref{sec:0N}, it is also shown that we get a mixture of Poisson processes also when the signal is in  a coherent state.
\end{remark}

\subsubsection{The photodiodes}\label{sec:responseF}
In the following we shall be interested in a generic signal and a strong LO; so we expect an intense flux of photons at the output ports and we cannot relay in single photon counters. We consider as detectors general photodiodes, whose output is some kind of time average of the photon arrivals in a time interval.
The output photocurrent of a photodiode can be seen as a smoothed version of the rate of arrival of the photons; this signal and its associated operator can be represented by
\begin{equation}\label{M:N}
M_j(t)=\int_0^tF_j(t,s)\rmd N_j(s),\qquad \hat M_j(t)=\int_0^tF_j(t,s)\rmd \hat N_j(s),
\end{equation}
where $F_j(t,s)$ is the \emph{response function} of the $j$-th detector.

The response functions contain an unavoidable smoothing on time; a good and general choice is to take
\begin{equation}\label{F:choice}
F_j(t,s) =\xi_jh(t-s), \qquad \xi_j> 0, \qquad h(t)\geq 0, \qquad \int_0^{+\infty}h(t)\rmd t=1.
\end{equation}
The function $h(t)$ represents the \emph{impulse response} of the photodiode; it decays as time grows and it is taken to be the inverse of a time and normalized as indicated in \eqref{F:choice}. Moreover, the conversion factors $\xi_j$ are dimensional coefficients, containing possible amplification contributions. The four detectors cannot be exactly equal; however, for simplicity, we took the same time behaviour of the four response functions, while we left the freedom of having different conversion factors $\xi_j$.

Again, the output signals $M_j(t)$ are represented by the commuting self-adjoint operators $\hat M_j(t)$, $ t\in(0,T)$, $j=1,\ldots,4$ and their probability law and moments are obtained by the usual rules of quantum mechanics. The characteristic operator of the family of the
$\hat M$-operators is presented in Appendix \ref{sec:PsiM}, while the moments can be obtained directly from \eqref{Nmoments} and \eqref{M:N}.
By using the field densities and putting them in normal order,  we get
\begin{equation}\label{mean:M}
\Ebb_P\left[M_{j}(t)\right]=\langle \hat M_{j}(t)\rangle_T= \int_0^tF_j(t,r)\langle\rmd \hat N_j(r)\rangle_T
=\xi_j \int_0^t\rmd r\,h(t-r)\langle d^\dagger_j(r)d_j(r)\rangle_T,
\end{equation}
\begin{multline}\label{corr:M}
\Ebb_P\left[M_{j}(t)M_{i}(s)\right]=\langle \hat M_{j}(t)\hat M_{i}(s)\rangle_T
=\delta_{ij}\xi_j^{\;2}\int_0^{t\wedge s}\rmd r\, h(t-r)h(s-r)\langle d^\dagger_j(r)d_j(r)\rangle_T
\\ {}+
\xi_j\xi_i \int_0^t\rmd r\int_0^s\rmd r'\,h(t-r)h(s-r') \langle d^\dagger_j(r)d^\dagger_i(r')d_i(r')d_j(r)\rangle_T;
\end{multline}
we are using the notation $t\wedge s\equiv \min\{t,s\}$.
Some more explicit expressions of means and correlations are given in Appendix \ref{sec:momM}.

\subsection{Postprocessing of the outputs}\label{sec:Xj}

To realize the balanced homodyne detection scheme \cite{ZolG97,WisM10,SYM79pII,YS80pIII,Bar06,Bar91} we need to subtract the two output photocurrents from each couple of photodiodes (1, 3 and 2, 4); then, the limit of strong LO will be considered (after a possible scaling of the output signals). So, we end with the two output signals
\begin{equation}\label{Xprocs}
X_j(t)=M_j(t)-M_{j+2}(t)=\xi_j\int_0^th(t-s)\rmd N_j(s)-\xi_{j+2}\int_0^th(t-s)\rmd N_{j+2}(s), \qquad j=1,2.
\end{equation}
By \eqref{M:N} we see that the two stochastic processes $X_j(\bullet)$ are linear combinations of the counting processes $N_i(\bullet)$; again they represent the observed values of the compatible quantum observables
\begin{equation}\label{hatX}
\hat X_j(t)=\int_0^tF_j(t,s)\rmd \hat N_j(s)-\int_0^tF_{j+2}(t,s)\rmd \hat N_{j+2}(s).
\end{equation}

\begin{remark} \label{F,Alm}  The usual scheme is to amplify the two difference currents, for instance by a \emph{transimpedence amplifier}, and to apply some suitable frequency filter \cite{APFPS20,YMC+11}. This means that the processes $X_j(t)$ are modified by a second convolution; however, this procedure is equivalent to a single convolution with a modified response function and its effects are included in the structure \eqref{Xprocs}. This electronic postprocessing gives the dimensions of a voltage to the observed processes; so, the physical dimension of the parameters $\xi_j$ is that of a time multiplied by a voltage.
The second amplification/deamplification stage can be used also to scale the total output signal.
If some amplification is introduced also before the difference is taken, we can partially tune the coefficients $\xi_j$, in order to get a better balancing in the final output.
\end{remark}

A typical choice is to include in $h(t)$ a Butterworth filter \cite{APFPS20}; as a simple example we shall use an exponential response function
\begin{equation}\label{hexp}
h(t)=\varkappa\rme^{-\varkappa t}.
\end{equation}

As discussed again in \cite{APFPS20}, the amplification process inside the photodiodes produces some extra additive noise.
Being additive and dependent only on the response function, we do not consider the electronic noise in the following, but we add it only in Sec.\ \ref{sec:totminentr}.

Again the probability law of the stochastic processes is uniquely determined by their characteristic functional (Appendix \ref{sec:PhiX}); here below we give means and correlations.

\subsubsection{Means and correlations of the observed processes}\label{sec:E+Cov}
The first moments can be obtained directly from the definition \eqref{Xprocs} and the moments of the photocurrents \eqref{mean:M+}, \eqref{corr:M+}.
Firstly, we define the following coefficients:
\begin{subequations}\label{kappas}
\begin{gather}
\kappa_{11}=\eta_1\left[\eta_3\epsilon_1\xi_1^{\,2}+\left(1-\eta_3\right)\epsilon_3\xi_3^{\,2}\right] ,
\qquad
\kappa_{21}=\left(1-\eta_1\right) \left[\eta_4\epsilon_2\xi_2^{\,2}+\left(1-\eta_4\right)\epsilon_4\xi_4^{\,2}\right],
\\
\kappa_{12}=\eta_2\left[\left(1-\eta_3\right)\epsilon_1\xi_1^{\,2}+\eta_3\epsilon_3\xi_3^{\,2}\right] ,
\qquad
\kappa_{22}=\left(1-\eta_2\right) \left[\left(1-\eta_4\right)\epsilon_2\xi_2^{\,2}+\eta_4\epsilon_4\xi_4^{\,2}\right] ,
\\
\kappa_{13}=\sqrt{\eta_1\eta_2\eta_3\left(1-\eta_3\right)}\left(\epsilon_1\xi_1+\epsilon_3\xi_3\right),
\qquad
\kappa_{23}=\sqrt{\left(1-\eta_1\right)\left(1-\eta_2\right)\eta_4\left(1-\eta_4\right)}\left(\epsilon_2\xi_2+\epsilon_4\xi_4\right),
\end{gather}
\end{subequations}
\begin{subequations}\label{Deltas}
\begin{gather}
\Delta_{11}=\eta_1\left[\eta_3\epsilon_1\xi_1-\left(1-\eta_3\right)\epsilon_3\xi_3\right] ,
\qquad
\Delta_{21}=\left(1-\eta_1\right) \left[\eta_4\epsilon_2\xi_2 -\left(1-\eta_4\right)\epsilon_4\xi_4\right],
\\
\Delta_{12}=\eta_2\left[\left(1-\eta_3\right)\epsilon_1\xi_1- \eta_3\epsilon_3\xi_3\right] ,
\qquad
\Delta_{22}=\left(1-\eta_2\right) \left[\left(1-\eta_4\right)\epsilon_2\xi_2 -\eta_4\epsilon_4\xi_4\right] ,
\\
\Delta_{13}=\sqrt{\eta_1\eta_2\eta_3\left(1-\eta_3\right)}\left(\epsilon_1\xi_1^{\;2}- \epsilon_3\xi_3^{\;2}\right),
\qquad
\Delta_{23}=\sqrt{\left(1-\eta_1\right)\left(1-\eta_2\right)\eta_4\left(1-\eta_4\right)}\left(\epsilon_2\xi_2^{\;2} -\epsilon_4\xi_4^{\;2}\right);
\end{gather}
\end{subequations}
Then, by using the field state introduced in Sec.\ \ref{sec:sysst}, the definitions \eqref{:gs}, and the notation $:\bullet :$ for normal order,  we get easily, for $i,j =1,2$,
\begin{equation}\label{EXprocs}
\Ebb_P\left[X_{j}(t)\right]=\int_0^t\rmd r\,h(t-r)\Bigl\{\kappa_{j3} \left(\rmi \rme^{\rmi \psi_j}\Ebb_f\left[f(r)\langle a_1^\dagger(r)\rangle_T^f\right] +\text{c.c.}\right)
+\Delta_{j1}\langle a_1^\dagger(r)a_1(r)\rangle_T
+ \Delta_{j2}\Ebb_f\left[\abs{f(r)}^2\right]\Bigr\},
\end{equation}
\begin{multline}\label{XXcorr}
\Ebb_P\left[X_{j}(t)X_{i}(s)\right]=\delta_{ij}\int_0^{t\wedge s}\rmd r\, h(t-r)h(s-r)\Bigl\{\kappa_{j1}\langle a_1^\dagger(r)a_1(r)\rangle_T
+\kappa_{j2} \Ebb_f\left[\abs{f(r)}^2\right]
\\ {} +\Delta_{j3} \left(\rmi \rme^{\rmi \psi_j}\Ebb_f\left[f(r)\langle a_1^\dagger(r)\rangle_T^f\right] +\text{c.c.}\right)\Bigr\}
\\ {}+
\int_0^t\rmd r\int_0^s\rmd r'\,h(t-r)h(s-r')  \Ebb_f\Big[\big\langle : \Bigl\{\Delta_{j1}a_1^\dagger(r)a_1(r)+ \Delta_{j2}\abs{f(r)}^2
+\kappa_{j3}\left(\rmi \rme^{\rmi \psi_j}f(r) a_1^\dagger(r) +\text{h.c.}\right)\Bigr\}
\\ {}\times \Bigl\{\Delta_{i1}a_1^\dagger(r')a_1(r')
+\Delta_{i2}\abs{f(r')}^2
+\kappa_{i3}\left(\rmi \rme^{\rmi \psi_i}f(r') a_1^\dagger(r') +\text{h.c.}\right)\Bigr\}: \big\rangle_T^f\Big].
\end{multline}
By these two equations one can write also the expression of $\Cov_P\left[X_{j}(t),\,X_{i}(s)\right]$.

\begin{remark}\label{rem:fasi+} Let us stress that the term starting by $\delta_{ij}$ in \eqref{XXcorr} comes out from the reordering of the creation and annihilation operators and, so, it represents the shot noise.
Moreover, note that the phases $\psi_j$ appear only in the terms containing the interference between signal and LO; so, they disappear when the signal vanishes.
\end{remark}

In the expression of the  means \eqref{EXprocs}, the last two terms are due to some imbalance in the circuit, while the first term is an indication  that the whole detection apparatus is a way to ``measure'' two quadratures of the signal field operator $a_1(t)$, even complementary quadratures (when $\phi\equiv\psi_2-\psi_1=\pm \pi/2$). The fact that these two observables do not commute implies the presence of some extra noise.

\subsubsection{The balanced case} \label{sec:perfectbc}
To gain some inside on the effects of imperfections it is useful to fix the perfectly balanced case:
\begin{equation}\label{perfectbc}
\eta_j=1/2, \qquad \epsilon_j=\epsilon, \qquad \xi_j= \xi.
\end{equation}
This gives
\begin{equation}
\kappa_{j1}= \kappa_{j2}=\epsilon\xi^2/2, \qquad \kappa_{j3}=\epsilon\xi/2, \qquad \Delta_{ij}=0.
\end{equation}

As an example, equations \eqref{EXprocs}, \eqref{XXcorr} become
\begin{equation}\label{EXprocsequi}
\Ebb_P\left[X_{j}(t)\right]=\frac{\epsilon\xi}2\,\rmi \rme^{\rmi \psi_j}\int_0^t\rmd r\,h(t-r) \Ebb_f\left[f(r)\langle a_1^\dagger(r)\rangle_T^f\right] +\text{c.c.},
\end{equation}
\begin{multline}\label{XXcorrequi}
\Ebb_P\left[X_{j}(t)X_{i}(s)\right]=\delta_{ij}\,\frac{\epsilon\xi^2}2\int_0^{t\wedge s}\rmd r\, h(t-r)h(s-r)\Bigl\{\langle a_1^\dagger(r)a_1(r)\rangle_T
+\Ebb_f\left[\abs{f(r)}^2\right]\Bigr\}
\\ {}+
\frac{\epsilon^2\xi^2}4 \int_0^t\rmd r\int_0^s\rmd r'\,h(t-r)h(s-r')  \Ebb_f\Big[\big\langle : \left(\rmi \rme^{\rmi \psi_j}f(r) a_1^\dagger(r) +\text{h.c.}\right)
\left(\rmi \rme^{\rmi \psi_i}f(r') a_1^\dagger(r') +\text{h.c.}\right): \big\rangle_T^f\Big].
\end{multline}

\begin{remark}[Rebalancing]\label{rebalance} An intermediate case, suggested in \cite{APFPS20}, is when we are able to fine tune the efficiencies $\epsilon_j$ and/or the coefficients $\xi_j$ (see Remarks \ref{effreg}, \ref{F,Alm}),  and we get $\Delta_{j2}=0$. In this way we would eliminate the problem of a mean growing with the LO intensity (the last term in \eqref{EXprocs}). At the same time the contribution of  the laser intensity noise $\Cov_f\left[\abs{f(r)}^2,\abs{f(r')}^2\right]$ would disappear from the  fluctuations of the observed processes.
\end{remark}

\section{Strong LO and double homodyne detection}\label{sec:Qj}

When the interest is in the measurement of the two signal quadratures or in the secure QRNG protocol introduced in \cite{Vill18}, we need to have a strong LO and to scale the observed processes.
\begin{assumption}\label{ass:constantint}
We assume the LO laser to have a constant mean intensity, as it holds for the model of Sec.\ \ref{sec:LOstate},
\begin{equation}\label{Yproc?}
\Ebb_f\left[\abs{f(t)}^2\right]=\abs\lambda^2,
\end{equation}
and we set
\begin{equation}\label{LOftilde}
\tilde f(t)= f(t)\big/\abs\lambda.
\end{equation}
\end{assumption}
Then, we scale the outputs with respect to this mean intensity:
\begin{equation}\label{Yproc}
Y_j(t)=\frac{X_j(t)}{\abs\lambda}=\int_0^t \frac{h(t-s)}{\abs\lambda}\left[\xi_j\rmd N_j(s) - \xi_{j+2}\rmd N_{j+2}(s)\right] .
\end{equation}
Recall that also the probability law $P$ of the counting processes $N_j(t)$ depends on $\lambda$, because the field state depends on $f$, see Sec.\ \ref{sec:sysst}.

From \eqref{EXprocs} we get
\begin{equation}\label{EYprocs}
\Ebb_P\left[Y_{j}(t)\right]=\int_0^t\rmd r\,h(t-r)\Bigl\{\kappa_{j3} \left(\rmi \rme^{\rmi \psi_j}\Ebb_f\left[\tilde f(r)\langle a_1^\dagger(r)\rangle_T^f\right] +\text{c.c.}\right)
+\frac{\Delta_{j1}}{\abs\lambda}\langle a_1^\dagger(r)a_1(r)\rangle_T
+ \Delta_{j2}\abs\lambda\Bigr\}.
\end{equation}
If $\Delta_{j2}\neq 0$, the last term explodes when the laser intensity grows.
We do not assume to be able to get a perfect rebalancing as in Remark \ref{rebalance}, but only to limit the growth with the laser intensity of the terms containing the expression $\Delta_{j2}$, so to have a moderate effect of the laser fluctuations at the working intensity of LO. By working at vanishing signal, we manage to tune the efficiency and transmissivity coefficients so to maintain a small mean up to the maximal LO power; this means to have $\Delta_{j2}\propto 1/\abs\lambda_{\max}$.

\begin{assumption}\label{assGj2} In mathematical terms, we set
\begin{equation}\label{:G_}
G_{j2}={\Delta_{j2}\abs\lambda}\big/{\kappa_{j3}}
\end{equation}
and we assume $G_{j2}$  to be independent  of $\lambda$. We also take  $\eta_j$ different from $0$ and $1$.
\end{assumption}

Now, we can take the limit $\abs\lambda\to +\infty$ and the mean \eqref{EYprocs} becomes
\begin{equation}\label{EYprocslim}
\Ebb_P\left[Y_{j}(t)\right]=\kappa_{j3}\int_0^t\rmd r\,h(t-r)\Bigl\{\left(\rmi \rme^{\rmi \psi_j}\Ebb_f\left[\tilde f(r)\langle a_1^\dagger(r)\rangle_T^f\right] +\text{c.c.}\right)+G_{j2}
\Bigr\},
\end{equation}
while \eqref{XXcorr}, \eqref{Yproc}, \eqref{:G_} give, by direct calculations,
\begin{multline}\label{CovYY}
\Cov_P\left[Y_{j}(t),\,Y_{i}(s)\right]=\delta_{ij}\kappa_{j2}
\int_0^{t\wedge s} h(t-r)h(s-r)\,\rmd r
\\ {}+
\kappa_{j3}\kappa_{i3}\int_0^t\rmd r\int_0^s\rmd r'\,h(t-r)h(s-r')\left[
C_f(j,r;i,r')+\mathtt{C}(j,r;i,r')\right],
\end{multline}
where the first term is the shot noise contribution and the two matrices in the double integral have the expressions
\begin{equation}\label{Cfmatrix}
C_f(j,r;i,r') =
\Cov_f\Big[G_{j2}\abs{\tilde f(r)}^2
+\left(\rmi \rme^{\rmi \psi_j}\tilde f(r) \langle a_1^\dagger(r)\rangle_T^f +\text{c.c.}\right),\;
G_{i2}\abs{\tilde f(r')}^2
+\left(\rmi \rme^{\rmi \psi_i}\tilde f(r')\langle  a_1^\dagger(r')\rangle_T^f +\text{c.c.}\right)\Big],
\end{equation}
\begin{multline}\label{Cmatrix}
\mathtt{C}(j,r;i,r')=\Ebb_f\Bigl[ \rme^{\rmi\left(\psi_j-\psi_i\right)}\tilde f(r) \overline{\tilde f(r')}\left(\langle a_1^\dagger(r)a_1(r')\rangle_T^f- \langle a_1^\dagger(r)\rangle_T^f\langle a_1(r')\rangle_T^f\right)\\ {}- \rme^{\rmi\left(\psi_i+\psi_j\right)}\tilde f(r) \tilde f(r')\left(\langle a_1^\dagger(r)a_1^\dagger (r')\rangle_T^f- \langle a_1^\dagger(r)\rangle_T^f\langle a_1^\dagger(r')\rangle_T^f\right) +\text{c.c.}\Bigr].
\end{multline}
The matrix \eqref{Cmatrix} vanishes when the signal is in a coherent state, and it can be not positive definite when the signal is in a squeezed state. It is important to stress that only the sum of the shot noise plus the contribution of \eqref{Cmatrix} is guaranteed to be a positive definite matrix.

\subsection{The reduced description and the probability law in the limit of strong LO}\label{sec:reddescr}
Up to now we have considered only observables represented by commuting self-adjoint operators and the associated pvm in the Fock space $\Gamma$, an Hilbert  space with the tensor product structure \eqref{Focksp}. Indeed, also to the processes $Y_j(t)$ we can associate the self-adjoint operators $\hat Y_j(t)$ by the operator version of \eqref{Yproc}; these are again compatible observables, as they are linear combinations of the compatible number operators, see Remark \ref{rem:compatible}.
POVMs enter into play when a reduced description is considered: the characteristic operator is reduced to the factor $\Gamma_1$, where the signal lives,  by using the fact that the state in the complementary factor is fixed. By this technique we can show that the limit for
$\abs\lambda\to +\infty$ exists for the whole law of the processes $Y_j(\bullet)$, not only for the first moments, and that this probability law is linked to a POVM describing the joint measurement of two quadratures of the signal field.

To have a clear picture of the underlying POVM, it is convenient to suitably decompose the coefficients $\kappa_{j2}$, which appear in the shot noise term in \eqref{CovYY}. By direct computations based on \eqref{kappas}, \eqref{Deltas}, one can check that the following equality holds:
\begin{equation}\label{j2decomp}
\kappa_{j2}=\kappa_{j3}^{\;2}\left(G_{j3}+V_j^{\,2}+\sigma_j^{\,2}+1\right),
\end{equation}
where
\begin{subequations}\label{G+V+s}
\begin{equation}\label{GsVs}
G_{13}=\frac{1-\eta_1}{\eta_1}, \qquad G_{23}
=\frac{\eta_1}{1-\eta_1},
\end{equation}
\begin{equation}
V_1^{\,2}= \frac{\Delta_{12}^{\;2}}{\eta_2\kappa_{13}^{\;2}}=\frac{\tilde V_1^{\,2}}{\eta_1},
\qquad V_2^{\,2}= \frac{\Delta_{22}^{\;2}}{\left(1-\eta_2\right)\kappa_{23}^{\;2}}=\frac{\tilde V_2^{\,2}}{1-\eta_1},
\end{equation}
\begin{equation}
\tilde V_j^{\,2}= \frac{\left[\left(1-\eta_{j+2}\right) \epsilon_j\xi_j -\eta_{j+2} \epsilon_{j+2}\xi_{j+2}\right]^2} {\eta_{j+2}\left(1-\eta_{j+2}\right)\left(\epsilon_j\xi_j+\epsilon_{j+2}\xi_{j+2}\right)^2},
\end{equation}
\begin{equation}\label{sigmas}
\sigma_1^{\,2}=\frac{\tilde \sigma_1^{\,2}}{\eta_1},\qquad \sigma_2^{\,2}=\frac{\tilde \sigma_2^{\,2}}{1-\eta_1},
\qquad  \tilde \sigma_j^{\,2}= \frac{ \frac{\epsilon_j\left(1- \epsilon_j\right)}{\eta_{j+2}}\xi_j^{\,2} +\frac{\epsilon_{j+2}\left(1- \epsilon_{j+2}\right)}{1-\eta_{j+2}}\xi_{j+2}^{\;2}} {\left(\epsilon_j\xi_j+\epsilon_{j+2}\xi_{j+2}\right)^2}.
\end{equation}
\end{subequations}
The contribution $\sigma_j^{\,2}$ represents some extra-noise due to the presence of optical losses, as it vanishes when $\epsilon_j=\epsilon_{j+2}=1$; this noise is of quantum origin because it is due to the need of preserving CCRs as discussed in Sec.\ \ref{sec:efficiency}. The contribution $V_j^{\,2}$ is due to the unbalancing and, by Assumption \ref{assGj2}, it  is small.

\begin{proposition} \label{prop:Y} With the above assumptions, the characteristic functional of the processes $Y_j(\bullet)$ admits  a limit for
$\abs\lambda\to +\infty$; so, we can write
\begin{equation}\label{PhiYlim}
\Phi_T^Y[\vec k]= \lim_{\abs\lambda \to +\infty} \Ebb_P\bigg[\exp\bigg\{\frac\rmi{\abs\lambda}\sum_{j=1}^2\int_0^T k_j(s)X_j(s)\rmd s\bigg\}\bigg].
\end{equation}
The structure of the characteristic functional \eqref{PhiYlim} turns out to be given by
\begin{subequations}\label{Yprob}
\begin{equation}\label{YintZ}
\Phi^Y_T[\vec k]= \Ebb_f\left[\Phi^Z_T[\vec k^T;f]\right],
\qquad k_j^T(t)=\kappa_{j3}\int_t^T h(s-t)k_j(s)\rmd s,
\end{equation}
\begin{equation}\label{GammaPsi}
\Phi^Z_T[\vec k;f]=\Phi^Q_T[\vec k;f]\Gamma_T[\vec k;f],
\end{equation}
\begin{equation}\label{PhiQ}
\Phi^Q_T[\vec k;f]=\Tr_{\Gamma_1}\left\{\hat \Psi^Q_T[\vec k;f]\rho^{f,T}_{1}\right\},
\end{equation}
\begin{equation}\label{Gamma}
\Gamma_T[\vec k;f]= \exp\biggl\{\int_0^T\rmd s\abs{\tilde f(s)}^2\sum_{j=1}^2\biggl[\rmi k_j(s)   G_{j2}
-\frac{\sigma_j^2+V_j^{\;2}}2\,k_j(s)^2\biggr]\biggr\},
\end{equation}
\end{subequations}
With a fixed $\tilde f(t)$, the quantity $\hat\Psi_T^Q[\vec k;f]$ is the characteristic operator of a POVM on the Hilbert space $\Gamma_1$, having the expression
\begin{subequations}\label{Psi=GP0}
\begin{equation}\label{Psif}
\hat\Psi_T^Q[\vec k;f]
= \exp\int_0^T\biggl\{\rmi \sum_{j=1}^2k_j(t)\rmd \hat Q_j(t)
-\frac12\sum_{i,j=1}^2k_j(t)\Xi_{ji}k_i(t)\abs{\tilde f(t)}^2\rmd t\biggr\},
\end{equation}
\begin{equation}\label{QjXi}
\rmd \hat Q_j(t)=\rmi\rme^{\rmi\psi_j}\tilde f(t)\rmd A_1^\dagger(t) +\text{\rm h.c.}, \qquad \Xi =\begin{pmatrix} G_{13} & -\cos\phi \\ -\cos \phi & G_{23}\end{pmatrix}.
\end{equation}
\end{subequations}
In the formulae above, $\tilde f$ and $G_{j2}$ are introduced in Assumptions \ref{ass:constantint} and \ref{assGj2}, the phase $\phi$ is defined in \eqref{phij},  the expressions of $\kappa_{j3}$ and $\sigma_j^{\,2}$ are given in \eqref{kappas}, \eqref{sigmas}; the matrix $\Xi$ is non-negative definite.
\end{proposition}
The proof of the proposition above is given in Appendix \ref{app:strLO}, where also the following Corollary is proved.

\begin{corollary}\label{corroll}
Let the signal state be a mixture of coherent states, given by $\rho^{f,T}_{1}\to\rho^{\fs,T}_{1}=|e_1(\fs)\rangle \langle e_1(\fs)|$, where $\fs(t)$ is a stochastic process and $P_{f}$ is the joint probability distribution of the processes $\fs$ and $\tilde f$, as in  Appendix \ref{sec:0N}. By using the notations of Proposition \ref{prop:Y}, we have
\begin{equation}\label{PhiZcoer}
\Phi_T^Z[\vec k;f]
= \exp\sum_{j=1}^2\int_0^T\rmd t \biggl\{\rmi k_j(t)\biggl[G_{j2}\abs{\tilde f(t)}^2+\left(\rmi \rme^{\rmi\psi_j}\overline{\fs(t)}\tilde f(t)+{\rm c.c.}\right)\biggr]
-\frac{\kappa_{j2}}{2\kappa_{j3}^{\;2}}\,k_j(t)^2\abs{\tilde f(t)}^2\biggr\}.
\end{equation}
\end{corollary}

\subsubsection{The structure of the observed processes}\label{sec:strY}

When $\abs{\tilde f(s)}$ is a non random given function, the expression \eqref{Gamma} is the characteristic functional of the increments of a bidimensional Gaussian process, which we denote by $L_j^f(t)$, which can be expressed as
\begin{equation}\label{Lproc}
L_j^f(t)= G_{j2}\int_0^t\abs{\tilde f(s)}^2 \rmd s +\sigma_j\int_0^t\abs{\tilde f(s)}\rmd W_{j1}(s)+V_j\int_0^t\abs{\tilde f(s)}\rmd W_{j2}(s),
\end{equation}
where the four processes $W_{ji}$ are independent, standard Wiener processes (their formal time derivatives are white noises).

Let us consider now the POVM (in continuous time) determined by the characteristic operator \eqref{Psi=GP0} and let us denote by $Q_j^f(t)$
the corresponding observables, measured in the signal state $\rho^{f,T}_{1}$. Then, the quantity $\Phi^Q_T[\vec k;f]$, given by \eqref{PhiQ}, is the characteristic functional of the random processes $Q_j^f(t)$.

As the product of two characteristic functionals is the characteristic functional of the sum process, the quantity
$\Phi^Z_T[\vec k;f]$ in \eqref{GammaPsi}  is the characteristic functional of the processes
\begin{equation}\label{Zstr}
Z_j^f(t) = Q_j^f(t)+L_j^f(t)
\end{equation}
Then, $\Phi^Z_T[\vec k^T;f]$, with $\vec k^T$ defined in \eqref{YintZ}, is the characteristic functional of the processes
\begin{equation}\label{Yfstr}
Y_j^f(t)=\kappa_{j3}\int_0^t h(t-s)\rmd Z_j^f(s).
\end{equation}

Let us summarize the physical meaning of these results.  The first equality in \eqref{YintZ} says that
the probability law of the observed processes $Y_j(t)$ is a \emph{mixture} of the probability laws of the processes $Y_j^f(t)$  with respect to the law of the random function $f$ (see Remark \ref{rem:P}), i.e. with respect to the LO fluctuations. Then, by the results \eqref{Yfstr} and \eqref{Zstr}, the processes $Y_j^f(t)$ are a smoothed and noisy versions of the processes $Q^f_j(t)$, whose characteristic functional is given by \eqref{PhiQ} and the associated characteristic operator by \eqref{Psif}. Finally, by the structure of this characteristic operator \eqref{Psif}, we can interpret the associated POVM as a joint measurement of the ``dilated quadratures'' \eqref{QjXi}. In some sense, the whole apparatus gives a joint noisy measurement of the field quadratures \eqref{QjXi}.

\subsubsection{The measured quadratures}
The measured field quadratures \eqref{QjXi} satisfy the commutation relations
\begin{equation}\label{Qjcommut}
\Big[\frac{\rmd \hat Q_1(t)}{\rmd t},\,\frac{\rmd \hat Q_2(s)}{\rmd s}\Big]=2\rmi \sin \phi \abs{\tilde f(t)}^2\delta(t-s), \qquad \left[ \hat Q_j(t),\, \hat Q_j(s)\right]=0.
\end{equation}
Note that these quadratures are not complementary when $\abs{\sin \phi}\neq 1$ and that they are random operators, as they contain the random process $\tilde f(t)$ \eqref{LOf+}, \eqref{LOftilde}.
For $\sin \phi \neq 0$ the two quadratures do not commute and a joint pvm does not exist; indeed, the second term in \eqref{Psif} represents some noise needed to have a true POVM. However, in general, this noise is not minimal. For instance,
for $\phi=n\pi$ the two quadratures \eqref{QjXi} are compatible quantum observables and the whole noise in \eqref{Psif} is due to the way the measurement is realized, but it is not necessary on a pure mathematical ground.

For $\phi=n\pi+\pi/2$, instead, the quadratures \eqref{QjXi} are orthogonal and \eqref{Psif} can be written as
\begin{equation}
\hat\Psi_T^Q[\vec k;f]= \exp\int_0^T\biggl\{\rmi \left[\tilde f(t)k(t)\rmi \rme^{\rmi \psi_2}\rmd A_1^\dagger(t) +\text{h.c.}\right]
-\frac12\abs{\alpha k(t)+ \alpha\left(1-2\eta_1\right) \overline{k(t)}}^2\abs{\tilde f(t)}^2\rmd t\biggr\},
\end{equation}
\begin{equation*}
k(t)=k_1(t)+\rmi k_2(t), \qquad \alpha= \frac 1 {\sqrt{4\eta_1\left(1-\eta_1\right)}} \qquad \Rightarrow \qquad \alpha\geq 1, \quad \alpha\abs{1-2\eta_1}=\sqrt{\alpha^2-1}.
\end{equation*}
The associated POVM was already introduced in  \cite[Sec.\ III]{Bar86}; the case with $\alpha=1$, i.e. $\eta_1=1/2$, it is a generalization to fields in continuous time of the POVM constructed from coherent states of a single mode, having the Husimi function as probability density.

\subsection{Homodyne and heterodyne detection}\label{sec:dyning}

As observed in Remark \ref{rem:fasi+}, the phases $\psi_j$ disappear from the moments of the observed processes in the case of vanishing signal. Similarly, also the contributions of the LO phase fluctuations disappear from the distributions of the processes $X_j(\bullet)$, $Y_j(\bullet)$, see \eqref{EXprocs}, \eqref{XXcorr}, \eqref{EYprocslim}--\eqref{Cmatrix}, \eqref{ZtoY}, \eqref{Psi=GP0}, \eqref{Qjcommut}. This is due to the fact that we are taking equal the optical paths arriving to the various interference points (the beam splitters). Things are different when the circuit is used as detection apparatus and there is interference between LO and signal. To detect squeezing in homodyne spectra, it is necessary to maintain phase coherence for a sufficiently long time; the LO must be phase locked to the signal \cite {RCCBS95} and this can be realized by taking as signal the light generated by some system stimulated by the same laser which produces the LO (mathematically, both LO and signal depend on $f$ \cite{BG12}, see Remark \ref{rem:phloc}).

To understand the effect of the phase fluctuations let us consider the case of a signal state totally independent of $f$; we take the laser model of Secs.\ \ref{sec:LOstate} and \ref{sec:LOprop} and we study means and fluctuations of the processes $Y_j(\bullet)$. In Eqs.\ \eqref{EYprocslim}--\eqref{Cmatrix} the $f$-dependence in the signal state disappears; then, by the expressions of the $f$-moments \eqref{Efff}, we get
\begin{subequations}\label{nof}
\begin{equation}\label{Enof}
\Ebb_P\left[Y_{j}(t)\right]=\kappa_{j3}\int_0^t\rmd r\,h(t-r)\Bigl\{w \rme^{-\gamma_0r}\left(\rmi \rme^{\rmi \left(\psi_j + \theta -\omega_0r\right)} \langle a_1^\dagger(r)\rangle_T +\text{c.c.}\right)+G_{j2} \Bigr\},
\end{equation}
\begin{multline}\label{Cfnof}
C_f(j,r;i,r')=G_{j2}G_{i2}\,\frac{v(r-r')}2 \left[2w^2+v(r-r')\right]+2\rmi G_{j2}v(r-r')w\rme^{\rmi\left(\psi_i +\theta -\omega_0 r'\right)-\gamma_0r'}\langle a_1^\dagger(r')\rangle_T
\\ {} +2\rmi G_{i2}v(r-r')w\rme^{\rmi\left(\psi_j+\theta -\omega_0 r\right)-\gamma_0r}\langle a_1^\dagger(r)\rangle_T
\\ {}+\rme^{\rmi \left(\psi_i+\psi_j+2\theta\right)-\left(\rmi\omega_0+\gamma_0\right)\left(r+r'\right)}\langle a_1^\dagger(r)\rangle_T \langle a_1^\dagger(r')\rangle_T
\left[w^2-\left(w^2+v(r-r')\right)\rme^{-2\gamma_0\left(r\wedge r'\right)}\right]
\\ {}+ \rme^{\rmi \left(\psi_j-\psi_i\right)+ \rmi\omega_0\left(r'-r\right)-\gamma_0\abs{r-r'}}\langle a_1^\dagger(r)\rangle_T \langle a_1(r')\rangle_T
\left[w^2+v(r-r')- w^2\rme^{-2\gamma_0\left(r\wedge r'\right)}\right] +\text{c.c.},
\end{multline}
\begin{multline}\label{Cnof}
\mathtt{C}(j,r;i,r')=\rme^{\rmi\left(\psi_j-\psi_i+\omega_0\left(r'-r\right)\right)-\gamma_0\abs{r-r'}}\left(w^2+v(r-r')\right)\left(\langle a_1^\dagger(r)a_1(r')\rangle_T- \langle a_1^\dagger(r)\rangle_T\langle a_1(r')\rangle_T\right)\\ {}- \rme^{\rmi\left[\psi_i+\psi_j+ 2\theta-\omega_0\left(r+r'\right)\right]- \gamma_0\left[4\left(r\wedge r'\right)+\abs{r-r'}\right]} \left(w^2+v(r-r')\right) \left(\langle a_1^\dagger(r)a_1^\dagger (r')\rangle_T- \langle a_1^\dagger(r)\rangle_T\langle a_1^\dagger(r')\rangle_T\right) +\text{c.c.}.
\end{multline}
\end{subequations}

Let us consider now the extreme case of an incoherent LO (the coherence time $1/\gamma_0$ is small); this means to take
the relevant times to be large: $t,s\gg 1/\gamma_0$, \ $t,s\gg \tau$, where $\tau$ is the decaying time of $h(t)$. From Eqs.\ \eqref{nof}, \eqref{CovYY} we get
\begin{subequations}\label{purehetero}
\begin{equation}\label{heteromean}
\Ebb_P\left[Y_{j}(t)\right]=\kappa_{j3}\int_0^t\rmd r\,h(t-r)G_{j2}\simeq \kappa_{j3}G_{j2},
\end{equation}
\begin{multline}\label{covhet}
\Cov_P\left[Y_{j}(t),Y_{i}(s)\right]\simeq\delta_{ij}\kappa_{j2}\int_0^{t\wedge s} h(t-r)h(s-r)\,\rmd r + \kappa_{j3}G_{j2}\kappa_{i3}G_{i2}G_v(t,s)
\\ {}+ \kappa_{j3}\kappa_{i3}
\int_0^t\rmd r\int_0^s\rmd r'\,h(t-r) h(s-r')
\rme^{-\gamma_0\abs{r-r'}}
\left[\rme^{\rmi\left(\psi_j-\psi_i+\omega_0\left(r'-r\right)\right)}\langle a_1^\dagger(r)a_1(r')\rangle_T \left(w^2+v(r-r')\right) +\text{c.c.} \right],
\end{multline}
\begin{equation}\label{def:Gv}
G_v(t,s)=\int_0^t\rmd r\int_0^s\rmd r'\,h(t-r)h(s-r') v(r-r') \left[2w^2+v(r-r')\right].
\end{equation}
\end{subequations}
In this limit, the terms with two annihilators or two creators disappear from the matrix \eqref{Cnof} and each of the three terms \eqref{covhet} is a  positive definite matrix for every choice of the signal state; so, it is not possible to detect squeezing. We can say that this is the limit of heterodyne detection: the phase fluctuations in the LO destroy any phase coherence \cite{BG12,WisM10}. Let us note that in \eqref{CovYY}, \eqref{Cmatrix} the contribution of
$\langle a_1^\dagger(r) a_1^\dagger(r')\rangle_T$  is cut again when its frequency spectrum is far from $\omega_0$; we can speak of heterodyne detection also in this case.

The other extreme case is pure homodyning. Without signal/LO correlations to attenuate the coherence losses, we need to have a highly coherent LO ($1/\gamma_0$ is very large), say $\gamma_0=0$ or $t,s \ll 1/\gamma_0$.
Now, the terms with two annihilators or two creators survive and the possible presence of negative eigenvalues of the matrix $\mathtt{C}(j,r;i,r')$ can allow to detect squeezing \cite{BarG13,BG21}.

Let us stress that we are speaking of homodyne detection when the measurement is phase sensitive, independently form the fact that a single quadrature or two orthogonal ones are monitored; essentially this terminology is used also in \cite{WisM10,BKS14,FOP05}, while in \cite{Vill18} they use the term ``heterodyne detection'' for the case of monitoring of two complementary quadratures.

An application of detection in continuous time is to construct a consistent quantum theory of  the various types of spectra for the signal.
In the heterodyne regime, by varying the frequency of the LO,  the  \emph{power spectrum} of the signal can be studied \cite{WalM94,Car08,BarG13}, while in  the homodyne regime we have  the \emph{homodyne spectrum} and the \emph{spectrum of squeezing} \cite{WalM94,Car08,BG21,BarG13}. By using the eight-port circuit we get  in the same run the spectra of two complementary quadratures.

\section{Discrete sampling} \label{dsampling}

The output of the apparatus of Fig.\ \ref{fig:optcir} is in continuous time, but in many experimental situations it is sampled at discrete times. This can happen both when it is used for QRNG \cite{Vill18,+AS+18}, and when it is used as a detection apparatus for the signal quadratures, as it is done for the outputs of various optical circuits \cite{Botta08}. Sometimes a pulsed laser can be used as LO \cite{Smith2019,RCCBS95,YMC+11,APFPS20}; in this situation, even a single detection per pulse is considered.

For instance, in the case of QRNG, the output of the difference photocurrents (the couple of processes $X_j(t)$) is sampled by an oscilloscope; the laser can be either CW or pulsed. The sampling rate will determine the maximum generation rate of the random numbers, while the ADC that digitize the signal determines the number of bits per sample and also limits the maximum level of output photocurrents that can be used without saturating the instrument.

In QRNG we need the samples at different times to be independent; eventually, this result can be obtained by undersampling when some correlation is detected in the experimental data. In view of this, we take the sampling times such that correlations for observations at different times can be due only to the signal. This choice is not necessary when the apparatus is used for signal detection, but even in this case it simplifies some mathematical expressions.

\begin{assumption} \label{ass:sampling} We assume that the sampling process is done at $m$ times $t_1, t_2,\ldots, t_m $ with $t_1>t_0\geq 0$ and
with an intertime
$t_l-t_{l-1}\geq \tau$, where $\tau $ is such that the response function and the intensity correlations are nearly completely decayed:
\begin{equation}\label{Covsimeq0}
h(t)\simeq 0, \qquad t\geq \tau,  \qquad \qquad \Cov_f[\abs{f(t)}^2,\abs{f(s)}^2]\simeq 0,\qquad \abs{t-s}\geq \tau.
\end{equation}
\end{assumption}
For the laser model of Sec.\ \ref{sec:LOstate}, by  Eq.\ \eqref{Efff3}, the decaying of the correlations  is equivalent to $v(t)\simeq 0$ for $\abs t\geq \tau$.

\subsection{Observed processes}\label{sec:sop}
We start by considering the sampling of the observed processes $X_j(\bullet)$, introduced in Sec.\ \ref{sec:Xj}.
Means and correlations of the observed samples can be obtained by Eqs.\ \eqref{EXprocs}, \eqref{XXcorr}. By Assumption \ref{ass:sampling}, the covariance of two observations at different times can be different from zero only due to correlations introduced by the signal state. In particular, when the signal state is a coherent one, the covariance of observations at different times vanishes.

For QRNG it is of main interest the case of pure noise; when the signal is in the vacuum state,  from \eqref{EXprocs}, \eqref{XXcorr}, \eqref{Yproc?}, \eqref{LOftilde}, \eqref{nRIN} we get
\begin{subequations}\label{vaX}
\begin{equation}\label{muX}
\Ebb_P\left[X_{j}(t_l)\right] \simeq \Delta_{j2}\abs\lambda^2\int_{t_l-\tau}^{t_l}\rmd r\,h(t_l-r) \simeq \Delta_{j2}\abs\lambda^2,
\end{equation}
\begin{equation}\label{CovC0}
\Cov_P[X_j(t_l),\,X_i(t_{l'})]\simeq
\delta_{ll'} \left(\delta_{ij}\kappa_{j2}\abs\lambda^2 S_0
+\Delta_{j2}\Delta_{i2}\abs{\lambda}^4 C_0\right),
\end{equation}
\begin{equation}\label{def:S_0}
S_0=\int_{t_l-\tau}^{t_l}\rmd r\, h(t_l-r)^2\simeq \int_0^{+\infty}\rmd r\, h(r)^2,
\end{equation}
\begin{equation}\label{def:C_0}
C_0=\int_{t_l-\tau}^{t_l}\rmd r\int_{t_l-\tau}^{t_l}\rmd r'\,h(t_l-r)h(t_l-r') \Ebb_f\left[n_{\rm RIN}(r)n_{\rm RIN}(r')\right].
\end{equation}
\end{subequations}
If the intensity fluctuations vanish, $X_1(t_l)$, $X_2(t_l)$ become uncorrelated. In the case of the laser model of Sec.\ \ref{sec:LOstate}, the RIN contribution $C_0$ becomes
\begin{multline}
C_0=2\int_{t_l-\tau}^{t_l}\rmd r\int_{t_l-\tau}^{t_l}\rmd r'\,h(t_l-r)h(t_l-r') v(r-r')\left(2w^2+v(r-r')\right)
\\ {} \simeq 2\int_0^{+\infty}\rmd s\int_0^{+\infty}\rmd s'\,h(s)h(s') v(s-s')\left(2w^2+v(s-s')\right).
\end{multline}

As an example, we can consider the case of an  exponential response function \eqref{hexp} and exponential correlations \eqref{vexp}; then, we have
\begin{equation}\label{S0C0}
S_0=\frac\varkappa 2, \quad C_0=\frac{2\varkappa\left(1-w^2\right) } {\varkappa+2\gamma_1}\left(1+w^2+\frac{2\gamma_1w^2}{\varkappa+\gamma_1}\right),
\end{equation}
where $w^2\leq 1$. Independently of the laser model, $S_0$ and $C_0$ are two parameters, to be estimated from the data.
As noted in Remark \ref{rem:fasi+}, phases no longer play any role. So, if we are only interested in the shot noise for random number generation, the relative phases do not have any effect on the statistics of the generated numbers.

In the case of signal in the vacuum state, the observables $X_j(t_l)$ comes from a mixture of distributions of linear combinations of Poisson processes (Remark \ref{rem:Phi0N}, Eq.\ \eqref{Xprocs}). However, by the effect of an LO intensity $\abs\lambda^2$ not too small and of  the smoothing on time due to the detector response function, we can relay in a normal approximation of the distribution of the observed processes. Therefore, the observations $\big(X_1(t_l), X_2(t_l)\big)$ can be considered a random sample from an (approximately) bi-variate normal distribution with means and covariances  given by \eqref{vaX}.

\subsection{Scaled processes}
We consider now the discrete sampling of the scaled processes $Y_j(t)$ in the case of strong LO; this is the situation considered, for instance, in \cite{Vill18,APFPS20}. Means and covariances of the random variables $Y_j(t_l)$ are obtained immediately by particularizing equations \eqref{EYprocslim}--\eqref{Cmatrix} and by taking into account Assumption \ref{ass:sampling}.
In the case of vanishing signal as above, we get means and covariances  simply by applying the scaling \eqref{Yproc} to Eqs.\ \eqref{vaX}:
\begin{equation}\label{EYl0}
\Ebb_P\left[Y_{j}(t_l)\right] \simeq \kappa_{j3}G_{j2},
\end{equation}
\[
\Cov_P[Y_j(t_l),\,Y_i(t_l')]\simeq
\delta_{ll'} \left(\delta_{ij}\kappa_{j2}S_0
+\kappa_{j3}G_{j2}\kappa_{i3}G_{i2} C_0\right).
\]
The coefficients $\kappa_{j2},\ \kappa_{j3},\ G_{j2}$ are defined in \eqref{kappas}, \eqref{:G_}.

In the case of the observables $Y_j(t_l)$, however, by the results of Sec.\ \ref{sec:reddescr}, we can obtain the characteristic function of these random variables, and, so, their full probability distribution.

\subsubsection{The discrete mode operators}\label{sec:randommodes}
To get explicitly the characteristic function of the observables $Y_j(t_l)$ and the associated POVM, it is convenient to introduce suitable bosonic mode operators. Moreover, this construction give a link between the quantum field approach to optical circuits and the more usual discrete mode approach.

We define the operators
\begin{subequations}\label{qjl}
\begin{gather}\label{themode}
a_l=\frac {-\rmi}{R_l(f)} \int_{t_l-\tau}^{t_l}h(t_l-s)\overline{\tilde f(s)} \, \rme^{-\rmi \psi_1 }\rmd A_1(s),
\\
\label{q1q2}
\hat q_1^l=\hat q_1^l(\phi)=\frac 1 {\sqrt{2}}\left(a_l^\dagger+a_l\right),
\qquad \hat q_2^l(\phi)=\frac 1 {\sqrt{2}}\left(\rme^{ \rmi \phi}a_l^\dagger+\rme^{- \rmi \phi }a_l\right),
\\
R_l(f)=\biggl( \int_{0}^{\tau}\rmd r \,h(r)^2\abs{\tilde f(t_l-r)}^2\biggr)^{1/2}.\label{def:RLF}
\end{gather}
\end{subequations}
Let us note that the normalization constant \eqref{def:RLF} is a random quantity and that, by \eqref{def:S_0}, \eqref{def:RLF}, \eqref{Eabsf2}, we have
\begin{equation}\label{ERl=S0}
\Ebb_f \left[R_l(f)^2\right]=S_0.
\end{equation}

\begin{remark}\label{rem:randop} The operators $a_l$, $a_l^\dagger$ are discrete mode operators, satisfying the CCRs,
while $\hat q_1^l$, $\hat q_2^l(\phi)$ are two quadratures, not complementary in general:
\begin{equation}
[a_{l'},\, a_l^\dagger]=\delta_{ll'},\quad [a_{l'},\, a_l]=0, \quad [\hat q_1^l,\,\hat q_2^l(\phi)]=\rmi \sin \phi.
\end{equation}
All these operators contain $\tilde f$: \emph{they are random operators}.
\end{remark}

The operators $\hat q_j^l(\phi)$ can be expressed in terms of the quantum observables \eqref{QjXi}, introduced in Sec.\ \ref{sec:Qj}; indeed, we have
\[
\hat q_j^l(\phi)=\frac 1 {\sqrt 2 R_l(f)}\int_{t_l-\tau}^{t_l} h(t_l-s)\rmd \hat Q_j(s).
\]

To give the structure of our  POVM, we shall need also squeezed coherent states; here we introduce some notation. Firstly, we decompose the Hilbert space $\Gamma_1$ and the vacuum state into the tensor product forms
\[
\Gamma_1=\Gamma_1^\bot\otimes {\prod_l}^\otimes \Gamma_1^l,\qquad e_1(0)=e_1^\bot(0) \otimes {\prod_l}^\otimes e_1^l0).
\]
The component $\Gamma_1^l$ is such that $\{a_l,\,a_l^\dagger,\; \Gamma_1^l\}$ gives  an irreducible representation of the CCRs for a single mode.
Then, we introduce the displacement operator and the squeezing operator
\begin{subequations}
\begin{equation}\label{displ}\begin{split}
&D_l(z)=\exp\left\{ z a_l^\dagger -\overline z \, a_l\right\}, \\ S_{\alpha,\beta}^l= &\exp \left\{\frac \xi 2 \,{a_l^\dagger}^2 -\frac{\overline\xi}2\, a_l^{\,2}\right\}, \quad \xi= \frac\beta{\abs\beta}\, \cosh^{-1}\alpha;
\end{split}\end{equation}
to have a well defined squeezing operator the coefficients $\alpha,\;\beta$ must satisfy
\begin{equation}\label{alpha2}
\alpha\in \Rbb, \qquad \beta\in \Cbb, \qquad \alpha^2-\abs \beta^2=1.
\end{equation}
\end{subequations}
By introducing also the unitary operator
\[
U_l(\phi)= \exp\left\{ \rmi \left(\phi -\frac \pi 2 \right)a_l^\dagger a_l\right\},
\]
we define the squeezed mode operator
\begin{equation}\label{a2b}
b_l=S_{\alpha,\beta}^lU_l(\phi) a_l U_l(\phi)^\dagger S_{\alpha,\beta}^{l\ \dagger}=\rmi\rme^{-\rmi \phi}\left(\alpha a_l+\beta a_l^\dagger\right).
\end{equation}
Finally, we introduce the coherent states for the  mode  $b_l$:
\begin{equation}\label{psil}\begin{split}
&\psi_l(z;\alpha,\beta)=S_{\alpha,\beta}^lU_l(\phi) D_l(z)e_1^l(0), \\ & b_l\psi_l(z;\alpha,\beta)=z\psi_l(z;\alpha,\beta), \qquad \psi_l(0;1,0)=e_1^l(0).
\end{split}\end{equation}

\subsubsection{The probability distribution in case of strong LO}\label{sec:Ylprob}

The \emph{characteristic function}  of the $2m$ random variables $Y_j(t_l)$,
\begin{equation}\label{cfjl}
\Phi^{\vec Y}(\vec k)=\Ebb_P\Big[\exp\Big\{\rmi \sum_{jl}k_j^lY_j(t_l)\Big\} \Big],
\end{equation}
is directly obtained from the \emph{characteristic functional}  $\Phi_T^Y[\vec k]$ \eqref{PhiYlim} of the stochastic process $Y_j(\bullet)$ by taking
$k_j(s)=\sum_{l=1}^m k_j^l \delta(s-t_l)$. To understand the structure of these random variables, we elaborate this characteristic function.

\begin{proposition}\label{prop:dsampl} Under Assumption \ref{ass:sampling}, the characteristic function \eqref{cfjl} of the random variables $Y_j(t_l)$
is given by
\begin{equation}\label{strcfjl}
\Phi^{\vec Y}(\vec k)=\Ebb_f\left[\Gamma^\Lcal(\vec k;f) \Phi^\Qcal(\vec k;f)\right], \qquad \Phi^\Qcal(\vec k;f)=\Tr_{\Gamma_1}\left\{\hat \Psi^q(\vec k;f)\rho^{f,T}_{1}\right\},
\end{equation}
\begin{equation}\label{prod}
\Gamma^\Lcal(\vec k;f)=\prod_{l=1}^m\Gamma^\Lcal_l(\vec k^l;f), \qquad \hat\Psi^q(\vec k;f)=\prod_{l=1}^m\hat\Psi^q_l(\vec k^l;f),
\end{equation}
\begin{subequations}\label{Gammasampl}
\begin{gather}\label{GammaL}
\Gamma^\Lcal_l(\vec k^l;f)=\exp\biggl\{\sum_{j=1}^2\biggl[\rmi k_j^l \, \mu_\Lcal^{jl}(f)
-\frac{{k_j^l}^2}2\, \sigma_\Lcal^{jl}(f)^2\biggr]\biggr\},
\\
\label{muLcal}
\mu_\Lcal^{jl}(f)=G_{j2}\kappa_{j3}\int_0^\tau\rmd s\abs{\tilde f(t_l-s)}^2h(s),
\end{gather}
\begin{equation} \label{sigmaLcal}
\sigma_\Lcal^{jl}(f)^2= \left(\sigma_j^{\,2}+V_j^{\,2}\right)\kappa_{j3}^{\;2}\int_0^\tau\rmd s\abs{\tilde f(t_l-s)}^2h(s)^2= \left(\sigma_j^{\,2}+V_j^{\,2}\right)\kappa_{j3}^{\;2}R_l(f)^2,
\end{equation}
\end{subequations}
\begin{multline}\label{hatPsiq}
\hat\Psi^q_l(\vec k^l;f)
= \exp\biggl\{\rmi \sqrt 2\, R_l(f)\left[\kappa_{13} k_1^l\hat q_1^l+\kappa_{23}k_2^l\hat q_2^l(\phi)\right]
\\ {}-\frac {R_l(f)^2}2\biggl|\sqrt{\frac{1-\eta_1}{\eta_1}}\,\kappa_{13} k_1^l -\rme^{-\rmi\phi}\sqrt{\frac{\eta_1}{1-\eta_1}}\,\kappa_{23}k_2^l\biggr|^2\biggr\}.
\end{multline}
For fixed $\tilde f$, the operator $\hat\Psi^q_l(\vec k^l;f)$ is the characteristic operator of a POVM on the Hilbert space $\Gamma_1$; the same statement holds for the product $\hat\Psi^q(\vec k;f)$.
\end{proposition}

The proof of Proposition \ref{prop:dsampl} is given in Appendix \ref{app:dsproof}.

When $\sin\phi=0$, i.e.\ $\phi = n\pi $, one has $\hat q_2^l(2n\pi)=\hat q_1^l$, \ $\hat q_2^l(2n\pi+\pi)=-\hat q_1^l$. In this case, a measurement of the two quadratures is trivially represented by a pvm; the part with the squared modulus in \eqref{hatPsiq} is some additive noise.

In the case $\sin \phi\neq 0$ instead, we shall see that the characteristic operator $\hat\Psi^q_l(\vec k^l;f)$  can be expressed through  the Fourier transform of a POVM-density based on  squeezed coherent states; this measure  is in a class which generalizes the POVM constructed starting from the Husimi transform \cite[Eqs.\ (3.4)--(3.10)]{Bar86}, \cite[Chapt.\ 4]{Hol82}.

\begin{proposition}\label{prop:V+b} Let us take the case $\sin\phi\neq 0$ and fix the squeezing parameters \eqref{alpha2} by
\begin{equation}\label{def:alpha,beta}
\alpha
=\frac{1}{2\sqrt{\eta_1\left(1-\eta_1\right)}\,\sin \phi},
\qquad \beta
=\frac{\rme^{2\rmi\phi}\left(1-\eta_1\right)+ \eta_1}{2\sqrt{\eta_1\left(1-\eta_1\right)}\,\sin \phi}.
\end{equation}
Then, the characteristic operator \eqref{hatPsiq} can be written as
\begin{subequations}
\begin{equation}\label{Vu}
\hat\Psi^q_l(\vec k^l;f)=\exp\left\{\rmi\left( u_lb_l^\dagger + \overline {u_l}\, b_l\right) - \frac 12\abs{u_l}^2\right\},
\end{equation}
where $b_l$ is defined in \eqref{a2b} and
\begin{equation}\label{k2ul}
u_l= R_l(f)\left[\sqrt{\frac{1-\eta_1}{\eta_1}}\,\kappa_{13} k_1^l -\rme^{-\rmi \phi }\sqrt{\frac{\eta_1}{1-\eta_1}}\,\kappa_{23} k_2^l\right].
\end{equation}
\end{subequations}
Moreover, we have
\begin{equation}\label{eqprop3}
\hat\Psi^q_l(\vec k^l;f)=\int_\Cbb \rmd^2 z \, \rme^{\rmi\left(u_l\,\overline z+\overline {u_l}\, z\right)}\hat g_{\alpha,\beta}^l( z), \qquad \hat g_{\alpha,\beta}^l(z)=\frac 1\pi |\psi_l(z;\alpha,\beta)\rangle\langle \psi_l(z;\alpha,\beta)|;
\end{equation}
$\hat g_{\alpha,\beta}^l(z)$ is a POVM-density on $\Cbb$ with respect to the Lebesgue measure $\rmd^2z$.

\end{proposition}

The proof of Proposition \ref{prop:V+b} is given in Appendix \ref{app:dsproof}.

\begin{remark}\label{al=bl} Let us note that, when $\eta_1=1/2$ and $\phi=\pi/2$ we get $(\alpha,\beta)=(1,0)$ and $b_l=a_l$; then, the POVM \eqref{eqprop3} reduces to the usual measure based on the coherent states of the ``random'' modes $a_l$, cf.\ Remark \ref{rem:randop}. So, squeezed states appear in the measurement operators when $\abs{\sin\phi}\neq 1$, which means that two not exactly complementary quadratures are involved, and/or that $\eta_1\neq 1/2$, which means an imbalance in the beam splitter which divides the signal into the two measured components.
\end{remark}

\subsection{Probability density}\label{sec:density}

By the results of Sec.\ \ref{sec:Ylprob}, we can write explicitly the POVM-density and the probability distribution of the observables $\big(Y_1(t_l),Y_2(t_l)\big)$, when $\sin \phi \neq 0$, \ $\eta_1\neq 0, \; 1$.

From equations \eqref{strcfjl}, \eqref{prod} we get the characteristic operator
\begin{equation}\label{COl}
\hat \Psi^{Y}_l(\vec k^l;f)=\Gamma^\Lcal_l(\vec k^l;f) \hat\Psi^q_l(\vec k^l;f),
\end{equation}
where the two factors are given
in \eqref{Gammasampl} and \eqref{eqprop3}.
We assume also  $\sigma_\Lcal^{jl}(f)^2>0$;
let us note that the electronic noise can be taken into account by increasing the variances \eqref{sigmaLcal}.

As proved in Appendix \ref{app:dsproof},
the POVM-density associated with the characteristic operator \eqref{COl} has the expression
\begin{multline}\label{YPOVMdens}
\hat g_Y^l(y_1,y_2;f)
=\frac 1 {2\pi \sigma_\Lcal^{1l}(f)\sigma_\Lcal^{2l}(f)}\int_\Cbb \rmd^2 z \, \hat g_{\alpha,\beta}^l( z)
\exp\biggl\{-\frac{\left(\mu_\Lcal^{2l}(f)+2K_2^l(f)\left(z_2\sin \phi -z_1\cos\phi\right)- y_2\right)^2}{2\sigma_\Lcal^{2l}(f)^2} \biggr\}\\ {}\times \exp\biggl\{-\frac {\left(\mu_\Lcal^{1l}(f)+2K_1^l(f)z_1- y_1\right)^2}{2\sigma_\Lcal^{1l}(f)^2} \biggr\},
\end{multline}
where $\mu_\Lcal^{jl}(f)$ and $\sigma_\Lcal^{jl}(f)$ are given in \eqref{Gammasampl}, $z=z_1+\rmi z_2$ and
\begin{equation}\label{def:x_j}
K_1^l(f)=R_l(f)\kappa_{13}\sqrt{\frac{1-\eta_1}{\eta_1}}\,, \qquad K_2^l(f)= R_l(f)\kappa_{23}\sqrt{\frac{\eta_1}{1-\eta_1}}\,.
\end{equation}
As expected, the density \eqref{YPOVMdens} is a convolution of a Gaussian  with the POVM introduced in Proposition \ref{prop:V+b}. Then, the probability density of the observables $Y_j(t_l)$, $j=1,2$, $l=1,\ldots,m$, is given by
\begin{equation}\label{mixtprobdens}
g_{\vec Y}(\vec y)=\Ebb_f\biggl[\Tr_{\Gamma_1}\biggl\{\rho^{f,T}_{1}\prod_l\hat g_Y^l(y_1^l,y_2^l;f)\biggr\}\biggr].
\end{equation}

\paragraph{Signal in the vacuum state.} An explicit expression of this density can be given when the signal is in a mixture of coherent states as in Corollary \ref{corroll}, see Sec.\ \ref{sec:s=cs}. When the signal is in the vacuum state the density \eqref{mixtprobdens}, \eqref{0mixtprobdens} takes the form
\begin{subequations}
\begin{gather}\label{00mixtprobdens}
g_{\vec Y}(\vec y)=\prod_{l=1}^m g_{Y}^l(y_1^l,y_2^l),\qquad g_{Y}^l(y_1^l,y_2^l)=\Ebb_f\bigl[g_Y^l(y_1^l,y_2^l;f)\bigr],
\\ \label{gYldens}
g_Y^l(y_1^l,y_2^l;f)=\prod_{j=1}^2 \frac{\exp\biggl\{-\frac{\left(y_j^l-\mu_\Lcal^{jl}(f)\right)^2 } {2\kappa_{j2}R_l(f)^2}\biggr\}} {\sqrt{2\pi\kappa_{j2}R_l(f)^2}}\,,
\end{gather}
\end{subequations}
where $\mu_\Lcal^{jl}(f)$ is given in \eqref{muLcal}.
In getting the product structure, Assumption \ref{ass:sampling} is involved.

\section{Min-entropy and random number generation}\label{sec:hmin}

In this section we discuss a specific application of the full quantum theory of a eight-port homodyne detection scheme presented in this work: random number generation.
As we stated in Sec.\ \ref{sec:sop}, for QRNG applications we are interested in pure noise (shot noise); in particular, we want to show the effects of unbalancing and   LO fluctuations with respect to the perfect case discussed in \cite{Vill18}.
To capture the characteristic features of a good random number generator we extract the random bits from the sampled joint distribution of $\big(X_1(t_l),X_2(t_l)\big)$, $l=1,\ldots,m$; when the signal is in the vacuum state, they are independent and identically distributed (i.i.d.) by Assumption \ref{ass:sampling}. If some dependence would be left, we can generate the random numbers after a suitable under-sampling.

We want the generated bits to be truly random and known only to the users of the QRNG.
The requirement of true randomness can be synthesized by the leftover Hash lemma, which relates the maximum number of extractable i.i.d.\ bits from a given string to an entropic quantity called min-entropy $H_{\rm min}$. Instead, to have secure and private random bits we have to relay on the quantum/classical conditional min-entropy, which takes into account possible side information hold by an eavesdropper \cite{KRS09,DattaR09,Vill18}. On the basis of the computed value of the min-entropy the truly random and secure bits are obtained from the raw data by employing a suitable algorithm, a randomness extractor \cite{Extractor,Vill18,Lupo+21}.

Given a discrete probability distribution $P=\{p_j,\, j\in I\}$  the (classical) \emph{min-entropy} is:
\begin{equation}\label{def:Hmin}
H_{\rm min}(P)=-\log_2 \max_{j\in I} p_j\geq 0;
\end{equation}
the quantity $P_{\rm guess}=\max_{j\in I} p_j$ is known as \emph{guessing probability}. Obviously $0<P_{\rm guess}\leq 1$ and the inequality in \eqref{def:Hmin} follows. Then, the maximal number of i.i.d.\ bits extractable per measurement is given by $H_{\rm min}(P)$   \cite{KRS09,Haw+15,Smith2019,DattaR09,Vill18}.

When the sampled quantity is a continuous one, it has to be discretized; this is automatically done because any real measuring apparatus has a finite resolution. To state the problem, let us think to the univariate case and fix one of our sampled observables, say $X_1(t_l)\equiv X_1$, and denote by $f_1(x)$ its probability density. An $n$-bit ADC as a finite range $2R_1$ and a resolution $\delta_1=2R_1/2^n$. If possible, the discretization range has to be placed symmetrically around the mean $\mu_1$. We set $x_0\simeq \mu_1-R_1$, and $x_j=x_0+j\delta_1$ for $j=1,\ldots, 2^n$; so, we get the discrete distribution
\[
p_0=\int_{-\infty}^{x_0} f_1(y)\rmd y, \qquad p_{2^n+1}=\int_{x_{2^n}}^{+\infty} f_1(y)\rmd y,
\qquad
p_j=\int_{x_{j-1}}^{x_j} f_1(y)\rmd y , \quad 1\leq j\leq 2^n.
\]
The length and the position of the discretization interval must be such that the \emph{saturation probabilities} $p_0$ and $p_{2^n+1}$ are negligible, cf.\ the discussion in \cite{OSSF13}.
Then, the guessing probability turns out to be
\begin{equation}
P_{\rm guess}(X_1,\delta_1)=\sup_{j=1,\ldots,2^n}\int_{x_{j-1}}^{x_{j}}f_1(y)\rmd y\leq \delta_1\times \sup_{x\in \Rbb} f_1(x),
\end{equation}
and for the min-entropy we get
\begin{equation}
H_{\rm min}(X_1,\delta_1)=-\log_2 P_{\rm guess}(X_1,\delta_1)\geq \log_2 \left(\delta_1 \sup_{x\in \Rbb} f_1(x)\right)^{-1}.
\end{equation}
The lower bound above could be also negative; in this case it gives no information, because the min-entropy is always positive. However, in usual situations $\delta_1$ is sufficiently small in order to have that the lower bound is positive and it is a good estimation of the min-entropy. In the case of a Gaussian distribution, reference \cite{Haw+15} contains also a discussion on how to optimize the choice of the discretization interval.

\subsection{The total min-entropy}\label{sec:totminentr} Let us denote by $p_X^l(x_1,x_2)$ the probability density of the observed sample
$\big(X_1(t_l),\, X_2(t_l)\big)$ at time $t_l$; the signal is in the vacuum state and we include also the electronic noise. We assume this distribution to be approximately Gaussian, as discussed at the end of Sec.\ \ref{sec:sop}. By  \eqref{muX}, \eqref{CovC0}, the means are given by $\mu_j=  \Delta_{j2}\abs\lambda^2$  and the covariance matrix  by
\begin{equation}\label{detC+el}
\mathbf C=\begin{pmatrix}\Sigma_1^{\,2}& \Delta_{12}\Delta_{22}\abs{\lambda}^4 C_0
\\ \Delta_{12}\Delta_{22}\abs{\lambda}^4 C_0& \Sigma_2^{\,2}
\end{pmatrix}, \qquad \Sigma_j^{\,2}=\kappa_{j2}\abs\lambda^2 S_0
+\Delta_{j2}^{\;2}\abs{\lambda}^4 C_0 + {\sigma^{\rm el}_{j}}^2.
\end{equation}
The quantities $S_0, \, C_0$ are given in \eqref{def:S_0}, \eqref{def:C_0}, and $\kappa_{j2}, \, \Delta_{j2}$ in \eqref{kappas}, \eqref{Deltas}.
The variances $\Sigma_j^{\,2}$ are expressed by the sum of the shot noise contribution $\kappa_{j2}\abs\lambda^2 S_0$, the RIN contribution $\Delta_{j2}^{\;2}\abs{\lambda}^4 C_0$, and the electronic noise contribution ${\sigma^{\rm el}_{j}}^2$. Here and in the following it is useful to have a short notation for the noise ratios; so, we set
\begin{equation}\label{noiseR}
\Upsilon_j=\frac{\Delta_{j2}^{\;2}\abs{\lambda}^2 C_0}{\kappa_{j2} S_0}, \qquad  \Theta_j=\frac{{\sigma^{\rm el}_{j}}^2}{\kappa_{j2}\abs\lambda^2 S_0}.
\end{equation}
Then, we have
$\big(X_1(t_l),\, X_2(t_l)\big) \sim \Ncal(\vec \mu; \mathbf C)$ and
\begin{equation}
\sup_{\vec x\in \Rbb^2}p_X^l(\vec x)=\left(2\pi\sqrt{\det \mathbf C}\right)^{-1},
\end{equation}
\begin{equation}\label{xHtot}
\det \mathbf C=\Sigma_1^{\,2}\Sigma_2^{\,2}- \Delta_{12}^{\;2}\Delta_{22}^{\;2}\abs{\lambda}^8 C_0^{\,2} =\kappa_{12}\kappa_{22}\abs\lambda^4 S_0^{\,2}E_{12},
\end{equation}
\begin{equation}\label{xHtot2}
E_{12}=\frac{\det \mathbf C}{\kappa_{12}\kappa_{22}\abs\lambda^4 S_0^{\,2}}=\left(1+ \Theta_1\right)\left(1+ \Theta_2\right)
+ \left(1+ \Theta_1\right)\Upsilon_2
+ \left(1+ \Theta_2\right)\Upsilon_1.
\end{equation}

As already said in Remark \ref{F,Alm}, the two signals are suitably filtered in order to eliminate the non-flat part of the spectrum, and, then, to digitize the two components of each sample, two $n$-bits ADCs are employed, with two ranges $2R_j$ and resolutions $\delta_j=2R_j/2^{n}$. The two ranges are placed around the means and are such that the saturation probability is negligible.
Possibly, the two ADCs are identical and this would give $\delta_1=\delta_2$.
Under these hypotheses, the guessing probability for the single sample is
\begin{equation}\label{Pguess1}
P_{\rm guess}(X,\delta)\simeq \sup_{x_1,x_2}\int_{x_1-\delta_1/2}^{x_1+\delta_1/2}\rmd y_1 \int_{x_2-\delta_2/2}^{x_2+\delta_2/2}\rmd y_2 \,p_X^l(\vec y)\leq \delta_1\delta_2\sup_{\vec x\in \Rbb^2}p_X^l(\vec x).
\end{equation}
We assume also $\delta_1$ and $\delta_2$ to be  small enough to have
$P_{\rm guess}(X,\delta)\simeq \delta_1\delta_2\sup_{\vec x\in \Rbb^2}p_X^l(\vec x)$.
Then, the min-entropy per sample is given by
\begin{equation}\label{Htot}
H_{\min}(X,\delta)= -\log_2P_{\rm guess}(X,\delta)\simeq \log_2\frac{2\pi\sqrt{\det \mathbf C}}{\delta_1\delta_2}.
\end{equation}
When the step lengths $\delta_j$ are not sufficiently small, the final expression on the right is only a lower bound and the guessing probability has to be expressed by using the error function \cite{Haw+15,Lupo+21,Smith2019}.

Let us stress that the min-entropy \eqref{Htot} is independent of $\eta_1$, as one can see from \eqref{kappas}, \eqref{Deltas}, where the expressions of the coefficients $\kappa_{j2}$ and $\Delta_{j2}$ are given. This is due to the fact that the signal is in the vacuum state and that $\eta_1$ is the transmissivity of the beam splitter which mixes the signal with another vacuum. Similarly, there is no dependence on $\phi$, because the probabilities do not depend on the phases when there is no interference between signal and LO (Remark \ref{rem:fasi+}). When $\Delta_{j2}=0$ as in case of  rebalancing (Remark \ref{rebalance}), the means and the terms with the RIN contribution $C_0$ disappear; in this case the placement of the discretization interval is easier, as it must be symmetric around zero due to the vanishing of the mean.

\subsubsection{Entropy losses due to correlations}

To  put in evidence the entropy losses due to the presence of correlations in the covariance matrix \eqref{detC+el}, it is useful to introduce the \emph{reference entropy}
\begin{equation}\label{def:Href}
H_{\rm ref}=\log_2 \frac{2\pi \Sigma_1\Sigma_2}{\delta_1\delta_2}.
\end{equation}
By comparing this formula with \eqref{Htot}, we see that we have replaced the determinant of the covariance matrix with the product of the variances; so, $H_{\rm ref}$ represents the min-entropy when the correlations are not taken into account.

Then, by equations \eqref{detC+el}, \eqref{noiseR}, \eqref{xHtot2}, \eqref{Htot}, \eqref{def:Href}, we obtain
\begin{equation}\label{Hdinam}
H_{\rm ref}-H_{\min}(X,\delta) \simeq -\frac 12 \log_2\left[1-\frac{\Upsilon_1\Upsilon_2}{\left(1+\Theta_1+\Upsilon_1\right)\left(1+\Theta_2+\Upsilon_2\right)}\right].
\end{equation}
This difference is positive and represents the entropy loss due to the correlations introduced by the RIN; this loss vanishes in the case of rebalancing, see Remark \ref{rebalance}.

It is important to note that the difference \eqref{Hdinam} does not depend on the resolution parameters $\delta_j$.

\subsubsection{Optimization of the discretization range}\label{sec:optim}

From \eqref{Htot} we see that the min-entropy increases when the discretization steps $\delta_j$ decrease. On the other side, this expression of the min-entropy is based on the hypothesis that the probability of saturation is negligible, but this probability increases with the decrease of the discretization range. We can try to quantify this tradeoff by saying that there is saturation when the signal $\big(X_1(t_l),\,X_2(t_l)\big)$ falls outside the discretization rectangle; this is a conservative choice, as we do not make a finer subdivision of the saturation region. Then, the saturation probability is given by
\begin{equation}\label{satprob}
P_{\rm saturation}(X,\delta)\simeq 1- \int_{\mu_1-R_1}^{\mu_1+R_1}\rmd y_1 \int_{\mu_2-R_2}^{\mu_2+R_2}\rmd y_2 \,p_X^l(\vec y).
\end{equation}
To be negligible, the saturation probability must satisfy
\begin{equation}\label{ass:negl}
P_{\rm saturation}(X,\delta) < P_{\rm guess}(X,\delta).
\end{equation}
By suitably tuning the ADC apparatus and the laser intensity, we can manage the range $R_j$ to be proportional to the standard deviation of the observed voltage (at least approximately); so, by \eqref{detC+el}, we can write
\begin{equation}\label{RproptoSigma}
R_j=x_j\Sigma_j, \qquad \delta_j=\frac {2R_j}{2^n}=\frac {x_j\Sigma_j}{2^{n-1}}.
\end{equation}
We have considered the same $n$ for both ADCs; eventually, also the proportionality constants $x_j$ could be independent from the index $j$.
To apply the condition \eqref{ass:negl} in a simple way, we consider the Gaussian approximation and we neglect the correlations; we add a tilde to  denote the guessing and saturation probabilities in this approximation:
\begin{equation}\label{Pguessapprox}
\tilde P_{\rm guess}(X,\delta)=\frac{\delta_1\delta_2}{2\pi \Sigma_1\Sigma_2 }=\frac{x_1x_2}{\pi\,2^{2n-1}},
\end{equation}
\begin{multline*}
\tilde P_{\rm saturation}(X,\delta)=1- P\left[-R_1<X_1(t_l)-\mu_1<R_1\right] P\left[-R_2<X_2(t_l)-\mu_2<R_2\right]\\ {}=  1-4\left(\Phi(x_1)-\frac12\right)\left(\Phi(x_2)-\frac12\right),
\end{multline*}
where $\Phi(x) $ is the cumulative distribution function of a standard Gaussian random variable.
Then,  condition \eqref{ass:negl} (approximately) becomes
\begin{equation}\label{sat<guess}
1-4\left(\Phi(x_1)-\frac12\right)\left(\Phi(x_2)-\frac12\right)<\frac{x_1x_2}{\pi \,2^{2n-1}}.
\end{equation}
Let us note that the quantity $H_{\rm ref}$ \eqref{def:Href} is just the min-entropy associated with the approximate guessing probability \eqref{Pguessapprox} and that we have
\begin{equation}\label{Href-n}
H_{\rm ref}=- \log_2 \tilde P_{\rm guess}(X,\delta)=2n-1 -\log_2\frac{x_1x_2}\pi.
\end{equation}

To have an idea of the values of the min-entropy and of good choices of $x_j$ and $n$, let us consider the case $x_1=x_2=x$. Table \ref{table1} gives the values of $H_{\rm ref}$ (the main contribution to the total min-entropy) as a function of the parameters $n$ and $x$.
\begin{table}[H]
\caption{The entropy contribution $H_{\rm ref}$ \eqref{Href-n} as a function of the proportionality parameter $x=x_1=x_2$ \eqref{RproptoSigma} and of the ADC number of bits $n$. A blank value means that the condition \eqref{sat<guess} is not satisfied.}\label{table1}
\begin{center}
\begin{tabular}{c||c|c|c|c|c|c|c|}
  $n\backslash x$&3.8 & 4.0 & 4.6&5.1& 6.0& 8.9 &9.5 \\
  \hline\hline
  8&\phantom{H} & 12.65 & 12.25 &11.95 & 11.48 & 10.34 &10.16\\
  \hline
 10& \phantom{H} & \phantom{H} & 16.25 & 15.95& 15.48& 14.34&14.16\\
  \hline
  12&\phantom{H} & \phantom{H} & \phantom{H} & 19.95&19.48&18.34 &18.16\\
  \hline
 16& \phantom{H} & \phantom{H} & \phantom{H} & &27.48 &26.34 &26.16\\
  \hline
32& \phantom{H} & \phantom{H} & \phantom{H} & & \phantom{H} &58.34 &58.16\\
  \hline
\end{tabular}
\end{center}
\end{table}

For a given $n$ the best choice for QRNG is to take the smallest $x$ compatible with condition \eqref{sat<guess}; however, if we want to use the apparatus  also for detection, or if we want to be more sure to avoid saturation, we need  $x$ to be bigger. In Table \ref{tablePP} we report the ratio $\tilde P_{\rm saturation}(X,\delta)/\tilde P_{\rm guess}(X,\delta)$. The good choice for QRNG is to take the parameters which give this ratio near 1; however, by comparing the two tables, we see that we can make this ratio very small without a strong decreasing of $H_{\rm ref}$.
\begin{center}
\begin{table}[H]
\caption{The ratio $\tilde P_{\rm saturation}(X,\delta)/\tilde P_{\rm guess}(X,\delta)$ as a function of the proportionality parameter $x=x_1=x_2$ \eqref{RproptoSigma} and of the ADC number of bits $n$. A value greater than 1 means that the condition \eqref{sat<guess} is not satisfied.}\label{tablePP}
\begin{center}
\begin{tabular}{c||c|c|c|c|c|c|c|}
  $n\backslash x$&3.8 & 4.0 & 4.6&5.1& 6.0& 8.9 &9.5 \\
  \hline\hline
  8&2.1 & 0.82 & $4.1\times 10^{-2}$ & $2.7\times 10^{-3}$& $1.1\times 10^{-5}$ & $1.5\times 10^{-15}$ & $4.8\times 10^{-18}$\\
  \hline
 10& 33 & 13 & 0.66 & $4.3\times 10^{-2}$& $1.8\times 10^{-4}$& $2.3\times 10^{-14}$&$7.7\times 10^{-17}$\\
  \hline
  12&$5.3\times 10^2$ & $2.1\times 10^2$ & 11 &0.69 & $2.9\times 10^{-3}$& $3.7\times 10^{-13}$&$1.2\times 10^{-15}$\\
  \hline
 16& $14\times 10^4$ & $5.3\times 10^4$ & $2.7\times 10^3$ & $1.8\times 10^2$ & 0.74 &$9.5\times 10^{-11}$ &$3.1\times 10^{-15}$\\
  \hline
32 & $5.8\times 10^{14}$ & $2.3\times 10^{14}$ & $1.2\times 10^{13}$ & $7.6\times 10^{11}$ & $3.2\times 10^9$ &0.41 &$1.3\times 10^{-3}$\\
  \hline
\end{tabular}
\end{center}
\end{table}
\end{center}

The computations of Tables \ref{table1} and \ref{tablePP} we have assumed that the ranges are centered on the means. When this is not possible, the saturation probability increases and it is convenient to enlarge the range with respect to the variance and to chose a value for $x$ giving a ratio well below 1 in Table \ref{tablePP}.

In \cite{Vill18} the equilibrated case is considered and experimentally implemented; by using a 10-bit ADC, they obtain a value of about 14 for the min-entropy. So, by looking at Table \ref{table1}, we can say that values from 14 to 26 for $H_{\rm ref}$ are experimentally reasonable, depending on the characteristics of the ADC and of the electronic part of the apparatus. A much higher value can be obtained, if a 32-bit ADC is available.

As already written,
to tune the ADC ranges to the noise variances one could increase the LO intensity. The shot noise intensity can be increased also by using two lasers, one at the LO port and one at the signal port. To use two similar lasers is nearly the same as doubling the LO intensity; explicit computations can be done by starting from the results of Appendix \ref{sec:s=cs}. In any case, the most important ingredient to increase the bit generation rate is the ADC resolution.

In the following, we assume that the instrumentation has been chosen and calibrated; so, $n$ and the ADC ranges are fixed, which means that also the values of $\delta_1$ and $\delta_2$ have been fixed.

\subsection{Side information: the classical noise}\label{sec:|cl}

When all the noise contributions, classical and quantum, are trusted, the randomness extractor can be calibrated on the value of the total min-entropy $H_{\min}(X,\delta)$. Fast random number generators based on various types of physical noise have been proposed and realized, see \cite{OSSF13} for an example based on laser noise. However, doubts have been raised on some of the noise components involved in homodyne-based random number generators, see for instance \cite{Vill17,Vill18,Smith2019,Huang+20,Lupo+21}. The presence of untrusted noise or of possible side information forces to calibrate the randomness extractor on suitable conditional min-entropies, as discussed here and in the next subsection.

In our formulation of the double homodyne detector we have included two sources of classical noise, the electronic noise and the laser fluctuations. The classical noise can be considered untrusted because not truly random and not well modeled, but even because it could be known to some intruder and it could convey some side information. To get secure random bits with respect to not certified noise and to side information, the \emph{average conditional min-entropy} \cite[Sec.\ C]{Haw+15} has to be used to calibrate the randomness extractor: the guessing probability has to be computed with respect to the probability distribution conditioned on laser fluctuations and electronic noise; then, the mean is taken.

Let us denote by $N_{\rm el}^{j,l}$ the contribution of the electronic noise to the observation at time $t_l$; we have assumed the electronic noise to be normal, $N_{\rm el}^{j,l}\sim\Ncal\left(0; {\sigma^{\rm el}_{j}}^2\right)$, and independent of any other noise contribution. Then, for fixed $f$, the sampled observables can be written as
\begin{equation}\label{XlY+N}
X_j^f(t_l)=\abs\lambda Y_j^f(t_l)+ N_{\rm el}^{j,l},
\end{equation}
where the $Y_j^f(t)$ are the scaled processes introduced in \eqref{Yproc}. We assume the LO-intensity to be sufficiently high, so that the probability density of the random variables $Y_j^f(t_l)$ is the one discussed in Sec.\ \ref{sec:density}. When the signal is in the vacuum state and $f$ and $N_{\rm el}^{j,l}$ are given, we get from \eqref{gYldens} that the conditional probability density for a single sample is
\[
p_X^l(x_1,x_2;f,N_{\rm el})=\frac1{\abs\lambda^2}\,g_Y^l\left((x_1^l-N_{\rm el}^{1,l})/\abs\lambda,(x_2^l-N_{\rm el}^{2,l})/\abs\lambda;f\right).
\]
In this situation the average guessing probability, conditional on the classical noise, is given by
\begin{multline}\label{Pguess2}
P_{\rm guess}(X,\delta|\Ecal_{\rm cl})
=\Ebb_{f,\, N_{\rm el}}\left[\sup_{x_1,x_2}\int_{x_1-\delta_1/2}^{x_1+\delta_1/2}\rmd y_1 \int_{x_2-\delta_2/2}^{x_2+\delta_2/2}\rmd y_2\, p_X^l(y_1,y_2;f,N_{\rm el})\right]
\\ {}\leq \frac{\delta_1\delta_2}{\abs\lambda^2}\, \Ebb_f\left[\sup_{y_1,y_2}g_Y^l\left(y_1^l/\abs\lambda,y_2^l/\abs\lambda;f\right)\right]
= \Ebb_f\left[\frac {\delta_1\delta_2}{2\pi \sqrt{\kappa_{12}\kappa_{22}}\abs\lambda^2R_l(f)^2}\right].
\end{multline}
Accordingly, the average conditional min-entropy per sample is given by
\begin{equation}\label{Hmin|}
H_{\min}(X,\delta|\Ecal_{\rm cl})=-\log_2P_{\rm guess}(X,\delta|\Ecal_{\rm cl})\simeq \log_2\frac {2\pi \abs\lambda^2\sqrt{\kappa_{12} \kappa_{22}}} {\delta_1\delta_2\Ebb_f \left[R_l(f)^{-2}\right]},
\end{equation}
where we have again simplified the computations of the Gaussian integrals by assuming the $\delta_j$ to be sufficiently small. From these  definitions we have immediately $P_{\rm guess}(X,\delta)\leq P_{\rm guess}(X,\delta|\Ecal_{\rm cl})$ and
$H_{\min}(X,\delta)\geq  H_{\min}(X,\delta|\Ecal_{\rm cl})$.

\begin{remark}\label{rem:RIN} Let us note that the electronic noise disappears from the expression \eqref{Hmin|} of the classically conditioned min-entropy, as this noise is purely additive. This is not the case of the RIN, which contributes through the expression $\Ebb_f \left[R_l(f)^{-2}\right]$. The reason is that the laser fluctuations are involved in the definition \eqref{themode} of the discrete mode operators, where $R_l(f)$ is the (random) normalization constant.
Moreover, we have
$\Ebb_f \left[R_l(f)^2\right]=S_0$, see \eqref{ERl=S0}, and
\begin{equation}\label{ER-2}
\Ebb_f \left[R_l(f)^{-2}\right]=\frac 1 {S_0}+\Ebb_f\left[\frac{\left(R_l(f)^{2}-S_0\right)^2} {R_l(f)^{2}S_0^{\,2}}\right]\geq \frac 1 {S_0}\,,
\end{equation}
which gives \ $S_0\Ebb_f \left[R_l(f)^{-2}\right]\geq 1$. \ \ Consistently with the laser models discussed in Sec.\ \ref{sec:LOstate}, we assume
\ \ $\Ebb_f \left[R_l(f)^{-2}\right]$ to be independent from $l$ and we set
\begin{equation}\label{def:S_-}
S_-=1\big/\Ebb_f \left[R_l(f)^{-2}\right]\leq S_0.
\end{equation}
\end{remark}
\begin{remark}\label{S-:S_0} By definition \eqref{def:RLF}, $R_l(f)$ turns out to be a time smoothing of the function $f$ realized through the response function $h(t)$.
If the involved integration time interval is not too short, it is realistic to have $R_l(f)^2\simeq S_0$, by ergodic properties of the process $f(t)$; in this case $S_-\simeq S_0$. In any case their difference should be small. From \eqref{S0C0} we see that to have the decay time of RIN correlations much shorter than the decay time of $h(t)$ gives also $C_0\ll1$.
\end{remark}

In the works on QRNG from homodyne detection it is usual to express the min-entropy by scaling the involved noise to the vacuum noise (1/2 in standard units) \cite{Smith2019,Vill18}. If we have a good estimate of $2\kappa_{j2}\abs\lambda^2S_0$ from the calibration stage, we can introduce the ``scaled'' resolutions and min-entropy:
\begin{equation}\label{scaleddeltas}
\delta_j^0=\frac {\delta_j}{\abs\lambda\sqrt{2\kappa_{j2}S_0}}, \qquad H_{0}=\log_2\frac {\pi } {\delta_1^0\delta_2^0} .
\end{equation}
The scaled resolutions $\delta_j^0$ turn out to be pure numbers; even in the case $\delta_1=\delta_2$, the presence of any imbalance should give $\delta_1^0\neq\delta_2^0$. From \eqref{kappas} and \eqref{scaleddeltas}, we can see that when the $\epsilon_j$ decrease (more losses) the parameters $\delta_j^0$ increase (worst scaled resolution).
Then, by comparing \eqref{Hmin|} with the reference min-entropy
\eqref{scaleddeltas}, we get
\begin{equation} \label{H0-Hcl}
H_{0}-H_{\min}(X,\delta|\Ecal_{\rm cl})\simeq \log_2\left(S_0/S_-\right)\geq 0.
\end{equation}
Under the conditions of Remark \ref{S-:S_0}, this entropy difference is small.
Also the total min-entropy \eqref{Htot} can be expressed in a similar way:
\begin{equation}\label{H-H}
H_{\min}(X,\delta) \simeq H_{0}+ \frac 12 \,\log_2\frac{\det \mathbf C }{\abs\lambda^4 \kappa_{12}\kappa_{22}S_0^{\,2}};
\end{equation}
only the min-entropy $H_0$ depends on the discretization steps $\delta_j$, not the last term.

The price in considering untrusted the classical noise is a loss of entropy quantified by
\begin{equation}\label{diff1}
H_{\min}(X,\delta) -H_{\min}(X,\delta|\Ecal_{\rm cl})\simeq \frac12\, \log_2E_{12}+\log_2\left(S_0/S_-\right),
\end{equation}
where $E_{12}$ is defined in \eqref{xHtot2} and depends on the noise ratios \eqref{noiseR}.
Note that the entropy  loss does not depend on the resolutions $\delta_j$.
By increasing the laser intensity $\abs\lambda^2$ and the coefficients $\epsilon_j$ (less optical losses), we can make the electronic noise contributions $\Theta_j$ to decrease, but one has to pay attention to the tradeoff with respect to the RIN (the terms with $\Upsilon_j$).
Indeed, these contributions increase with $\abs\lambda^2$, while they can be made to decrease by some rebalancing which can be obtained by decreasing some of the $\epsilon_j$, see Remark \ref{rebalance}. This trade off is more evident if we consider the entropy loss with respect to $H_{\rm ref}$, because, by Eqs.\ \eqref{Hdinam} and \eqref{diff1}, this loss takes the simpler expression
\begin{equation}\label{diffxxx}
H_{\rm ref} -H_{\min}(X,\delta|\Ecal_{\rm cl})\simeq \log_2\left(S_0/S_-\right)+\log_2 \sqrt{\left(1+\Theta_1+\Upsilon_1\right)\left(1+\Theta_2+\Upsilon_2\right)}.
\end{equation}

If the laboratory is ``secure'' and we trust in our detection apparatus and in the LO laser, we can relay on the conditional min-entropy $H_{\min}(X,\delta|\Ecal_{\rm cl})$ to calibrate the randomness extractor. To be sure that no intruder can violate the privacy of the generated random bits, we can physically block the vacuum input ports, represented in Figure \ref{fig:optcir} by the channel 1 (signal) and the channels 2 and 4 (vacuum); to block the unused ports is suggested also in \cite{+AS+18,Haw+15}.

\subsection{Side information: the signal} \label{sec:paran}

In \cite{Vill18} one of the reasons to propose double homodyne detection for QRNG was that one can obtain secure random bits even if an intruder can manipulate the signal (but in the same article this possibility is referred to as a ``paranoid scenario''). Indeed, in QRNG the 8-port circuit is not used to detect a signal coming from the external world, as in QKD, and it can be blocked to external influences, as noticed above. Moreover, if the intruder is sending too strong signals, as in a blinding attack \cite{Qin2018},  the intrusion is easily detected. On another side, to conduct a successful eavesdropping attack, one needs to send a signal phase locked to the LO laser \cite{Thewes2019} and this means to have access to the laboratory. In any case, it is possible to extend the approach of \cite{Vill18} and to take into account the quantum side information which an intruder could gain by manipulating the signal.

From \eqref{XlY+N} and \eqref{mixtprobdens}, the probability density for the $m$-sample turns out to be
\begin{equation}\label{pX}
p_{\vec X}(\vec x)=\Ebb_{f,N_{\rm el}} \biggl[\Tr_{\Gamma_1}\biggl\{\rho^{f,T}_{1}\prod_{l=1}^m\frac1{\abs\lambda^2}\,\hat g_Y^l\left((x_1^l-N_{\rm el}^{1,l})/\abs\lambda,(x_2^l-N_{\rm el}^{2,l})/\abs\lambda;f\right)\biggr\}\biggr],
\end{equation}
where $\hat g_Y^l(y_1,y_2;f)$ is given by \eqref{YPOVMdens}. By setting
\[
\hat P_l(x_1^l,x_2^l;f, N_{\rm el},\delta)=\frac1{\abs\lambda^2}\int_{x_1^l-\delta_1/2}^{x_1^l+\delta_1/2} \rmd x_1' \int_{x_2^l-\delta_1/2}^{x_2^l+\delta_1/2}\rmd x_2'\,\hat g_Y^l\left((x_1'-N_{\rm el}^{1,l})/\abs\lambda,(x_2'-N_{\rm el}^{2,l})/\abs\lambda;f\right),
\]
we can introduce a classical/quantum ``worst-case'' conditional guessing probability for the full $m$-sample
\begin{equation}\label{Pguess3}
P_{\rm guess}^{\rm full}(\vec X,\delta|\Ecal_{\rm cl \& qu})=\Ebb_{f,N_{\rm el}} \biggl[\sup_{\rho_{1}}\sup_{\vec x_1,\vec x_2}\Tr_{\Gamma_1}\biggl\{\rho_{1}\prod_{l=1}^m \hat P_l(x_1^l,x_2^l;f, N_{\rm el},\delta)\biggr\}\biggl]
\end{equation}
(cf.\ \cite{KRS09,DattaR09,Haw+15,Vill18,Smith2019}); then, the related min-entropy is
\begin{equation}\label{HTOT}
H_{\min}^{\rm full}(\vec X,\delta|\Ecal_{\rm cl\& qu})=-\log_2 P_{\rm guess}^{\rm full}(\vec X,\delta|\Ecal_{\rm cl \& qu})\geq 0.
\end{equation}
By comparing these definitions with \eqref{Pguess1}, \eqref{Htot}, \eqref{Pguess2}, \eqref{Hmin|}, we get
\[
mH_{\min}(X,\delta)\geq mH_{\min}(X,\delta|\Ecal_{\rm cl})\geq H_{\min}^{\rm full}(\vec X,\delta|\Ecal_{\rm cl\& qu}).
\]

Now, let us consider the (normalized) squeezed coherent states $|\psi_l(z;\alpha,\beta)\rangle$ \eqref{psil} and the POVM-density
$\hat g_{\alpha,\beta}^l(z)$, which is defined by the second equality in \eqref{eqprop3} and is involved in the definition \eqref{YPOVMdens} of $\hat g_Y^l(y_1,y_2;f)$. The operator
$\pi \hat g_{\alpha,\beta}^l(z)=|\psi_l(z;\alpha,\beta)\rangle\langle \psi_l(z;\alpha,\beta)|$ is a rank one orthogonal projection (for every choice of $\alpha$, $\beta$, $z$) and, so,
\begin{equation}\label{boundhatg}
\hat g_{\alpha,\beta}^l(z)\leq \openone/\pi.
\end{equation}
The results about secure QRNG developed in \cite{Vill18} are based on the bound \eqref{boundhatg} in the particular case $\alpha=1$, $\beta=0$. As it is proved in Appendix \ref{sec:densbound}, this bound implies
\begin{equation}\label{boundpX}
p_{\vec X}(\vec x)\leq  \frac {1} {\left(4\pi \kappa_{13} \kappa_{23}\abs\lambda^2\abs{\sin\phi}S_-\right)^m}.
\end{equation}
By introducing the \emph{entropy lower bound} per sample
\begin{equation}\label{Hbound}
H_{\text{lb}}( X,\delta|\Ecal_{\rm cl\& qu})=\log_2\frac{ 4\pi \kappa_{13} \kappa_{23}\abs{\sin\phi}\abs\lambda^2S_-}{\delta_1\delta_2}\,,
\end{equation}
from \eqref{Pguess3}, \eqref{HTOT}, \eqref{boundpX}, we obtain
\begin{equation}\label{HTOTineq}
H_{\min}^{\rm full}(\vec X,\delta|\Ecal_{\rm cl\& qu})\geq m \,H_{\text{lb}}( X,\delta|\Ecal_{\rm cl\& qu}).
\end{equation}
Even in the case of a possible variant of the definition of conditional min-entropy with respect to the quantum side-information, inequality \eqref{HTOTineq} would be valid, because the lower bound is independent of any kind of signal the intruder could send. Then,  the entropy lower bound per sample \eqref{Hbound} can be safely used to calibrate the randomness extractor; this bound holds even in the case of a coherent attack. However, let us note that the quantity \eqref{Hbound} is not guaranteed to be positive; so, to be  useful it needs $\abs{\sin\phi}$ to be not too small.

The dependence on $\abs{\sin\phi}$ implies a decrease of the min-entropy bound \eqref{Hbound} when the two observed quadratures are not perfectly complementary.
Moreover, with respect to the  transmissivities, the bound \eqref{Hbound} is maximum for $\eta_j=1/2$; indeed, by \eqref{kappas} we have
\begin{multline}
4\kappa_{13} \kappa_{23}=4
\sqrt{\prod_{j=1}^4\eta_j\left(1-\eta_j\right)}\left(\epsilon_1\xi_1+\epsilon_3\xi_3\right)
\left(\epsilon_2\xi_2+\epsilon_4\xi_4\right)
\\ {}\leq 4\kappa_{13} \kappa_{23}\big|_{\eta_j=1/2}=\frac{\epsilon_1\xi_1+\epsilon_3\xi_3}2\times \frac{\epsilon_2\xi_2+\epsilon_4\xi_4}2\,.
\end{multline}

With respect to the entropy losses \eqref{diff1}, to ask for independence from any quantum intrusion gives a further loss of min-entropy:
\begin{multline}\label{diff2}
H_{\min}(X,\delta|\Ecal_{\rm cl})-H_{\text{lb}}( X,\delta|\Ecal_{\rm cl\& qu})\simeq \log_2\frac {\sqrt{\kappa_{12} \kappa_{22}}} {2 \kappa_{13} \kappa_{23}\abs{\sin\phi}}
\\ {}=\log_2\frac {\sqrt{\left(1+\tilde V_1^{\,2} + \tilde \sigma_1^{\,2}\right)\left(1+\tilde V_2^{\,2} + \tilde \sigma_2^{\,2}\right)}} {\sqrt{4\eta_1\left(1-\eta_1\right)} \abs{\sin\phi}}\geq 0;
\end{multline}
here we have used equations \eqref{Hmin|}, \eqref{Hbound}, \eqref{j2decomp}, \eqref{G+V+s}. In a more explicit way we can write
\begin{equation}\label{1+tt}
1+\tilde V_j^{\,2} + \tilde \sigma_j^{\,2}=\frac{\left(1-\eta_{j+2}\right)\epsilon_j\xi_j^{\,2}+ \eta_{j+2}\epsilon_{j+2}\xi_{j+2}^{\;2}} {\eta_{j+2}\left(1-\eta_{j+2}\right)\left(\epsilon_j\xi_j+ \epsilon_{j+2}\xi_{j+2}\right)^2}.
\end{equation}
In the entropy loss \eqref{diff2} we can identify a first contribution, $\log_2 \left(\sqrt{4\eta_1\left(1-\eta_1\right)} \abs{\sin\phi}\right)^{-1}\geq 0$, due to the fact that the intruder is allowed to send any kind of squeezed states, such as the eigen-states of the squeezed modes $b_l$ \eqref{a2b}, see Remark \ref{al=bl}. The second contribution is
$\frac 12\sum_{j=1}^2\log_2 \left(1+\tilde V_j^{\,2} + \tilde \sigma_j^{\,2}\right)\geq 0$; note that $\tilde V_1^{\,2}$ and $\tilde V_2^{\,2}$ vanish in the rebalancing case of Remark \ref{rebalance}, while $\tilde \sigma_1^{\,2}$ and $\tilde  \sigma_2^{\,2}$ vanish when there are no optical losses, i.e.\ $\epsilon_j=1$. So, by using the entropy lower bound \eqref{Hbound}, one gets secure random bits also with respect to possible intrusions through the optical losses, which is one of the points raised in \cite{Lupo+21}.

\subsection{Examples} \label{sec:examp}
An interesting question is to understand the effects on the min-entropies of the various imperfections with respect to the balanced case of Sec.\ \ref{sec:perfectbc}.

\begin{remark} In the case of the perfect balancing \eqref{perfectbc}, equations \eqref{kappas}, \eqref{Deltas}, \eqref{G+V+s}, \eqref{noiseR}, \eqref{xHtot2} give
\[%\begin{gather*}
\tilde V_j^{\,2}=0, \qquad \tilde \sigma_j^{\,2}= \frac {1-\epsilon}\epsilon, \qquad \Upsilon_j=0, \qquad  \Theta_j=\frac{2{\sigma^{\rm el}_{j}}^2}{\epsilon\xi^2\abs\lambda^2 S_0}, \qquad E_{12}=\left(1+ \Theta_1\right)\left(1+ \Theta_2\right) .
\]%\end{gather*}
Then, \eqref{Htot}, \eqref{Hdinam}, \eqref{diffxxx}, \eqref{diff2} reduce to
\[
H_{\min}(X,\delta)=H_{\rm ref}\simeq \log_2\frac{\pi\epsilon\abs{\lambda\xi}^2 S_0\sqrt{\left(1+\Theta_1\right)\left(1+\Theta_2\right)}}{\delta_1\delta_2},
\]
\[
H_{\rm ref} -H_{\min}(X,\delta|\Ecal_{\rm cl})\simeq \log_2\left(S_0/S_-\right)+\log_2 \sqrt{\left(1+\Theta_1\right)\left(1+\Theta_2\right)},
\]
\[
H_{\min}(X,\delta|\Ecal_{\rm cl})-H_{\text{lb}}( X,\delta|\Ecal_{\rm cl\& qu})\simeq -\log_2 \left(\epsilon\abs{\sin\phi}\right).
\]
If, in addition, we have perfectly complementary quadratures, $\abs{\sin\phi}=1$, and
totally efficient detectors, $\epsilon=1$, the conditional min-entropy $H_{\min}(X,\delta|\Ecal_{\rm cl})$ and the lower bound
$H_{\text{lb}}( X,\delta|\Ecal_{\rm cl\& qu})$ become equal, which is the case  of \cite{Vill18}.
\end{remark}

To have an idea of the effects of unbalancing and losses, we particularize the choice of the various parameters. Firstly, we take $\epsilon_j=\epsilon$ and $\xi_j=\xi$, which means that the quantum efficiencies and the conversion factors of the four detectors are equal. We take also the same unbalancing in the two detection channels, i.e.\ $\eta_3=\eta_4=\eta$, while the first two beam splitters are taken well balanced, $\eta_1=\eta_2=1/2$. We take equal also the variances of the two electronic noises,
${\sigma^{\rm el}_{j}}^2={\sigma_{\rm el}}^2$. Then, equations \eqref{kappas}, \eqref{Deltas}, \eqref{detC+el},  \eqref{noiseR}, \eqref{1+tt}  give
\begin{equation}\label{simplified}\begin{split}
&\kappa_{j2}=\frac\epsilon 2 \,\xi^2,\quad \Delta_{j2}=\frac\epsilon2\,\xi\left(1-2\eta\right), \qquad \Upsilon_j=\epsilon\Upsilon, \qquad \Theta_j=\frac \Theta\epsilon,\qquad  \Upsilon=\left(1-2\eta\right)^2\frac{\abs{\lambda}^2 C_0}{2 S_0},
\\ & \Theta=\frac{2{\sigma_{\rm el}}^2}{\abs{\xi\lambda}^2 S_0} , \qquad 1+ \tilde V_j^{\,2}+ \tilde \sigma_j^{\,2}= \frac 1{4\eta\left(1-\eta\right)\epsilon}, \qquad \Sigma_j^{\,2}= \Sigma^2=\frac {\abs{\xi\lambda}^2}2\,S_0 \left(\Theta+\epsilon+\epsilon^2\Upsilon\right).
\end{split}\end{equation}
Finally, equations \eqref{def:Href}, \eqref{Hdinam}, \eqref{diffxxx}, \eqref{diff2} reduce to
\begin{equation}\label{red:Href}
H_{\rm ref}=\log_2 \frac{2\pi \Sigma^2}{\delta_1\delta_2}= \log_2 \left[\frac{\pi \abs{\xi\lambda}^2 S_0 }{\delta_1\delta_2}\left(\Theta+\epsilon+\epsilon^2\Upsilon\right)\right],
\end{equation}
\begin{equation}\label{red:Hdinam}
H_{\rm ref}-H_{\min}(X,\delta) =
-\frac 12 \log_2\left[1-\Big(1+\frac 1 {\epsilon\Upsilon}+\frac\Theta{\epsilon^2\Upsilon}\Big)^{-2}\right],
\end{equation}
\begin{equation}\label{diffxx}
H_{\rm ref} -H_{\min}(X,\delta|\Ecal_{\rm cl})\simeq \log_2\left(S_0/S_-\right)+\log_2 \left(1+\epsilon\Upsilon+\Theta/\epsilon\right),
\end{equation}
\begin{equation}\label{red:diff2}
H_{\min}(X,\delta|\Ecal_{\rm cl})-H_{\text{lb}}( X,\delta|\Ecal_{\rm cl\& qu})\simeq -\log_2\left[4\eta\left(1-\eta\right) \epsilon \abs{\sin\phi}\right].
\end{equation}

The reference entropy \eqref{red:Href} decreases when $\epsilon$ or $\abs{1-2\eta}$ decrease. However, its value can be kept constant by tuning the instrumentation, as discussed in Sec.\ \ref{sec:optim}, and reasonable values for $H_{\rm ref}$ go
from 14 to 26 bits, see Table \ref{table1}. So, here we study the behaviour of the entropy differences \eqref{red:Hdinam}--\eqref{red:diff2}.

We take the parameters $\epsilon$ and $\eta$ free, and we fix some reasonable values for the other parameters; let us recall that $\abs{1-2\eta}$ measures the unbalancing in the detectors, while $\epsilon$ is the effective quantum efficiency of the photodiodes. The parameter $S_0$, given in \eqref{def:S_0}, \eqref{S0C0}, is linked to the detection bandwidth and we can take
$S_0=10 \; \text{GHz}=10^{10}\;\text{sec}^{-1}$; we take also $S_-\simeq S_0$, see Eq.\ \eqref{def:S_-} and Remark \ref{S-:S_0}. With lasers of a power of 1.25 mW and wavelength around 1550 nm,  we can get the mean number of photons per unit of time $\abs\lambda^2 = 10^{16}\;\text{sec}^{-1}$; higher values are also possible. By assuming the decay time of the RIN correlations much shorter of the decay time of the detector response function $h(t)$, from Eqs.\ \eqref{def:C_0}, \eqref{S0C0} and Remark \ref{S-:S_0} we see that $C_0$ must be small and we take $C_0=0.01$. Finally, we can take $\Theta=\frac{2\sigma_{\rm el}^{\;2}}{\abs{\lambda\xi}^2S_0}=0.12$, which comes out by taking a variance for the electronic noise compatible with the values reported in \cite{Vill18}. With these choices we get also $\frac{\abs\lambda^2 C_0}{2S_0}=0.5 \times 10^4$.

\begin{figure}[H]
\begin{center}
\includegraphics[scale=0.9]{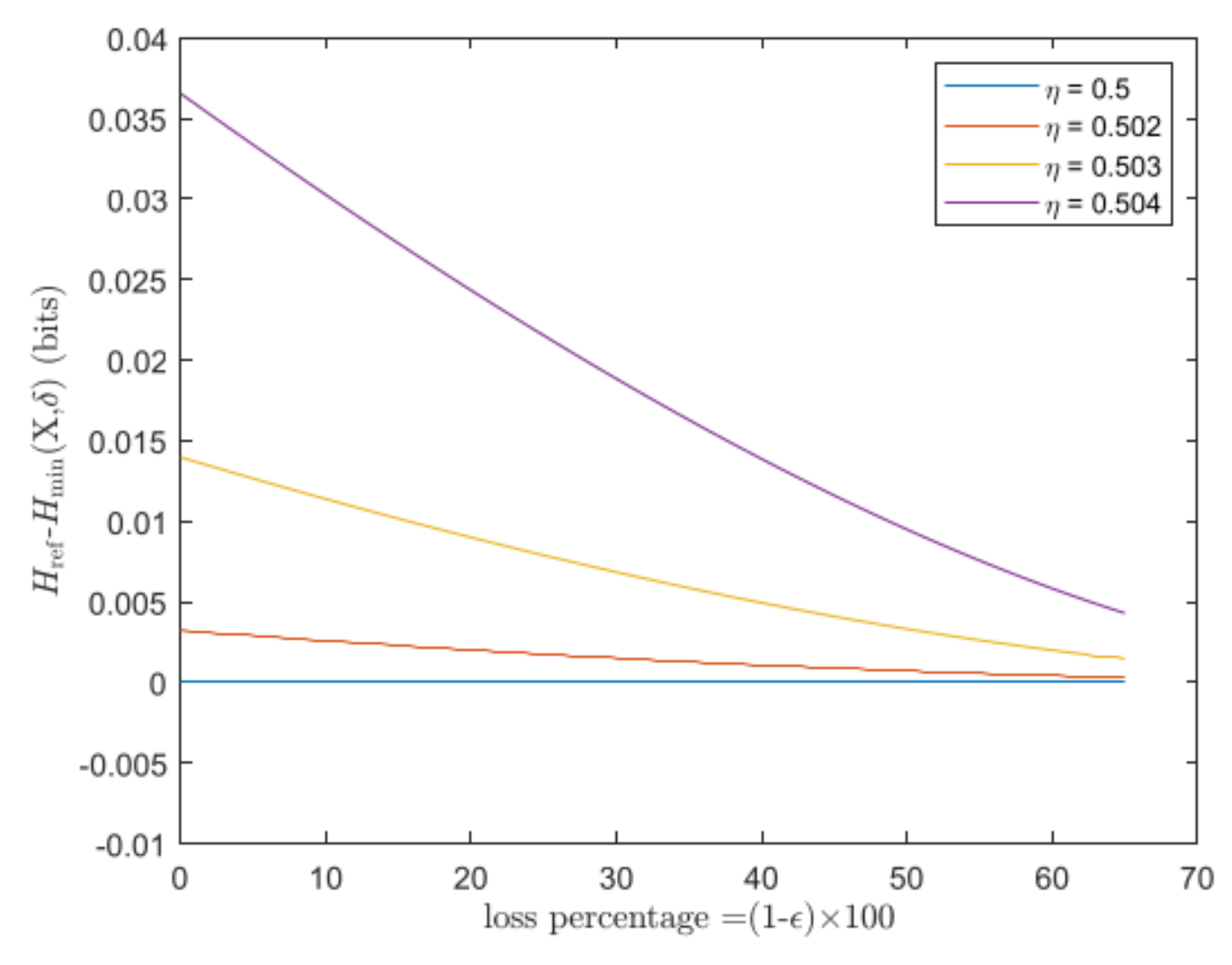}
\end{center}
\caption{Entropy loss $H_{\rm ref}-H_{\min}(X,\delta)$, Eq.\ \eqref{red:Hdinam}, for small unbalancing $\abs{1-2\eta}\leq 0.008$.}\label{fig:DH1+}
\end{figure}

\begin{figure}[H]
\begin{center}
\includegraphics[scale=0.9]{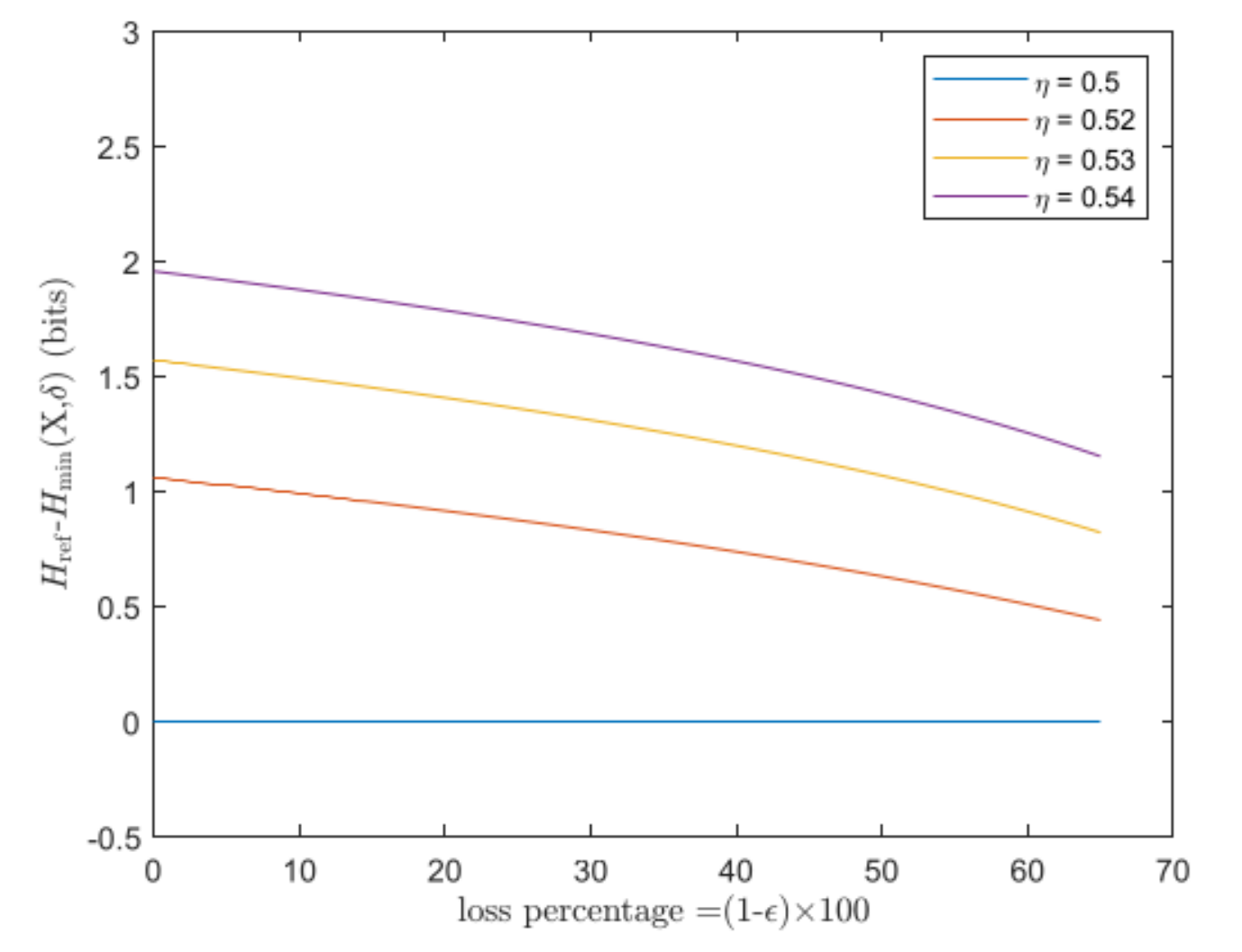}
\end{center}
\caption{Entropy loss $H_{\rm ref}-H_{\min}(X,\delta)$, Eq.\ \eqref{red:Hdinam}, for big unbalancing $0.04\leq\abs{1-2\eta}\leq 0.08$.}\label{fig:DH1++}
\end{figure}

Let us start with the entropy losses \eqref{red:Hdinam}, due to the correlations introduced by the RIN. We plot this entropy loss with respect to the ``quantum loss percentage $= 100\left(1-\epsilon\right)$'' of the photodiodes for small values of the detector unbalancing $\abs{1-2\eta}$ in Figure \ref{fig:DH1+}, and for bigger unbalancing in Figure \ref{fig:DH1++}.
From Figures \ref{fig:DH1+} and \ref{fig:DH1++} we see that the entropy loss \eqref{red:Hdinam} can be made to decrease by increasing the detector losses $1-\epsilon$ (apart from the balanced case $\eta=1/2$), a phenomenon which is evident also from formula \eqref{red:Hdinam}, but not a priori expected. Note a change of curvature in going from the cases of Fig.\ \ref{fig:DH1+} to the cases of Fig.\ \ref{fig:DH1++}.
We see also that the entropy loss decreases with the decreasing of the distance of $\eta$ from 1/2. Moreover, this loss is acceptably small for small unbalancing as in Fig.\ \ref{fig:DH1+}, while it can take values of even two bits per sample in the cases of Fig.\ \ref{fig:DH1++}.

\begin{figure}[H]
\begin{center}
\includegraphics[scale=0.9]{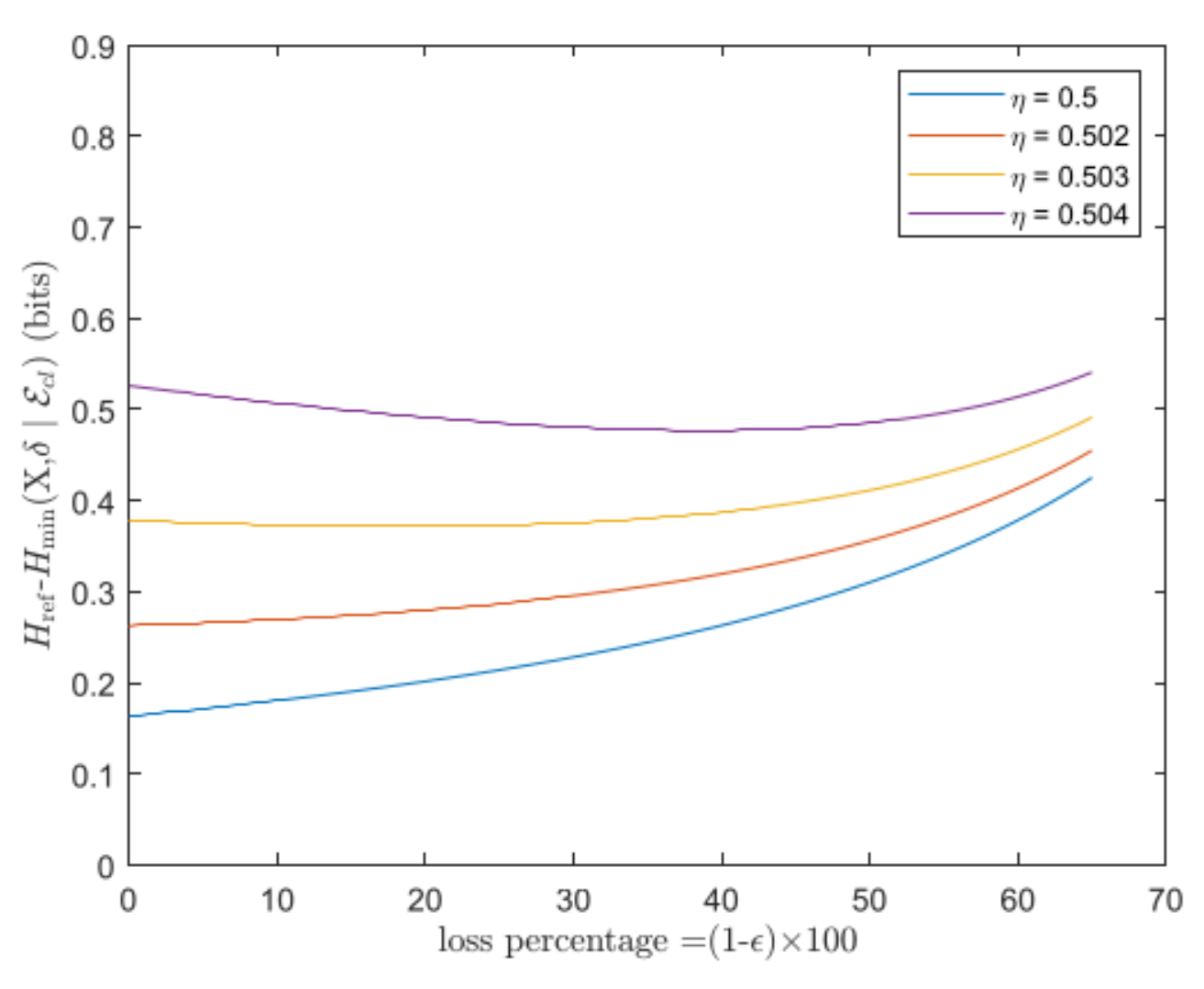}
\end{center}
\caption{Entropy loss $H_{\rm ref} -H_{\min}(X,\delta|\Ecal_{\rm cl})$, Eq.\ \eqref{diffxx},  for small unbalancing $\abs{1-2\eta}\leq 0.008$. }\label{fig:DH2+}
\end{figure}

\begin{figure}[H]
\begin{center}
\includegraphics[scale=0.9]{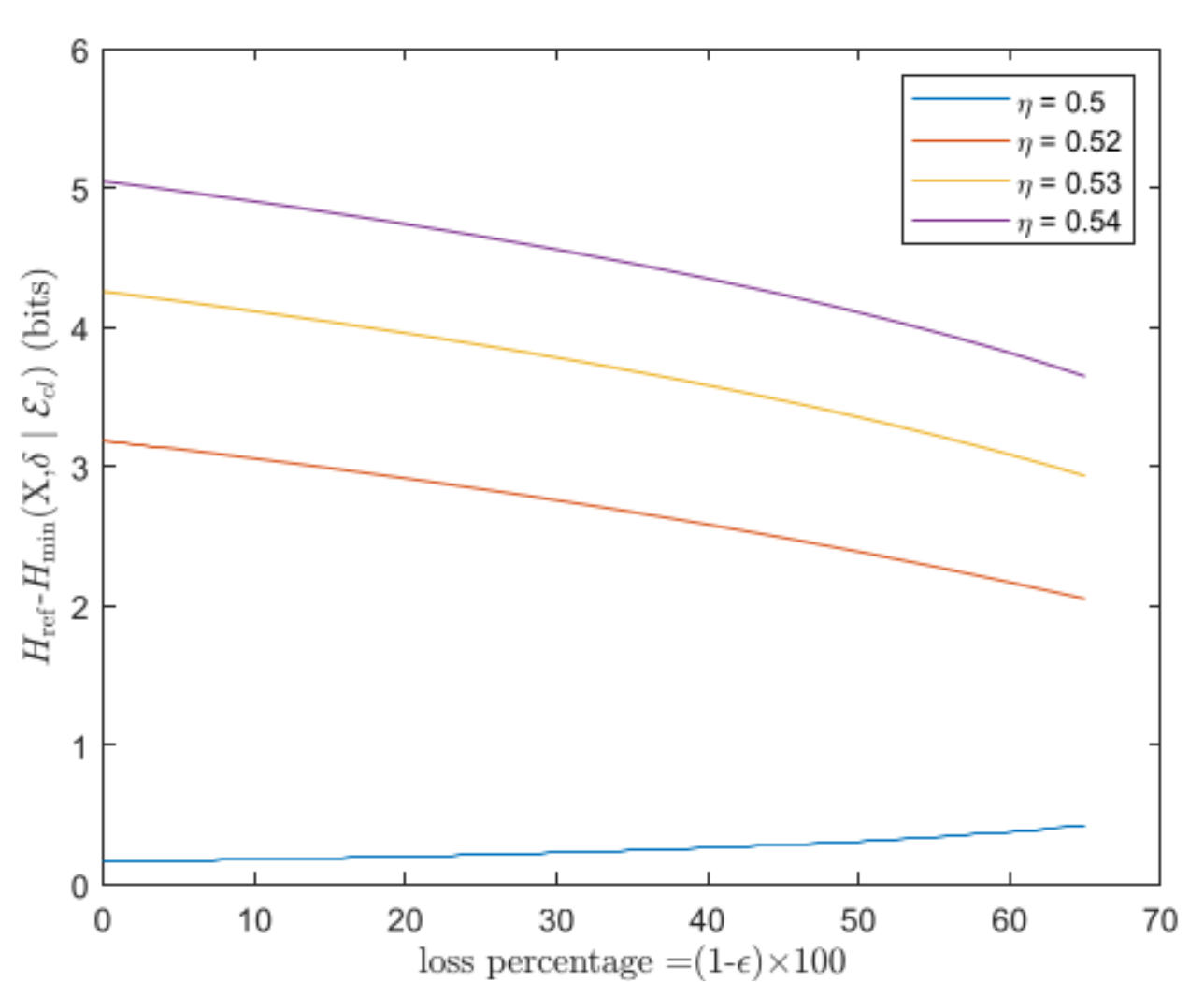}
\end{center}
\caption{Entropy loss $H_{\rm ref} -H_{\min}(X,\delta|\Ecal_{\rm cl})$, Eq.\ \eqref{diffxx}, for big unbalancing $0.04\leq\abs{1-2\eta}\leq 0.08$.} \label{fig:DH2++}
\end{figure}

When the contributions of the classical noises (RIN and electronic noise) are considered not secure, the entropy loss is given by
\eqref{diffxx}; the behaviour is plotted in Figures \ref{fig:DH2+} and \ref{fig:DH2++} for the same values of $\eta$ as before.
Now we have again a decrease of the entropy loss with the increase of $1-\epsilon$ in the case of large imbalance, see Fig.\ \ref{fig:DH2++}. However, for small imbalance (Fig.\ \ref{fig:DH2+}), the behaviour is more complex. For $\eta =0.5$ and $\eta = 0.502$, the entropy loss monotonically grows with $1-\epsilon$, while for $\eta=0.503$ and $\eta=0.504$ we have a non-monotonic behaviour, first a decreasing and then an increasing of the entropy loss.

To avoid losses of some bits in the case of large unbalancing, as in the case of Figs.\ \ref{fig:DH1++} and \ref{fig:DH2++}, one has to relay on a partial rebalancing, see Remark \ref{rebalance}, or on a change of the detector response in order to decrease the coefficient $C_0$.

In the extreme case of Sec.\ \ref{sec:paran}, when we consider not secure even the signal port, we have to add a further entropy loss represented by \eqref{red:diff2}. Now the behaviour is very simple: minus the logarithm of a product. For
$4\eta\left(1-\eta\right) \epsilon \abs{\sin\phi}=0.5$ the entropy loss is 1 bit; we can think to arrive to 2 bits by increasing a little the imperfections (and a such loss is not small). However, with reasonable values of the involved parameters, we can remain below 1 bit.

\section{QRNG from single homodining}\label{sec:singlehom}
The whole construction of the previous sections can be particularized to the case of a single homodyning and this allows to see the effect of imperfections either when the apparatus works as a detector, as in \cite{ZolG97,WisM10,Bar06,Bar91,YMC+11}, either when it is used for QRNG, as in \cite{Haw+15,Smith2019,Thewes2019,+AS+18,APFPS20,Huang+20,Lupo+21,Extractor,Qin2018}.

One can get single homodyning from the circuit of Figure \ref{fig:optcir} by eliminating the beam splitters BS1 and BS2, which means to take $\eta_1=\eta_2=1$. Then, no light arrives to the photodiodes PD2 and PD4 and we can eliminate them; so, we are left with a single detected process $X_1(t)$ \eqref{Xprocs}, and the scaled process $Y_1(t)$ \eqref{Yproc}.

The first two moments of the process $X_1(t)$ are given by Eqs.\ \eqref{EXprocs}, \eqref{XXcorr}, where the coefficients, defined in \eqref{kappas}, \eqref{Deltas}, reduce to
\begin{equation}\label{kappasSH}
\begin{split}
&\kappa_{11}=\eta_3\epsilon_1\xi_1^{\,2}+\left(1-\eta_3\right)\epsilon_3\xi_3^{\,2},
\qquad \kappa_{12}=\left(1-\eta_3\right)\epsilon_1\xi_1^{\,2}+\eta_3\epsilon_3\xi_3^{\,2} ,
\\
&\kappa_{13}=\sqrt{\eta_3\left(1-\eta_3\right)}\left(\epsilon_1\xi_1+\epsilon_3\xi_3\right),
\qquad\Delta_{11}=\eta_3\epsilon_1\xi_1-\left(1-\eta_3\right)\epsilon_3\xi_3 ,
\\ &\Delta_{12}=\left(1-\eta_3\right)\epsilon_1\xi_1- \eta_3\epsilon_3\xi_3 ,
\qquad\Delta_{13}=\sqrt{\eta_3\left(1-\eta_3\right)}\left(\epsilon_1\xi_1^{\;2}- \epsilon_3\xi_3^{\;2}\right).
\end{split}
\end{equation}
Moreover, the characteristic functional of the process $Y_1(t)$ in the limit of strong LO is given by Proposition \ref{prop:Y}, where one has to take $k_2(s)=0$. The various quantities introduced in Eqs.\ \eqref{G+V+s} reduce to
\begin{equation*}%\begin{split}
G_{13}=0,  \qquad  V_1^{\,2}=\tilde V_1^{\,2}= \frac{\Delta_{12}^{\;2}}{\kappa_{13}^{\;2}}, \qquad \sigma_1^{\,2}=\tilde \sigma_1^{\,2}= \frac{\left(1-\eta_{3}\right) \epsilon_1\left(1- \epsilon_1\right)\xi_1^{\,2} +\eta_{3} \epsilon_{3}\left(1- \epsilon_{3}\right)\xi_{3}^{\;2}} {\eta_{3}\left(1-\eta_{3}\right)\left(\epsilon_1\xi_1+\epsilon_{3}\xi_{3}\right)^2}.
%\end{split}
\end{equation*}
Let us stress that the POVM with characteristic operator \eqref{Psif} becomes the pvm associated to the single quadrature $\hat Q_1(t)$.

Then, one can consider the discrete sampling as in Sec.\ \ref{dsampling}; if interested in QRNG, we have to consider the case of no signal as in the double homodyne set up. From \eqref{vaX}, \eqref{detC+el} we get
\begin{subequations}\label{momXsingle}
\begin{equation}
\Ebb_P\left[X_{1}(t_l)\right]  \simeq \Delta_{12}\abs\lambda^2,
\end{equation}
\begin{equation}
\Cov_P[X_1(t_l),\,X_1(t_l')]\simeq
\delta_{ll'}\Sigma_1^{\,2}= \delta_{ll'}\kappa_{12}\abs\lambda^2 S_0\left(1
+\Upsilon_1 + \Theta_1\right),
\end{equation}
\end{subequations}
where $\Upsilon_1$ and $\Theta_1$ are given in \eqref{noiseR}. However, the coefficients take the expressions \eqref{kappasSH}, which means that their values are nearly two times the values in the double homodyne scheme.

\subsection{The total min-entropy}

For a sufficiently intense LO and signal in the vacuum, $X_1(t_l)$ has a distribution which is nearly Gaussian with mean $\mu_1=\Delta_{12}\abs\lambda^2$ and variance $\Sigma_1^{\,2}$, given in \eqref{momXsingle}. Let us denote by $p_{X_1}(x)$ the density of this Gaussian distribution. Now the guessing probability and the associated min-entropy per sample
\eqref{Pguess1}, \eqref{Htot} become
\begin{equation}\label{Pguess1sin}
P_{\rm guess}(X_1,\delta_1)\simeq \sup_{x}\int_{x-\delta_1/2}^{x+\delta_1/2}\rmd y \,p_{X_1}^l(y)\simeq \frac{\delta_1}{\sqrt{2\pi}\,\Sigma_1},
\end{equation}
\begin{equation}\label{Htotsin}
H_{\min}(X_1,\delta_1)= -\log_2P_{\rm guess}(X_1,\delta_1)\simeq \log_2\frac{\sqrt{2\pi}\,\Sigma_1}{\delta_1}.
\end{equation}
In the univariate case correlations are not involved; so, the analogous of $H_{\rm ref}$ \eqref{def:Href} coincides with the total min-entropy \eqref{Htotsin}.

As discussed in Sec.\ \ref{sec:optim}, we can try to optimize the discretization range $2R_1$ by acting on the ADC and on the laser power. Analogously to \eqref{RproptoSigma}, we can write $R_1=x\Sigma_1$, which gives
\begin{equation*}
\delta_1=\frac{x\Sigma_1}{2^{n-1}},\qquad  \tilde P_{\rm guess}(X_1,\delta_1)\simeq\frac{x}{\sqrt{\pi}\,2^{n-1/2}},\qquad H_{\min}(X_1,\delta_1)\simeq n-\frac 12+ \log_2\frac{\sqrt \pi}x;
\end{equation*}
these expressions are the analog of \eqref{Pguessapprox}, \eqref{Href-n}. Now, the saturation probability \eqref{satprob} and the condition \eqref{ass:negl} become
\begin{equation}\label{satprobsingle}
P_{\rm saturation}(X_1,\delta_1)=1- P\left[-R_1<X_1(t_l)-\mu_1<R_1\right] =  2\left[1-\Phi(x)\right] < P_{\rm guess}(X_1,\delta_1).
\end{equation}

As done in Table \ref{table1} for $H_{\rm ref}$, we can give the min-entropy $H_{\min}(X_1,\delta_1)$ for some values of $x$ and $n$:
\begin{table}[H]
\caption{The min-entropy $H_{\min}(X_1,\delta_1)$ as a function of the proportionality parameter $x$  and of the ADC number of bits $n$. A blank value means that the inequality in \eqref{satprobsingle} is not satisfied.}\label{table3}
\begin{center}
\begin{tabular}{c||c|c|c|c|c|c|c|}
  $n\backslash x$&3.0 & 3.4 & 4.0&4.6& 6.1& 8.9 &9.5 \\
  \hline\hline
  8&6.74 & 6.56 & 6.33 &6.12 & 5.72 & 5.17 &5.08\\
  \hline
 10& \phantom{H} & 8.56 & 8.33 & 8.12& 7.72& 7.17&7.08\\
  \hline
  12&\phantom{H} & \phantom{H} & 10.33 & 10.12&9.72&9.17 &9.08\\
  \hline
 16& \phantom{H} & \phantom{H} & \phantom{H} & 14.12& 13.72&13.17 &13.08\\
  \hline
32& \phantom{H} & \phantom{H} & \phantom{H} & \phantom{H}& 29.72&29.17 &29.08\\
  \hline
\end{tabular}
\end{center}
\end{table}

By comparing Table \ref{table3} with Table \ref{table1}, we see that, for the same values of $x$ and $n$, the min-entropy $H_{\min}(X_1,\delta_1)$ is half of $H_{\rm ref}$, as expected because in $H_{\rm ref}$ the correlations between the two outputs are not taken into account. Moreover, by the positions of the blank values we see that the inequality in \eqref{satprobsingle} is a less stringent condition than \eqref{ass:negl}.

\subsection{Side information}
When the classical noise is not trusted, as in Sec.\ \ref{sec:|cl}, we have to relay on the conditional min-entropy. The analogs of
\eqref{Pguess2} and \eqref{Hmin|} are now
\begin{equation}
P_{\rm guess}(X_1,\delta_1|\Ecal_{\rm cl})
\simeq \Ebb_f\left[\frac {\delta_1}{ \sqrt{2\pi\kappa_{12}}\abs{\lambda} R_l(f)}\right], \qquad H_{\min}(X_1,\delta_1|\Ecal_{\rm cl})\simeq \log_2\frac { \abs\lambda\sqrt{2\pi\kappa_{12} }} {\delta_1\Ebb_f \left[R_l(f)^{-1}\right]}.
\end{equation}
As in \eqref{diff1}, \eqref{diffxxx}, by considering unreliable the classical noise we have the entropy loss
\begin{equation}\label{Hdiffsing}
H_{\min}(X_1,\delta_1) -H_{\min}(X_1,\delta_1|\Ecal_{\rm cl})\simeq \frac12 \log_2\left(1
+\Upsilon_1+\Theta_1\right)+\log_2\left(\Ebb_f \left[R_l(f)^{-1}\right]\sqrt{S_0}\right),
\end{equation}
where the expressions of $\Upsilon_1$ and $\Theta_1$ are given in \eqref{noiseR}.

\begin{figure}[H]
\begin{center}
\includegraphics[scale=0.9]{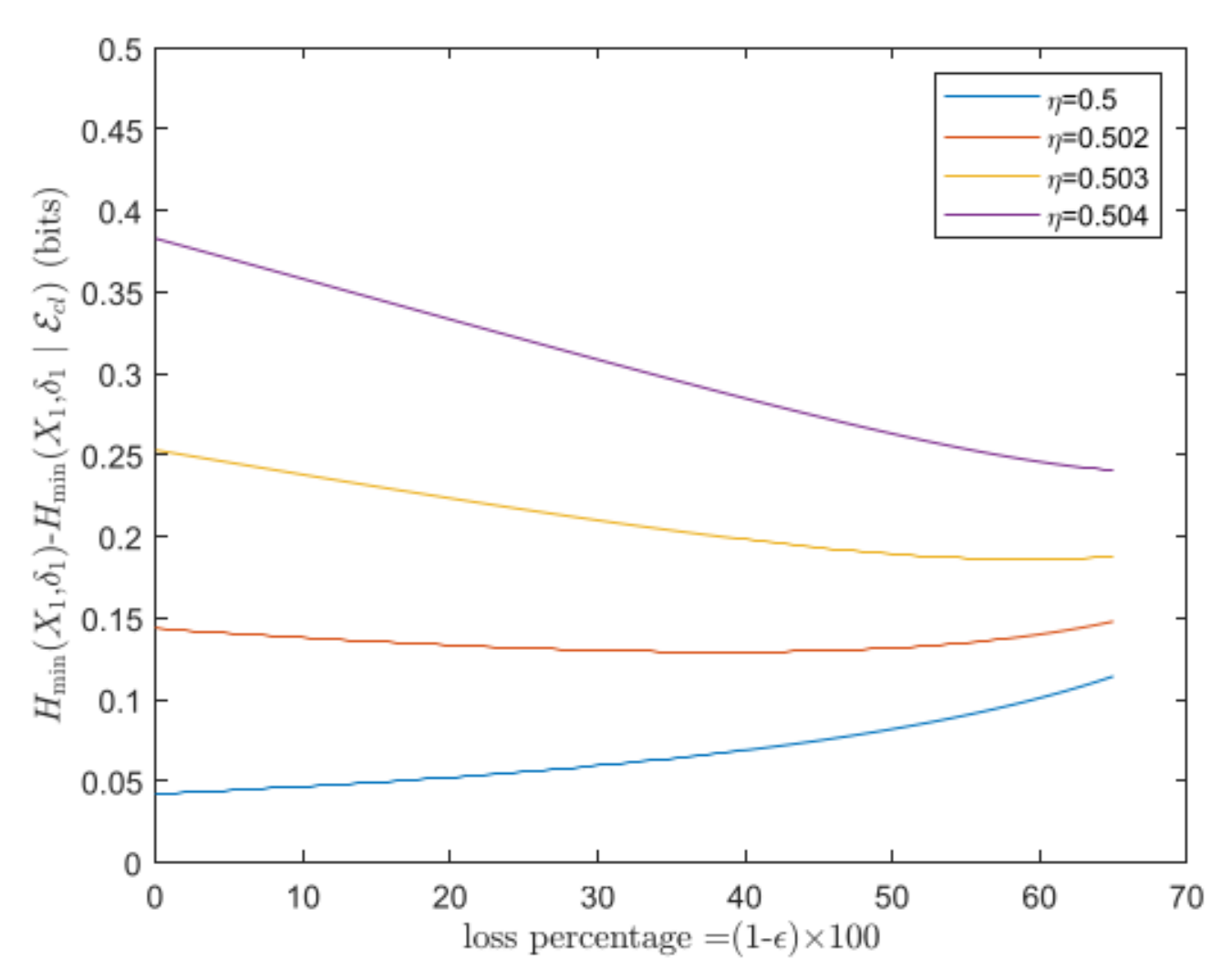}
\end{center}
\caption{Entropy loss $H_{\min}(X_1,\delta_1) -H_{\min}(X_1,\delta_1|\Ecal_{\rm cl})$, Eq.\ \eqref{Hdiffsing2}. \  Small unbalancing $\abs{1-2\eta}\leq 0.008$. }\label{fig:1S6}
\end{figure}

\begin{figure}[H]
\begin{center}
\includegraphics[scale=0.9]{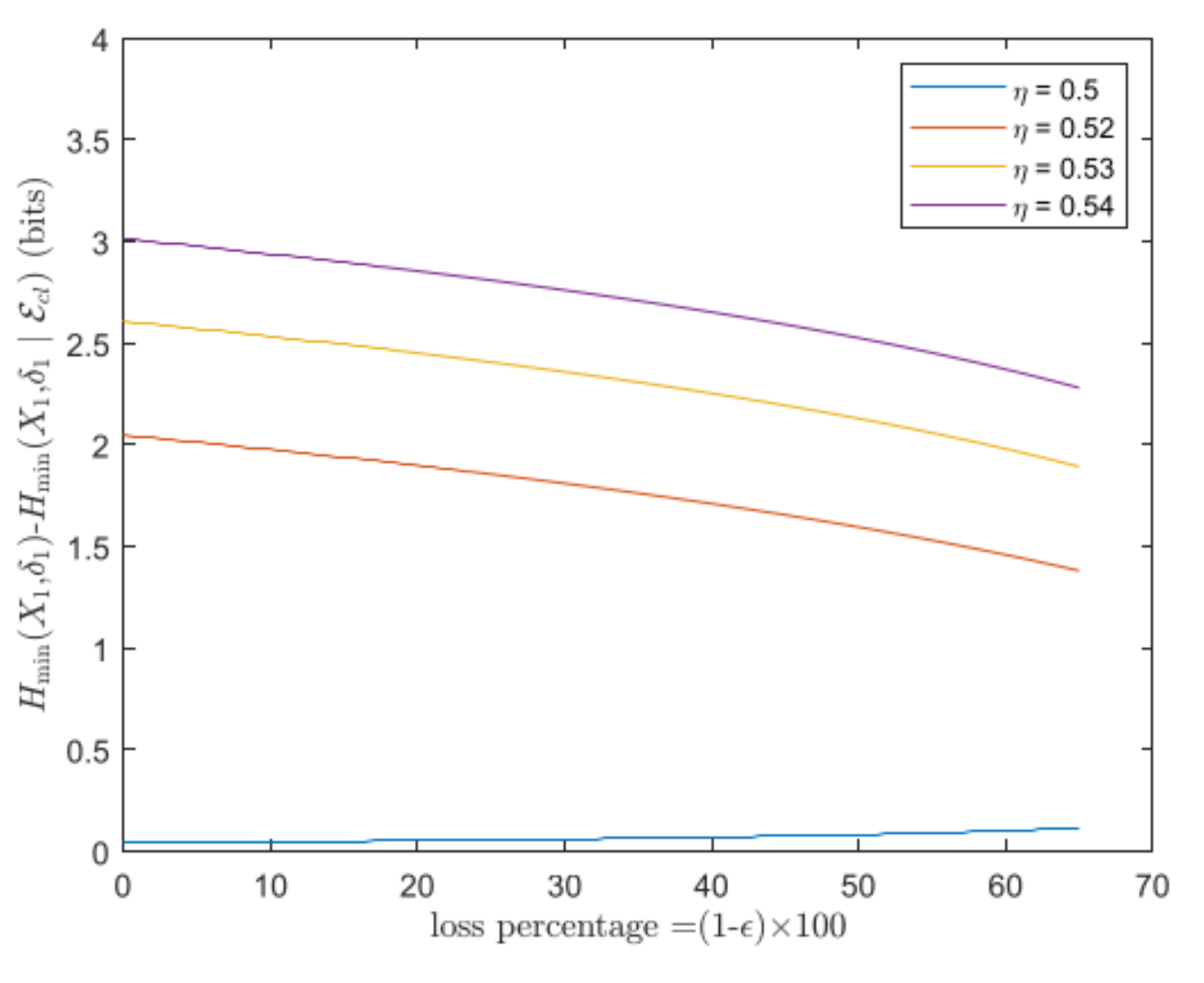}
\end{center}
\caption{Entropy loss $H_{\min}(X_1,\delta_1) -H_{\min}(X_1,\delta_1|\Ecal_{\rm cl})$, Eq.\ \eqref{Hdiffsing2}. \ Big unbalancing $0.04\leq\abs{1-2\eta}\leq 0.08$.} \label{fig:2S6}
\end{figure}

To have examples of the effects of the various imperfections on this entropy loss, we take
$\epsilon_1=\epsilon_3=\epsilon $ and $\xi_1=\xi_3=\xi$ as in Sec.\ \ref{sec:examp} and we set also $\eta=\eta_3$.
By assuming $\log_2\left(\Ebb_f \left[R_l(f)^{-1}\right]\sqrt{S_0}\right)\simeq 0$ and by using the expressions \eqref{kappasSH} of the various parameters, we get
\begin{equation}\label{Hdiffsing2}
H_{\min}(X_1,\delta_1) -H_{\min}(X_1,\delta_1|\Ecal_{\rm cl})\simeq \frac12\, \log_2\left(1
+2\epsilon\Upsilon+\Theta/2\epsilon\right),
\end{equation}
where $\Upsilon$ and $\Theta$ are defined in \eqref{simplified}. With the numerical choices for the various parameters discussed in Sec.\ \ref{sec:examp} we get $2\epsilon\Upsilon=10^4\epsilon\left(1-2\eta\right)^2$ and $\Theta/2\epsilon=0.06/\epsilon$. The analogous entropy loss for the double homodyne case is given in  \eqref{diffxx}, when $\log_2\left(S_0/S_-\right)\simeq 0$. Note the different expressions of the coefficients in front of the RIN contribution $\Upsilon$ and the electronic noise contribution $\Theta$.
In Figs.\ \ref{fig:1S6} and \ref{fig:2S6} we give the plots of the entropy loss \eqref {Hdiffsing2} with the same choice for the values of the involved parameters as in the analogue figures  \ref{fig:DH2+}, \ref{fig:DH2++}. The qualitative behaviour is very similar in the two cases of double and single homodyne scheme.

By comparing Table \ref{table3} with Figures \ref{fig:1S6} and \ref{fig:2S6}, again we see that the most important parameter for the rate of random bit production is $n$, giving the ADC resolution.

The single homodyne scheme is considered experimentally simpler than the double scheme \cite{Smith2019,Lupo+21}, but now we have not some analog of what is done in Sec.\ \ref{sec:paran}. However, as already discussed at the end of Sec.\ \ref{sec:|cl}, we can relay on the conditional min-entropy $H_{\min}(X_1,\delta_1|\Ecal_{\rm cl})$ to calibrate the randomness extractor, because  we can physically block the vacuum input port. Even in the case in which an intruder can send a signal through the signal port we can follow the strategy suggested in \cite{Smith2019}. By using a pulsed laser we can make the measured quadrature at time $t_l$ to be characterized by a new random phase (see $\theta$ in \eqref{LOf+}) not known to the intruder. This prevents the intruder to be able of sending eigenstates of the measured quadrature and to have some knowledge of the produced random numbers.

An alternative strategy, which can be applied also in our general case, is the one proposed in \cite{Vill17}: one samples a fixed quadratures, but, to avoid a possible eavesdropping, the orthogonal quadrature is sampled at \emph{random times}; then, an entropy lower bound is estimated by an entropic version of the uncertainty relations. A variant of this strategy is discussed in
\cite{Kra22}, where also the detector inefficiency is taken into account. However, this strategy is not so simpler than the use of the double homodyning.  The 8-port circuit needs two ADC components and four photodiodes; the random sampling strategy needs only an ADC and two photodiodes, but a component which switches the phases at random times, with its own source of randomness, has to be added.

\section{Discussion}\label{sec:conclusion}

By using the example of the eight-port optical circuit, we have shown how to use QSC to give a fully quantum treatment of travelling waves in the circuit and of direct, homodyne, heterodyne detection in continuous time. A key point is that the number operators of the quantum fields used in QSC are a family of commuting selfadjoint operators and their joint pvm can be introduced, see Remark \ref{rem:compatible}. Then, the POVMs related to the field quadratures enter into play when the Hilbert space component associated with the LO field is traced out and the system is reduced to the Hilbert component associated with the signal alone. Moreover, we have shown how to introduce imperfections such as imbalanced beam splitters, phase and intensity noise in the LO laser, inefficiency of the photodetectors, electronic noise. In this way we have a complete quantum description of the apparatus of Fig.\ \ref{fig:optcir} monitoring in continuous time the quantum field entering the signal port.

We consider also the case in which the continuous output is sampled at discrete times. Now discrete mode operators can be introduced, but they result to be random operators, because the LO fluctuations are involved in the definition of such operators, see Remark \ref{rem:randop}. So, even in this case, the description in continuous time is an essential step, as already suggested in \cite{APFPS20}.

Finally, we study the case in which the apparatus is used as QRNG, both in the cases of double and single homodyne detection. Our treatment allows for taking into account the various imperfections on the detected noise. By the use of the conditional min-entropy the effect of untrusted parts of the noise can be eliminated. While the electronic noise is treated as additive noise as usual, a very peculiar role is played by the LO intensity noise which is involved in the construction of the discrete quadrature operators and its effect is less straightforward, see Remark \ref{rem:RIN}. We show how the random bit generation rate decreases when we consider that side information could be gained by an intruder through the classical noise or when he could have also access to the input signal port (quantum side information). Some experimentally reasonable examples are treated and the entropy losses are numerically computed, depending on some parameters characterizing the imperfections in the circuit and in the detection part of the apparatus.

Our quantum analysis of the apparatus of Fig.\ \ref{fig:optcir} could be applied when it is employed as a detector, for instance in problems of quantum teleportation and quantum dense coding, see \cite{BKS14}, or of QKD, see \cite{APFPS20}. Note that in QKD and QRNG the security instances are completely different. In QRNG the unused ports can be physically blocked, see the discussion at the end of Sec.\ \ref{sec:|cl}, while QKD is a problem of secure communication and an intruder could have access to the transmitted signal.

\appendix

\section{Properties of the Bose fields and some quantum expectations}\label{app:fields}
Let us recall some properties of the Bose fields $a_j(t)$ introduced in Sec.\ \ref{sec:ocpd} \cite{Parthas92}.
We work in the Fock representation, which means that the CCRs are realized in  the Hilbert space
\begin{equation}\label{Focksp}
\Gamma={\prod_{j=1}^d}^\otimes \Gamma_j, \qquad \Gamma_j= \Cbb\oplus \sum_{n=1}^\infty L^2({\mathbb R})^{\otimes_s n};
\end{equation}
$\Gamma_j$ is the \emph{symmetric Fock space} over the one-particle space $L^2({\mathbb R})$ and the direct sum on the right is its decomposition in the $n$-particle spaces.
In all developments, a key role is played by the coherent vectors $e_j(f)\in \Gamma_j$, or normalized \emph{exponential vectors}, which can be introduced by giving their components in the $n$-particle spaces: for $f\in L^2({\mathbb R})$,
\begin{equation}\label{expV}
e_j(f) =\rme^{
-\frac{1}{2}\,\norm{f}^2}\left(1,f,\frac{f\otimes f}{\sqrt{2!}},\ldots,\frac{ f^{
\otimes n}}{\sqrt{n!}},\ldots \right).
\end{equation}
These vectors are completely analogous to the coherent vectors of the case of discrete modes, as one sees by comparing the representations in the spaces with fixed number of photons.

Finally, let us recall that QSC \cite{Parthas92} is an It\^o-type calculus involving the integral form \eqref{fdensity} of the  quantum fields; by this calculus a theory of quantum stochastic differential equations has been developed. In handling the ``stochastic differentials'' a ``promemoria'' is given by the It\^o table:
\begin{equation}\label{Itotab}
\begin{split}
\rmd A_k(t)\rmd A_l^\dagger(t)=\delta_{kl}\rmd t,
\qquad &\rmd A_i(t) \rmd \Lambda^A_{kl}(t)= \delta_{ik} \rmd A_l(t),
\\
\rmd \Lambda^A_{kl}(t)\rmd A_i^\dagger(t)=\delta_{li}\rmd A_k^\dagger(t),
\qquad &\rmd \Lambda^A_{kl}(t)\rmd \Lambda^A_{ij}(t)=\delta_{li}\rmd \Lambda^A_{kj}(t),
\end{split}
\end{equation}
all the other products vanish. This table has the same role as the heuristic rule $(\rmd W(t))^2=\rmd t$ in classical It\^o stochastic calculus. The rigorous definition of field and gauge operators \eqref{fdensity} is through their action on the exponential vectors.

\subsection{Output fields}
By combining Eqs.\ \eqref{AtoB}, \eqref{BtoC}, \eqref{CtoD} we express the output fields $D_j(t)$ in terms of the fields $A_j(t)$, $A_{j+}(t)$; in terms of field densities we get
\begin{subequations}\label{fieldsdj}
\begin{multline}
d_1(t)= \sqrt{\eta_3\epsilon_1}\left[\sqrt{\eta_1}\,a_1(t)+\rmi \sqrt{1-\eta_1}\, a_2(t) \right]\\ {}+ \rme^{\rmi \psi_1}\sqrt{\left(1-\eta_3\right)\epsilon_1}\left[\rmi\sqrt{\eta_2}\,a_3(t)-  \sqrt{1-\eta_2}\,a_4(t)\right] +\rmi \sqrt{1-\epsilon_1}\, a_{1+}(t),
\end{multline}
\begin{multline}
d_3(t)
=\sqrt{\left(1-\eta_3\right)\epsilon_3}\left[\rmi \sqrt{\eta_1}\,a_1(t)- \sqrt{1-\eta_1}\, a_2(t) \right]\\ {}+ \rme^{\rmi \psi_1}\sqrt{\eta_3\epsilon_3}\left[\sqrt{\eta_2}\,a_3(t)+\rmi   \sqrt{1-\eta_2}\,a_4(t)\right]+\rmi \sqrt{1-\epsilon_3}\, a_{3+}(t),
\end{multline}
\begin{multline}
d_2(t)
=\sqrt{\eta_4\epsilon_2}\left[\rmi\sqrt{1-\eta_1}\, a_1(t)+ \sqrt{\eta_1}\, a_2(t)\right]\\ {}+ \rme^{\rmi \psi_2} \sqrt{\left(1-\eta_4\right)\epsilon_2}\left[-   \sqrt{1-\eta_2}\,a_3(t)+ \rmi \sqrt{\eta_2}\,a_4(t)\right]+\rmi \sqrt{1-\epsilon_2}\, a_{2+}(t),
\end{multline}
\begin{multline}
d_4(t)
=\sqrt{\left(1-\eta_4\right)\epsilon_4}\left[-\sqrt{1-\eta_1}\, a_1(t)+\rmi \sqrt{\eta_1}\, a_2(t)\right]\\ {}+ \rme^{\rmi \psi_2} \sqrt{\eta_4\epsilon_4}\left[\rmi   \sqrt{1-\eta_2}\,a_3(t)+ \sqrt{\eta_2}\,a_4(t)\right]+\rmi \sqrt{1-\epsilon_4}\, a_{4+}(t).
\end{multline}
\end{subequations}

\subsection{Properties of the laser}\label{sec:LOprop}

Let us collect here some features of the laser model of Sec.\ \ref{sec:LOstate}.
By the definition of the various processes we have easily \eqref{Eabsf2} and
\begin{subequations}\label{Efff}
\begin{gather}
\Ebb_f[f(t)]= \lambda w \rme^{-\rmi\omega_0t-\gamma_0t},
\qquad
\Ebb_f\left[\overline{f(s)}f(t)\right]=\abs\lambda^2 \exp\left\{\rmi\omega_0\left(s-t\right)-\gamma_0\abs{t-s}\right\}\left(w^2+v(t-s)\right), \label{Efff2}
\\
\Ebb_f\left[f(s)f(t)\right]=\lambda^2\rme^{-\left(\rmi \omega_0+\gamma_0\right)\left(t+s\right)- 2\gamma_0\left(t\wedge s\right)}\left(w^2+v(t-s)\right),
\\
\Ebb_f\left[\abs{f(t)}^2f(s)\right]=\abs\lambda^2\lambda w \left[1+2v(t-s)\right]\rme^{-\left(\rmi \omega_0+\gamma_0\right)s},
\\
\Cov_f\left[\abs{f(t)}^2,\abs{f(s)}^2\right]=2\abs{\lambda}^4 v(t-s)\left(2w^2+v(t-s)\right). \label{Efff3}
\end{gather}
\end{subequations}

The relative intensity noise \eqref{nRIN} reduces to $n_{\rm RIN}(t)=u(t)^2-1$, which has zero mean and covariance \eqref{RINcov}.
The RIN is not normally distributed, but it is the square of a Gaussian process, due to the $u(t)^2$ contribution. By the choice of the function $v(t)$ we can make its correlations to decay very fast. As a measure of this noise, an effective RIN coefficient can be introduced by integrating the correlations \eqref{RINcov} over time:
\[
{\rm RIN}_{\rm eff}=\int_\Rbb\rmd s\,\Ebb_f\left[n_{\rm RIN}(t)n_{\rm RIN}(t-s)\right]=4w^2 \tilde v(0)+\frac 1 \pi\int_\Rbb \tilde v(\nu)^2\,\rmd \nu,
\]
where we have used \eqref{RINcov} and \eqref{LOf3}.

A simple choice is to take an exponential behaviour for the correlations $v(t)$:
\begin{equation}\label{vexp}
v(t)=v(0)\rme^{-\gamma_1 \abs t},\qquad v(0)\geq 0, \quad \gamma_1>0.
\end{equation}
This gives
\begin{equation}
\Ebb_f\left[n_{\rm RIN}(t)n_{\rm RIN}(s)\right]=4w^2v(0)\rme^{-\gamma_1\abs{t-s}}+2v(0)^2\rme^{-2\gamma_1\abs{t-s}} ,
\end{equation}
\begin{equation}\label{tildev}
\tilde v(\nu)=\frac{2v(0)\gamma_1}{\gamma_1^{\,2}+\nu^2},\qquad {\rm RIN}_{\rm eff}
= \frac 2 {\gamma_1}\left(1-w^2\right)\left(1+3w^2\right).
\end{equation}
Moreover, the intensity spectrum, introduced in \eqref{LOspectrum}, turns out to be
\begin{equation}\label{LOpower}
\Pi_f(\mu)=\frac {2\abs\lambda^2w^2 \gamma_0}{\gamma_0^{\,2} + \left(\mu-\omega_0\right)^2}+ \frac {2\abs\lambda^2\left(1-w^2\right)\left( \gamma_0+\gamma_1\right)}{\left(\gamma_0+\gamma_1\right)^{\,2} + \left(\mu-\omega_0\right)^2}.
\end{equation}
Finally, the RIN spectrum takes the expression
\begin{equation}
\Pi_{\rm RIN}(\mu)=8\left(1-w^2\right)\gamma_1\biggl[ \frac {w^2}{\gamma_1^{\,2}+\mu^2}+\frac {1-w^2}{4\gamma_1^{\,2}+\mu^2}\biggr].
\end{equation}

\subsection{Moments of the photocurrents}\label{sec:momM}

Let us introduce the following combinations of transmissivity and efficiency parameters:
\begin{subequations}\label{:gs}
\begin{gather}
g_{11}=\eta_1\eta_3\epsilon_1, \qquad g_{22}= \left(1-\eta_2\right)\left(1-\eta_4\right)\epsilon_2,
\qquad
g_{12}=\eta_2\left(1-\eta_3\right)\epsilon_1, 
\\ g_{21}=\left(1-\eta_1\right)\eta_4\epsilon_2,
\qquad
g_{31}=\eta_1\left(1-\eta_3\right)\epsilon_3, \qquad  g_{32}=\eta_2\eta_3\epsilon_3,
\\
g_{j3}=\sqrt{g_{j1}g_{j2}}, \qquad g_{41}=\left(1-\eta_1\right)\left(1-\eta_4 \right)\epsilon_4 , 
\qquad g_{42}= \left(1-\eta_2\right)\eta_4\epsilon_4.
\end{gather}
\end{subequations}
By using the input/output expression of the fields
\eqref{fieldsdj}, the state \eqref{rhoT} and the notation \eqref{:gs} we get
\begin{subequations}\label{Dreduced}
\begin{gather}
d_1(r)\rho^T_f= \left[\sqrt{g_{11}}\,a_1(r) + \rmi\rme^{\rmi \psi_1}\sqrt{g_{12}}\,f(r)\right] \rho^T_f,
\qquad
d_3(r)\rho^T_f
=\left[\rmi\sqrt{g_{31}} \,a_1(r)+ \rme^{\rmi \psi_1}\sqrt{g_{32}}\,f(r)\right]\rho^T_f,
\\
d_2(r)\rho^T_f
= \left[\rmi\sqrt{g_{21}}\, a_1(r)- \rme^{\rmi \psi_2} \sqrt{g_{22} }  \,f(r)\right]\rho^T_f,
\qquad
d_4(r)\rho^T_f
=\left[-\sqrt{g_{41}}\, a_1(r)+ \rmi  \rme^{\rmi \psi_2} \sqrt{g_{42}} \,f(r)\right]\rho^T_f.
\end{gather}
\end{subequations}
From these equations, we can compute the quantum expectation of any normal ordered function of the fields $d_i^\dagger(\cdot)$, $d_j(\cdot)$.

By using \eqref{rho13}, \eqref{qexpect}, \eqref{mean:M}, \eqref{corr:M}, \eqref{Efff}, \eqref{Dreduced} we get easily
\begin{equation}\label{mean:M+}
\Ebb_P\left[M_{j}(t)\right]=\xi_j\int_0^t\rmd r\,h(t-r)\Bigl\{g_{j1}\langle a_1^\dagger(r)a_1(r)\rangle_T+g_{j2}\Ebb_f\left[\abs{f(r)}^2\right]
+s_j g_{j3} \left(\rmi \rme^{\rmi \psi_j}\Ebb_f\left[f(r)\langle a_1^\dagger(r)\rangle_T^f\right] +\text{c.c.}\right)\Bigr\},
\end{equation}
\begin{multline}\label{corr:M+}
\Ebb_P\left[M_{j}(t)M_{i}(s)\right]=\delta_{ij}\xi_j^{\,2}\int_0^{t\wedge s}\rmd r\, h(t-r)h(s-r)\Bigl\{g_{j1}\langle a_1^\dagger(r)a_1(r)\rangle_T+g_{j2}\Ebb_f\left[\abs{f(r)}^2\right]
\\ {}+s_j g_{j3} \left(\rmi \rme^{\rmi \psi_j}\Ebb_f\left[f(r)\langle a_1(r)\rangle_T^f\right] +\text{c.c.}\right)\Bigr\}
+ \xi_j\xi_i
\int_0^t\rmd r\int_0^s\rmd r'\,h(t-r)h(s-r')\\ {}\times \Ebb_f\Big[\big\langle : \Bigl\{g_{j1}a_1^\dagger(r)a_1(r)+g_{j2}\abs{f(r)}^2
+s_j g_{j3} \left(\rmi \rme^{\rmi \psi_j}f(r) a_1^\dagger(r) +\text{h.c.}\right)\Bigr\}
\\ {}\times \Bigl\{g_{i1}a_1^\dagger(r')a_1(r') +g_{i2}\abs{f(r')}^2
+s_i g_{i3} \left(\rmi \rme^{\rmi \psi_i}f(r') a_1^\dagger(r') +\text{h.c.}\right)\Bigr\}: \big\rangle_T^f\Big];
\end{multline}
the constants $g_{ij}$ are defined in \eqref{:gs} and $s_1=s_2=+1$, $s_3=s_4=-1$; the notation $: \bullet :$ means normal order.

\section{Probability law and characteristic operator}\label{P:Phi}
For stochastic processes the probability law is uniquely determined by its Fourier transform, the \emph{characteristic functional} \cite{DVJ08}; the same holds for pvms and POVMs, whose Fourier transform is called \emph{characteristic operator} \cite{Bar86,ZolG97,Bar06}.

\subsection{Characteristic operator for the counts of photons and of the output photocurrents}\label{Phi:counts}
In the case of the increments  of the commuting selfadjoint number operators \eqref{numop}, the characteristic operator in the time interval $(0,t)$ and the characteristic functional of the associated counting processes $N_j(t)$ in the interval $(0,T)$ are given by
\begin{subequations}\label{PhiN}
\begin{equation}
\hat\Phi_t^N[\vec k]=\exp \bigg\{\rmi\sum_{j=1}^4\int_0^t k_j(s)\rmd \hat N_j(s)\bigg\} =\prod_{j=1}^4\exp \bigg\{\rmi\int_0^t k_j(s)\rmd \hat N_j(s)\bigg\},
\end{equation}
\begin{equation} \label{PhiN2}
\Phi_T^N[\vec k]=\Ebb_P \bigg[\exp \bigg\{\rmi\sum_{j=1}^4\int_0^T k_j(t)\rmd N_j(t)\bigg\}\bigg]=\langle \hat\Phi_T^N[\vec k]\rangle_T;
\end{equation}
\end{subequations}
we have used the notation \eqref{qexpect} for the quantum expectations. The functions $k_j(t)$
are called test functions and are the analogue of the variables which are introduced in the definition of an usual Fourier transform. A very important point is that the characteristic operator $\hat\Phi_t^N[\vec k]$ satisfies a closed evolution equation, represented by the quantum stochastic equation
\begin{equation}\label{dPhiN}
\rmd \hat\Phi_t^N[\vec k]= \hat\Phi_t[\vec k] \sum_{j=1}^4\left(\rme^{ \rmi k_j(t)}-1\right)\rmd \hat N_j(t),
\end{equation}
which can be obtained by using the heuristic rules \eqref{Itotab}.

\subsubsection{The case of a coherent signal}\label{sec:0N}
Let us consider the case of a coherent state for the signal or a mixture of coherent states. The system state is given by \eqref{rhoT}--\eqref{qexpect} with $\rho^{f,T}_{1}\to\rho^{\fs,T}_{1}=|e_1(\fs)\rangle \langle e_1(\fs)|$, where $\fs(t)$ is a stochastic process. Now $P_{f} $ denotes the joint probability law of the two processes, so that \eqref{rho13} becomes
\begin{equation}\label{rhoalphaf}
\rho_{13}^T= \Ebb_{f}\left[\rho_{13}^{\vec f,T}\right],\qquad  \rho_{13}^{\vec f,T}= |e_1(\fsT)\rangle\langle e_1(\fsT)| \otimes |e_3(f_T)\rangle\langle e_3(f_T)|.
\end{equation}

Given $f$ and $\fs$ fixed, by the factorization properties of the exponential vectors and of the Fock space \cite{Parthas92,Bar06}, equations \eqref{PhiN2}, \eqref{dPhiN}, \eqref{Dreduced} give
\[
\frac{\rmd \Phi_t^N[\vec k;\vec f]}{\rmd t}= \Phi_t[\vec k;\vec f] \sum_{j=1}^4\left(\rme^{ \rmi k_j(t)}-1\right)J_j(t;\vec f),
\]
\begin{equation}\label{def:J}
J_j(t;\vec f)=\abs{\sqrt{g_{j1}}\fs(t)+s_j\rmi \rme^{\rmi\psi_j}\sqrt{g_{j2}}f(t)}^2,\qquad s_1=s_2=+1, \quad s_3=s_4=-1, \quad \psi_{j+2}=\psi_j.
\end{equation}
Then, we have
\begin{equation}
\Phi_T^N[\vec k;\vec f]=\exp \biggl\{\sum_{j=1}^4\int_0^T\left(\rme^{ \rmi k_j(t)}-1\right)J_j(t;\vec f)\rmd t \biggr\},
\end{equation}
which is the characteristic functional of four Poisson processes of intensities $J_j(t;\vec f)$. When $f$ and $\fs$ are random, the total characteristic functional turns out to be
\begin{equation}\label{PhiNalpha}
\Phi_T^N[\vec k]=\Ebb_{f}\left[\Phi_T^N[\vec k;\vec f]\right]=\Ebb_{ f}\biggl[\exp \biggl\{\sum_{j=1}^4\int_0^T\left(\rme^{ \rmi k_j(t)}-1\right)J_j(t;\vec f)\rmd t \biggr\}\biggr],
\end{equation}
and we have a mixture of Poisson processes, as reported in Remark \ref{rem:Phi0N}, where the notation $J_j^f(t)=J_j(t;f,0)$ is used.

\subsubsection{The output currents}\label{sec:PsiM}
In the case of the processes $M_j(t)$ and of the pvm of the commuting selfadjoint operators $\big\{\hat M_j(t)$, $j=1,\ldots, 4,$ $t\in(0,T)\big\}$, the characteristic operator and the characteristic functional are given by
\begin{subequations}\label{PhiM}
\begin{equation}
\hat\Phi_T^M[\vec k]=\exp \bigg\{\rmi\sum_{j=1}^4\int_0^T \rmd s\, k_j(s) \hat M_j(s)\bigg\},
\end{equation}
\begin{equation}
\Phi_T^M[\vec k]=\Ebb_P \bigg[\exp \bigg\{\rmi\sum_{j=1}^4\int_0^T \rmd s\,k_j(s)M_j(s)\bigg\}\bigg]=\langle \hat\Phi_T^M[\vec k]\rangle_T.
\end{equation}
\end{subequations}
By inserting the expressions \eqref{M:N} into \eqref{PhiM} we immediately find
\begin{equation}\label{charatteristicO,NtoM}
\hat\Phi_T^M[\vec k]=\hat\Phi_T^N[\vec k], \qquad \Phi_T^M[\vec k]= \Phi_T^N[\vec l],
\qquad
l_j(r)=\int_r^T\rmd s \,F_j(s,r)k_j(s)=\xi_j\int_r^T\rmd s \,h(s-r)k_j(s),
\end{equation}
where the characteristic operator of the number operators and the characteristic functional of the counting processes are given by \eqref{PhiN}.

\subsection{Characteristic operator of the observed processes}\label{sec:PhiX}
Similarly, the characteristic functional of the observed processes $X_j(t)$ \eqref{Xprocs} is defined by
\begin{subequations}\label{20}
\begin{equation}
\Phi_T^X[\vec k]= \Ebb_P\bigg[
\exp\Bigl\{\rmi \sum_{j=1}^2\int_0^T\rmd t \,k_j(t)  X_j(t)\Bigr\}\biggr],
\end{equation}
while the characteristic operator of the compatible operators \eqref {hatX} is
\begin{equation}
\hat\Phi_T^X[\vec k]=
\exp\Bigl\{\rmi \sum_{j=1}^2\int_0^T\rmd t \,k_j(t) \hat X_j(t)\Bigr\}.
\end{equation}
\end{subequations}
Then, we get
\begin{subequations}\label{21}
\begin{equation}
\Phi_T^X[\vec k]= \Phi_T^N[\vec l], \qquad \hat\Phi_T^X[\vec k]=\hat \Phi_T^N[\vec l],\qquad \Phi_T^X[\vec k]=\langle \hat\Phi_T^X[\vec k]\rangle_T;
\end{equation}
$\Phi_T^N$ and $\hat \Phi_T^N$ are given by Eqs.\ \eqref{PhiN}, while
the functions $l_j(r)$ are defined by Eqs.\ \eqref{charatteristicO,NtoM} together with
\begin{equation}
k_3(s)=-k_1(s),
\qquad k_4(s)=-k_2(s).
\end{equation}
\end{subequations}

\paragraph{Characteristic functional in the case of signal in a coherent state.}
In the case of the mixture of coherent states as in Appendix \ref{sec:0N}, we get from \eqref{21} the characteristic functional of the observed processes $X_j(t)$:
\begin{equation}\label{PhiXcoer}
\Phi_T^X[\vec k]= \Ebb_{f}\biggl[\exp \biggl\{\sum_{j=1}^4\int_0^T\left(\rme^{ \rmi l_j(t)}-1\right)J_j(t;\vec f)\rmd t \biggr\}\biggr],
\end{equation}
where $l_j(t)$ is given in \eqref{charatteristicO,NtoM}
and $J_j(t;\vec f)$ in \eqref{def:J}.

\subsection{Characteristic operator in the limit of strong LO}\label{app:strLO}

To analyze the structure of the processes $Y_j(\bullet)$ and to get their probability law in the strong LO limit, we introduce the processes
\begin{equation}\label{NtoZ}
Z_j(t)=\frac{\xi_jN_j(t) - \xi_{j+2}N_{j+2}(t)}{\abs{\lambda}{\kappa_{j3}}}, \qquad j=1,2.
\end{equation}
By \eqref{Xprocs}, \eqref{Yproc} we have
\begin{equation}\label{ZtoY}
Y_j(t)= \kappa_{j3}\int_0^t h(t-r)\rmd Z_j(r).
\end{equation}

By using the processes $Z_j(\bullet)$ we can prove Proposition \ref{prop:Y} and Corollary \ref{corroll}:
\begin{proof}
The probability law of the processes $Z_j(\bullet)$ is uniquely determined by the characteristic functional of their increments, defined by
\begin{equation}\label{PhiZ}
\Phi^Z_t[\vec k]=\Ebb_P\biggl[\exp\biggl\{\rmi \sum_{j=1}^2\int_0^tk_j(s)\rmd Z_j(s)\biggr\}\biggr].
\end{equation}
By \eqref{NtoZ}, this functional can be expressed in terms of the characteristic functional and characteristic operator \eqref{PhiN} of the increments of the four counting processes $N_j(\bullet)$
\begin{equation}\label{:ell}
\Phi^Z_t[\vec k]=\Phi_T^N[\vec \ell/\abs\lambda]=\langle \hat\Phi_T^N[\vec \ell/\abs\lambda]\rangle_T,
\qquad \ell_j(s)=\frac{\xi_jk_j(s)}{\kappa_{j3}} , \qquad \ell_{j+2}(s)=-\frac{\xi_{j+2}k_j(s)} { \kappa_{j3}}.
\end{equation}
By using the structure \eqref{rhoT}--\eqref{qexpect} of the field state, we can introduce the \emph{reduced characteristic operator} of the $Z$-observables:
\begin{equation}\label{Zchop}
\hat\Psi^Z_t[\vec k;f]=\Tr_{\Gamma_3\otimes\Gamma^\bot}\left\{\hat \Phi^N_t[\vec \ell/\abs\lambda]\left(\rho^{f,T}_{3}\otimes \rho^\bot\right)\right\}.
\end{equation}
Then, the $Z$-functional \eqref{PhiZ} can be written as
\begin{equation}\label{PhiZPsi}
\Phi^Z_t[\vec k]=\Ebb_f\left[\Tr_{\Gamma_1}\left\{\hat \Psi^Z_t[\vec k;f]\rho^{f,T}_{1}\right\}\right].
\end{equation}
By construction, $\hat\Psi^Z_t[\vec k;f]$ is the Fourier transform of a POVM on the signal Hilbert space $\Gamma_1$.

By using the differential of the characteristic operator of the number operators \eqref{dPhiN} and the factorization properties of Fock spaces and exponential vectors \cite{Parthas92,Bar06}, we get a quantum stochastic differential equation for the reduced characteristic operator \eqref{Zchop}:
\begin{equation}\label{dPsi1}
\rmd \hat\Psi^Z_t[\vec k;f]= \hat\Psi_t^Z[\vec k;f] \sum_{j=1}^4\left(\rme^{ \rmi \ell_j(t)/\abs\lambda}-1\right)\Tr_{\Gamma_3\otimes\Gamma^\bot}\left\{\rmd \hat N_j(t)\left(\rho^{f,T}_{3}\otimes \rho^\bot\right)\right\},
\end{equation}
where the $\ell$-functions are given in \eqref{:ell}. Then, we can expand the exponential up to the second order in $1/\abs\lambda$. By
the assumption of $G_{j2}$ \eqref{:G_} independent of $\lambda$,  we get, in the limit $\abs\lambda\to +\infty$,
\begin{multline*}
\frac {\rmi \ell_j(t)}{\abs\lambda}\Tr_{\Gamma_3\otimes\Gamma^\bot}\left\{\rmd \hat N_j(t)\left(\rho^{f,T}_{3}\otimes \rho^\bot\right)\right\} + \frac {\rmi \ell_{j+2}(t)}{\abs\lambda}\Tr_{\Gamma_3\otimes\Gamma^\bot}\left\{\rmd \hat N_{j+2}(t)\left(\rho^{f,T}_{3}\otimes \rho^\bot\right)\right\}
\\ {}\simeq \rmi k_j(t)\left[G_{j2}\abs{\tilde f(t)}^2\rmd t+ \left( \rmi \rme^{\rmi \psi_j}\tilde f(t) \rmd A_1^\dagger(t) + \text{h.c.}\right)\right],
\end{multline*}
\begin{multline*}
-\frac { \ell_j(t)^2}{2\abs\lambda^2}\Tr_{\Gamma_3\otimes\Gamma^\bot}\left\{\rmd \hat N_j(t)\left(\rho^{f,T}_{3}\otimes \rho^\bot\right)\right\} - \frac {\ell_{j+2}(t)^2}{2\abs\lambda^2}\Tr_{\Gamma_3\otimes\Gamma^\bot}\left\{\rmd \hat N_{j+2}(t)\left(\rho^{f,T}_{3}\otimes \rho^\bot\right)\right\}
\\ {}\simeq -k_j(t)^2\,\frac{\kappa_{j2}}{2\kappa_{j3}^{\;2}}\abs{\tilde f(t)}^2\rmd t .
\end{multline*}
By using these expressions, we have that the limit for $\abs\lambda\to +\infty$ of Eq.\ \eqref{dPsi1} exists and, by using \eqref{QjXi}, it is given by
\begin{equation}\label{diffPsi}
\rmd \hat\Psi^Z_t[\vec k;f]= \hat\Psi_t^Z[\vec k;f] \sum_{j=1}^2\biggl\{\rmi k_j(t) \left[ G_{j2}\abs{\tilde f(t)}^2\rmd t+ \rmd \hat Q_j(t) \right]-\frac{\kappa_{j2}}{2\kappa_{j3}^{\;2}}\abs{\tilde f(t)}^2k_j(t)^2\rmd t\biggr\}.
\end{equation}
By the rules of QSC, this equation can be integrated and we get
\begin{multline}\label{integratedPsi}
\hat\Psi^Z_t[\vec k;f]= \exp\int_0^t\biggl\{ \sum_{j=1}^2\rmi k_j(s) \left( G_{j2}\abs{\tilde f(s)}^2\rmd s+ \rmd \hat Q_j(s) \right)
\\ {}- \frac 12\biggl[\sum_{j=1}^2 \biggl(\frac{\kappa_{j2}}{\kappa_{j3}^{\;2}}-1\biggr)k_j(s)^2-2 k_1(s)k_2(s)\cos\phi\biggr]\abs{\tilde f(s)}^2\rmd s
\biggr\}.
\end{multline}
To check this result one has to differentiate \eqref{integratedPsi} by using the rules \eqref{Itotab}; in this way \eqref{diffPsi} is obtained. These computations prove also the existence of the limit in \eqref{PhiYlim}.

By using  \eqref{integratedPsi}, \eqref{ZtoY}, \eqref{PhiZPsi},  we obtain the expressions \eqref{Yprob}, \eqref{Psi=GP0}.
The expression  $\hat\Psi_T^Q[\vec k;f]$ \eqref{Psif} is the characteristic operator which we would have obtained in the case of perfect efficiency, $\epsilon_j=1$, balanced outputs, i.e.\ $ G_{j2}=0$ and $\Delta_{j2}=0$, and non random laser; as these parameters are arbitrary, also  $\hat\Psi_T^Q[\vec k;f]$ is the characteristic operator of a POVM. This ends the proof of Proposition \ref{prop:Y}.

To prove Corollary \ref{corroll} the easiest way is to take the characteristic functional \eqref{PhiXcoer},
\eqref{charatteristicO,NtoM}, \eqref{def:J} and to compute the limit \eqref{PhiYlim}. Then, the characteristic functional of the $Y$-processes turns out to be given by \eqref{YintZ} and \eqref{PhiZcoer}.
\end{proof}

\subsection{Characteristic operator and probability density for the case of discrete sampling}\label{app:dsproof}
In this appendix we prove Propositions \ref{prop:dsampl} and \ref{prop:V+b}, giving the structure of the characteristic function of the random variables $Y_j(t_l)$.

\paragraph{Proof of Proposition \ref{prop:dsampl}:}

\begin{proof} To get the characteristic function $\Phi^{\vec Y}(\vec k)$ \eqref{cfjl} we insert $k_j(s)=\sum_l k_j^l \delta(s-t_l)$ into the characteristic functional $\Phi_T^Y[\vec k]$ \eqref{PhiYlim}. From \eqref{YintZ}, \eqref{GammaPsi} and Assumption \ref{ass:sampling} we get
\[
\Phi^Y_T(\vec k)= \Ebb_f\left[\Phi^Q_T[\vec u;f]\Gamma_T[\vec u;f]\right], \qquad u_j(s)=\kappa_{j3}\sum_l k_j^lh(t_l-s)\ind_{(t_l-\tau,t_l)}(s).
\]
From \eqref{Gamma} we obtain
\[
\Gamma_T[\vec u;f]=  \exp\biggl\{\sum_{j,l}\int_{t_l-\tau}^{t_l}\rmd s\abs{\tilde f(s)}^2\biggl[\rmi k_j^l \, G_{j2}\kappa_{j3}h(t_l-s)
-\frac{\sigma_j^2+V_j^{\,2}}{2}\,\kappa_{j3}^{\;2}{k_j^l}^2h(t_l-s)^2\biggr]\biggr\};
\]
this expression gives \eqref{prod}, \eqref{Gammasampl}.
Moreover, from \eqref{PhiQ}, \eqref{Psi=GP0}, we get
\begin{equation*}
\Phi^Q_T[\vec u;f]=\Tr_{\Gamma_1}\left\{\hat \Psi^Q_T[\vec u;f]\rho^{f,T}_{1}\right\},
\end{equation*}
\[%\begin{multline*}
\hat\Psi_T^Q[\vec u;f]
= \exp\sum_l\biggl\{\rmi \sum_{j=1}^2\int_{t_l-\tau}^{t_l}\kappa_{j3} k_j^lh(t_l-s)\,\rmd\hat Q_j(s)
-\frac 12 \int_{t_l-\tau}^{t_l}h(t_l-s)^2\abs{\tilde f(s)}^2\rmd s
\sum_{i,j=1}^2 k_j^l \kappa_{j3}\Xi_{ji}\kappa_{i3}k_i^l\biggr\}.
\]%\end{multline*}
By using equations \eqref{qjl} and the fact that the quadrature operators commute for different values of $l$, we get the product structure and by expressing the quantity in square brackets as a squared modulus, this equation gives \eqref{hatPsiq}.

Once again, the expression  $\hat\Psi^q(\vec k;f)$ \eqref{hatPsiq} is the characteristic operator which we would have obtained in the case of perfect efficiency, $\epsilon_j=1$, balanced outputs, i.e.\ $ G_{j2}=0$ and $\Delta_{j2}=0$, and non random laser; as these parameters are arbitrary, also  $\hat\Psi^q(\vec k;f)$ is the characteristic operator of a POVM. By taking a single time $t_l$ we have that also each one of the factors is the characteristic operator of a POVM.
\end{proof}

\paragraph{Proof of Proposition \ref{prop:V+b}:}

\begin{proof} Firstly, we define the parameter
\begin{equation*}
v_l= R_l(f)\left[\kappa_{13} k_1^l +\rme^{\rmi \phi }\kappa_{23} k_2^l\right].
\end{equation*}
By using the parameters $v_l$ and $\alpha, \,\beta$ \eqref{def:alpha,beta}, we can check by direct computations that the characteristic operator \eqref{hatPsiq} can be written as
\begin{equation*}
\hat\Psi^q_l(\vec k^l;f)=\exp\left\{\rmi\left( v_l a_l^\dagger + \overline{ v_l}\, a_l\right) - \frac 12\abs{\alpha v_l-\beta \overline{v_l}}^2\right\}.
\end{equation*}
Again by direct computations, by using \eqref{a2b} and \eqref{k2ul}, we can verify that
\[
a_l=-\rmi\left(\rme^{\rmi \phi}\alpha b_l+\rme^{-\rmi \phi}\beta b_l^\dagger\right),
\qquad
u_l=\rmi\rme^{-\rmi \phi}\left(\alpha v_l - \beta \,\overline{v_l}\right).
\]
This gives
$v_l a_l^\dagger +\overline{v_l}\, a_l=u_l b_l^\dagger +\overline{u_l}\, b_l$ and $\abs{\alpha v_l-\beta \overline{v_l}}^2=\abs{u_l}^2$ and \eqref{Vu} is proved.

Then, by using CCRs and the over-completeness property of the coherent states for the mode $b_l$ we have
\begin{multline*}
\hat\Psi^q_l(\vec k^l;f) =\rme^{\rmi \, \overline {u_l}\, b_l}\rme^{\rmi  u_l b_l^\dagger}=\rme^{\rmi \, \overline {u_l}\, b_l}\frac 1\pi\int_{\Cbb} \rmd^2 \zeta\,  |\psi_l(\zeta;\alpha,\beta)\rangle\langle \psi_l(\zeta;\alpha,\beta)|\rme^{\rmi  u_l b_l^\dagger} \\ {}= \frac 1\pi\int_{\Cbb} \rmd^2 \zeta\, \rme^{\rmi \, \overline {u_l}\, \zeta} |\psi_l(\zeta;\alpha,\beta)\rangle\langle \psi_l(\zeta;\alpha,\beta)|\rme^{\rmi  u_l \,\overline \zeta};
\end{multline*}
this proves \eqref{eqprop3}.
\end{proof}

\paragraph{Proof of equation \eqref{YPOVMdens}.}

\begin{proof} Firstly,
the parameter \eqref{k2ul}, needed in \eqref{eqprop3}, can be written as
\begin{equation*}
u_l= K_1^l(f)k_1^l -\rme^{-\rmi \phi } K_2^l(f) k_2^l.
\end{equation*}
Then, we can compute the anti-Fourier transform of \eqref{COl}:
\begin{multline*}
\hat g_Y^l(y_1,y_2;f)=\frac 1{4\pi^2}\int_{\Rbb^2}\rmd k_1 \rmd k_2\, \rme^{-\rmi\left(y_1k_1+ y_2k_2\right)}\hat \Psi^{\vec Y}_l(k_1,k_2) \\ {}= \frac 1{4\pi^2}\int_\Cbb \rmd^2 z \, \hat g_{\alpha,\beta}^l( z)
\int_{\Rbb^2}\rmd k_1 \rmd k_2\, \exp\Big\{-\rmi\left(y_1k_1+ y_2k_2\right)+ \rmi\left(u_l\,\overline z+\overline {u_l}\, z\right)+ \sum_{j}\left[\rmi k_j \mu_\Lcal^{jl}(f)
-\frac12 \,\sigma_\Lcal^{jl}(f)^2 \,{k_j}^2\right]\Big\}
\\ {}= \frac 1{4\pi^2}\int_\Cbb \rmd^2 z \, \hat g_{\alpha,\beta}^l( z)
\int_{\Rbb^2}\rmd k_1 \rmd k_2\, \exp\biggl\{
-\frac12 \,\sum_{j}\sigma_\Lcal^{jl}(f)^2 \,{k_j}^2\biggr\} \\ {}\times \exp\Big\{\rmi k_1\left(\mu_\Lcal^{1l}(f)+2z_1x_1^l(f)- y_1\right) + \rmi k_2\left(\mu_\Lcal^{2l}(f)+2x_2^l(f)\left(z_2\sin \phi -z_1\cos\phi\right)- y_2\right)\Big\},
\end{multline*}
where we have used  $z=z_1+\rmi z_2$. By computing the Gaussian integral in $\rmd k_1 \rmd k_2$, we get the POVM-density $\hat g_Y^l(y_1,y_2;f)$ \eqref{YPOVMdens}.
\end{proof}

\section{The POVM and the probability density of the $Y$-observables}\label{app:probdens}

\subsection{Signal in a mixture of coheret states} \label{sec:s=cs}
The density \eqref{mixtprobdens} can be explicitly computed, for instance, in the case of the signal in a mixture of coherent states as in Corollary \ref{corroll}. By using \eqref{PhiZcoer}, \eqref{Yprob} we get the characteristic function, from which we see that the density turns out to be a mixture of normal distributions.
We define
\[
\mu_j^{l}(\vec f)=\mu_\Lcal^{jl}(f)+\kappa_{j3}\int_0^\tau \rmd t \, h(t) \left(\rmi\rme^{\rmi\psi_j} \overline{\fs(t_l-t)}\tilde f(t_l-t)+{\rm c.c.} \right),
\]
where $\mu_\Lcal^{jl}(f)$ is given in \eqref{muLcal}. Then, the probability density can be written as
\begin{equation}\label{0mixtprobdens}
g_{\vec Y}(\vec y)=\Ebb_f\biggl[\prod_{l=1}^m\prod_{j=1}^2\frac{1} {\sqrt{2\pi\kappa_{j2}R_l(f)^2}} \exp\biggl\{-\frac{\left(y_j^l-\mu_j^{l}(\vec f)\right)^2 } {2\kappa_{j2}R_l(f)^2}\biggr\}\biggr].
\end{equation}
Let us note that \eqref{j2decomp}, \eqref{GsVs}, \eqref{sigmaLcal}  give the following decomposition of the variances:
\begin{equation}\label{k2sigma}
\kappa_{j2}R_l(f)^2=\sigma_\Lcal^{jl}(f)^2 +\kappa_{j3}^{\;2}\left(G_{j3}+1\right)R_l(f)^2, \qquad G_{13}+1 =\frac 1 {\eta_1}, \quad G_{23}+1 =\frac 1 {1-\eta_1}.
\end{equation}

\subsection{A density bound}\label{sec:densbound}

\begin{proposition} The POVM-density \eqref{YPOVMdens} is bounded by
\begin{equation}\label{boundYPOVM}
\hat g_Y^l(y_1,y_2;f)\leq \frac {\mathds{1}} {4\pi R_l(f)^2\kappa_{13} \kappa_{23}\abs{\sin\phi}}.
\end{equation}
Moreover, the total probability density \eqref{mixtprobdens} is bounded by
\begin{equation}\label{boundYY}
g_{\vec Y}(\vec y)\leq \frac {1} {\left(4\pi \kappa_{13} \kappa_{23}\abs{\sin\phi}\right)^m}\prod_{l=1}^m\Ebb_f \left[R_l(f)^{-2}\right],
\end{equation}
$\forall \vec y\in \Rbb^{2m}$. This bound holds for any choice of the signal state $\rho_1^f$ in the expression \eqref{mixtprobdens}.
\end{proposition}
\begin{proof}
By using the bound \eqref{boundhatg} we have
\begin{multline*}
\hat g_Y^l(y_1,y_2;f)\leq \frac 1 {2\pi^2 \sigma_\Lcal^{1l}(f)\sigma_\Lcal^{2l}(f)}\int_\Cbb \rmd^2 z \, \exp\biggl\{-\frac {\left(\mu_\Lcal^{1l}(f)+2z_1K_1^l(f)- y_1\right)^2}{2\sigma_\Lcal^{1l}(f)^2} \biggr\}
\\ {}\times \exp\biggl\{-\frac{\left(\mu_\Lcal^{2l}(f)+2K_2^l(f)\left(z_2\sin \phi -z_1\cos\phi\right)- y_2\right)^2}{2\sigma_\Lcal^{2l}(f)^2} \biggr\}\openone
\\ {} =\frac 1 {2\pi\sqrt{2\pi}\, \sigma_\Lcal^{1l}(f)\abs{K_2^l(f)\sin\phi}}\int_\Rbb \rmd z_1 \,\exp\biggl\{-\frac {\left(\mu_\Lcal^{1l}(f)+2z_1K_1^l(f)- y_1\right)^2}{2\sigma_\Lcal^{1l}(f)^2} \biggr\}\openone
=\frac 1 {4\pi\abs{K_1^l(f)K_2^l(f)\sin\phi}}\,\openone.
\end{multline*}
By inserting the expressions of $K_j^l(f)$ \eqref{def:x_j} we get \eqref{boundYPOVM}.
As the POVM-densities \eqref{YPOVMdens} act, for different values of $l$, on different factors of the Hilbert space, the analogous bound holds for the total probability density \eqref{mixtprobdens}, for any choice of the signal state (even if not factorized), and this proves \eqref{boundYY}.
\end{proof}

Let us note that, by inserting the bound \eqref{boundYPOVM} into \eqref{pX}, we get \eqref{boundpX}.


\begin{thebibliography}{99}
\bibitem{Stefanov00}  A. Stefanov, N. Gisin, O. Guinnard, L. Guinnard, H. Zbinden,
\textsl{Optical quantum random number generator}, \href{https://doi.org/10.1080/09500340008233380}{Journal of Modern Optics \textbf{47} (2000) 595-598}.

\bibitem{Extractor}  Xiongfeng Ma, Feihu Xu, He Xu, Xiaoqing Tan, Bing Qi, Hoi-Kwong Lo, \textsl{Postprocessing for quantum random-number generators: Entropy evaluation and randomness extraction}, \href{https://doi.org/10.1103/PhysRevA.87.062327}{Phys. Rev. A \textbf{87} (2013) 062327.}

\bibitem{Haw+15}  J.Y. Haw, S.M. Assad, A.M. Lance, N.H.Y. Ng, V. Sharma, P.K. Lam, T. Symul,
\textsl{Maximization of extractable randomness in a quantum random-number generator}, \href{https://journals.aps.org/prapplied/abstract/10.1103/PhysRevApplied.3.054004}{Phys. Rev. Applied \textbf{3} (2015) 054004}.

\bibitem{Vill17}  D.G. Marangon, G. Vallone, P. Villoresi, \textsl{Source-device-independent ultrafast quantum random number generation},
\href{https://journals.aps.org/prl/abstract/10.1103/PhysRevLett.118.060503}{Phys. Rev. Lett. \textbf{118} (2017) 060503.}

\bibitem{+AS+18}  F. Raffaelli , G. Ferranti, D.H. Mahler, P. Sibson, J.E. Kennard,
A. Santamato, G. Sinclair, D. Bonneau, M.G. Thompson, J.C.F. Matthews, \textsl{A homodyne detector integrated onto a photonic chip for measuring
quantum states and generating random numbers}, \href{https://doi.org/10.1088/2058-9565/aaa38f}{Quantum Sci. Technol. \textbf{3} (2018) 025003}.

\bibitem{Smith2019}  P.R. Smith, D.G. Marangon, M. Lucamarini, Z.L. Yuan, A. J. Shields, \textsl{Simple source device-independent continuous-variable quantum random number generator}, \href{https://doi.org/10.1103/PhysRevA.99.062326}
{Phys. Rev. A \textbf{99} (2019)  062326}.

\bibitem{Thewes2019}  J. Thewes, C. L\"uders, M. Assmann, \textsl{Eavesdropping attack on a trusted continuous-variable quantum random-number generator}, \href{https://doi.org/10.1103/PhysRevA.100.052318}{Phys. Rev. A \textbf{100} (2019)   052318. }

\bibitem{APFPS20}  M. Almeida, D. Pereira, M. Fa\~cao, A.N. Pinto, N.A. Silva, \textsl{Impact of imperfect homodyne detection on measurements of vacuum states shot noise}, \href{https://doi.org/10.1007/s11082-020-02622-z}{Opt. Quant. Electron. (2020) 52:503}.

\bibitem{Huang+20}  Weinan Huang, Yichen Zhang, Ziyong Zheng, Yang Li, Bingjie Xu, Song Yu, \textsl{Practical security analysis of a continuous-variable quantum random-number generator with a noisy local oscillator}, \href{https://journals.aps.org/pra/abstract/10.1103/PhysRevA.102.012422}{Phys. Rev. A \textbf{102} (2020) 012422.}

\bibitem{Lupo+21}  T. Gehring, C. Lupo, A. Kordts, D. Solar Nikolic, N.J.T. Rydberg, T.B. Pedersen, S. Pirandola, U.L. Andersen, \textsl{Homodyne-based quantum random number generator at 2.9 Gbps secure against quantum side-information}, \href{https://www.nature.com/articles/s41467-020-20813-w}{Nature Communications  (2021) 12:605.}

\bibitem{Vill18}  M. Avesani, D.G. Marangon, G. Vallone, P. Villoresi, \textsl{Source-device-independent heterodyne-based
quantum random number generator at 17 Gbps}, \href{https://doi.org/10.1038/s41467-018-07585-0}{Nature Commun. \textbf{9} (2018) 5365}.

\bibitem{FOP05}  A. Ferraro, S. Olivares, M.G.A. Paris, \href{https://bibliopolis.it/shop/gaussian-states-in-quantum-information/}{\textit{Gaussian States in Quantum Information} (Bibliopolis, Naples, 2005).}

\bibitem{KiuL08b}  J. Kiukas, P. Lahti,  \textsl{A note on the measurement of phase space observables with an eight-port homodyne detector}, \href{https://www.tandfonline.com/doi/abs/10.1080/09500340701864718?journalCode=tmop20}{J. Mod. Optics \textbf{55} (2008) 1891-1898.}

\bibitem{Leo10}  U. Leonhardt, \textit{Essential Quantum Optics. From Quantum Measurements to
Black Holes} (Cambridge University Press, Cambridge, 2010)

\bibitem{LPS10}  P. Lahti, J.-P. Pellonp\"a\"a,  J.\ Schultz, \textsl{Realistic eight-port homodyne detection and covariant phase space observables}, \href{https://www.tandfonline.com/doi/abs/10.1080/09500340.2010.503013}{J. Mod. Opt. \textbf{57} (2010) 1171--1179.}

\bibitem{Dmo19}  G. Dmochowski, A. Feizpour, X. Xing, \textsl{8-port homodyne detection of optical fields using IQ demodulation}, \href{https://doi.org/10.1088/1361-6501/ab150d}{Meas. Sci. Technol. \textbf{30} (2019) 095201}.

\bibitem{KiuL08a}  J. Kiukas, P. Lahti, \textsl{On the moment limit of quantum observables, with an application to the balanced homodyne detection}, \href{https://www.tandfonline.com/doi/full/10.1080/09500340701624658}{J. Mod. Optics \textbf{55} (2008) 1175-1198}.

\bibitem{BKS14}  M. Ban, S. Kitajimaa, F. Shibataa, \textsl{Quadrature operators with arbitrary phase and
applications to phase-space distribution and quantum communication},  \href{https://doi.org/10.1080/09500340.2014.900826}{Journal of Modern Optics 61:7 (2014) 582--607}.

\bibitem{RCCBS95}  M.G. Raymer , J. Cooper, H.J. Carmichael, M. Beck, D.T. Smithey, \textsl{Ultrafast measurement of optical-field statistics by dc-balanced homodyne detection},  \href{https://opg.optica.org/josab/abstract.cfm?uri=josab-12-10-1801}{J. Opt. Soc. Am. B \textbf{12} (1995) 1801.}

\bibitem{Bar86}  A. Barchielli, \textsl{Measurement theory and stochastic differential equations in quantum mechanics}, \href{https://journals.aps.org/pra/abstract/10.1103/PhysRevA.34.1642}{Phys.\ Rev.\ A \textbf{34} (1986) 1642--1649.}

\bibitem{ZolG97}   P. Zoller, C.W. Gardiner,   \textsl{Quantum noise in quantum optics: the
    stochastic Schr\"odinger equation}. In S. Reynaud, E. Giacobino \& J. Zinn-Justin eds., \textit{Fluctuations quantiques, (Les Houches 1995)} (North-Holland, Amsterdam, 1997) pp. 79--136.

\bibitem{Bar06}  A. Barchielli, \textsl{Continual  Measurements in Quantum Mechanics and  Quantum Stochastic Calculus}.
    In S. Attal, A. Joye, C.-A. Pillet (eds.),
    \textit{Open   Quantum Systems III}, \href{http://www.springer.com/mathematics/dynamical+systems/book/978-3-540-30993-2}{Lect.\ Notes Math.\ \textbf{1882}  (Springer, Berlin, 2006)}, pp. 207--291.

\bibitem{ZGVit15}  S. Zippilli, G. Di Giuseppe, D. Vitali, \textsl{Entanglement and
squeezing of continuous-wave stationary light}, \href{https://doi.org/10.1088/1367-2630/17/4/043025}{New J. Phys. \textbf{17} (2015) 043025}.


\bibitem{HudP84}  R.L. Hudson and K.R. Parthasarathy, \textsl{Quantum It\^o's formula and  stochastic evolutions}, \href{https://link.springer.com/article/10.1007/BF01258530}{Commun. Math. Phys. \textbf{93} (1984) 301--323.}

\bibitem{GarC85}  C.W. Gardiner, M.J. Collet, \textsl{Input and output in damped quantum systems: Quantum stochastic differential equations and the master equation}, \href{https://journals.aps.org/pra/abstract/10.1103/PhysRevA.31.3761}{Phys. Rev. A \textbf{31} (1985) 3761--3774.}

\bibitem{Parthas92}  K.R. Parthasarathy: \textit{An Introduction to Quantum Stochastic  Calculus} (Birkh\"{a}user, Basel, 1992).

\bibitem{ASth} A. Santamato, \textsl{A quantum theory of photodetection and other optical devices}, master thesis, University of Milan (2010). \href{http://dx.doi.org/10.13140/RG.2.2.36655.48801}{DOI 10.13140/RG.2.2.36655.48801}.

\bibitem{BG21}  A. Barchielli, M. Gregoratti, \textsl{Quantum optomechanical system in a Mach-Zehnder interferometer}, \href{https://doi.org/10.1103/PhysRevA.104.013713}{Phys. Rev. A \textbf{104} (2021)  013713}.

\bibitem{Bar90}  A.~Barchielli,
    \textsl{Direct and heterodyne detection and other applications of quantum stochastic calculus to quantum optics}, \href{https://iopscience.iop.org/article/10.1088/0954-8998/2/6/002}{Quantum Opt.\ {\bf 2} (1990) 423--441.}

\bibitem{Bar91}  A.\ Barchielli, \textsl{Detection theory in quantum optics and quantum stochastic calculus}. In C.\
    Bendjaballah, O.\ Hirota, S.\ Reynaud (eds.), \textit{Quantum Aspects    of Optical Communications}, \href{https://link.springer.com/chapter/10.1007/3-540-53862-3_178}{Lect.\ Notes Phys.\ vol.\ 378 (Springer,
    Berlin, 1991) pp.\ 179--189.}

\bibitem{Hol82}  A.S. Holevo, \textit{Probabilistic and Statistical Aspects of Quantum Theory} (North-Holland, Amsterdam, 1982); second edition: Edizioni della Normale  \href{https://edizioni.sns.it/prodotto/probabilistic-and-statistical-aspects-of-quantum-theory/}{(Scuola Normale Superiore, Pisa, 2011).}

\bibitem{WisM10}  H.M. Wiseman,  G.J.  Milburn, \textit{Quantum Measurement and Control} (Cambridge University Press, Cambridge  2010).

\bibitem{BG12}  A.\ Barchielli, M.\ Gregoratti, \textsl{Quantum measurements in continuous time, non-Markovian evolutions and feedback}, \href{http://rsta.royalsocietypublishing.org/content/370/1979/5364.abstract}{Phil. Trans.    R. Soc. A \textbf{370} no. 1979 (2012) 5364--5385}.

\bibitem{BarG13}  A. Barchielli, M. Gregoratti, \textsl{Quantum continuous measurements: The stochastic Schr\"o\-dinger equations and the spectrum of the output},  \href{http://dx.doi.org/10.2478/qmetro-2013-0005}{Quantum Measurements and Quantum Metrology \textbf{1} (2013) 34--56}.

\bibitem{Raym03} C. Dorrer, D.C. Kilper, H.R. Stuart, G. Raybon, M.G. Raymer, \textsl{Linear Optical Sampling}, IEEE Photonis Technology Letters \textbf{15} (2003) 1746-1748.

\bibitem{Qin2018}  H. Qin, R. Kumar, V. Makarov, R. All\'eaume, \textsl{Homodyne-detector-blinding attack in continuous-variable quantum key distribution}, \href{https://doi.org/10.1103/PhysRevA.98.012312}
{Phys. Rev. A \textbf{98} (2018)  012312.}
\bibitem{YMC+11}  Yue-Meng Chi, Bing Qi, Wen Zhu, Li Qian, Hoi-Kwong Lo, Sun-Hyun Youn, A.I. Lvovsky, Liang Tian,
\textsl{A balanced homodyne detector for high-rate Gaussian-modulated coherent-state quantum key distribution}, \href{https://iopscience.iop.org/article/10.1088/1367-2630/13/1/013003}{New J. Phys. \textbf{13} (2011) 013003.}
\bibitem{KRS09}  R. Konig, R. Renner, C. Schaffner, \textsl{The operational meaning of min- and max-entropy}, \href{http://dx.doi.org/10.1109/TIT.2009.2025545}{IEEE Trans. Inf. Theory \textbf{55} (2009) 4337--4347}.

\bibitem{SYM79pII}  J.H. Shapiro, H.P. Yuen, J.A. Machado Mata, \textsl{Optical communication with two-photon coherent states --- Part II: Photoemissive detection and structured receiver performance}, \href{http://dx.doi.org/10.1109/TIT.1979.1056033}{IEEE Trans. Inf. Theory \textbf{25} (1979) 179--192}.

\bibitem{YS80pIII}  H.P. Yuen, J.H. Shapiro, \textsl{Optical communication with two-photon coherent states --- Part III: Quantum measurements realizable with photoemissive detectors}, \href{http://dx.doi.org/10.1109/TIT.1980.1056132}{IEEE Trans. Inf. Theory \textbf{26} (1980) 78--92}.



\bibitem{AZVit16}  M. Asjad, S. Zippilli, D. Vitali, \textsl{Mechanical Einstein-Podolsky-Rosen entanglement with a finite-bandwidth squeezed reservoir}, \href{https://doi.org/10.1103/PhysRevA.93.062307}{Phys. Rev. A \textbf{93} (2016) 062307}.

\bibitem{ScuZ97}  M.O. Scully, M.S. Zubairy, \textit{Quantum Optics} (Cambridge University Press, Cambridge, 1997).

\bibitem{Botta08} S. Bottacchi, \textit{Noise and Signal Interference in Optical Fiber Transmission Systems} (Wiley, 2008)

\bibitem{Mesc07}  D. Meschede, \textit{Optics, Light and Lasers} (Wiley, 2007).

\bibitem{WalM94}  D.F. Walls, G.J. Milburn, \textit{Quantum Optics} (Springer, Berlin, 1994).

\bibitem{Car08}  H.J. Carmichael, \textit{Statistical Methods in Quantum    Optics}, Vol 2 (Springer, Berlin, 2008).

\bibitem{DattaR09}  N. Datta, R. Renner, \textsl{Smooth entropies and the quantum information spectrum}, \href{https://ieeexplore.ieee.org/document/4957652}{IEEE Trans. Inform. Theory \textbf{55} (2009) 2807--2815.}

\bibitem{OSSF13}  N. Oliver, M.C. Soriano, D.W. Sukow, I. Fischer, \textsl{Fast Random Bit Generation Using a Chaotic Laser: Approaching the Information Theoretic Limit},  \href{https://doi.org/10.1109/JQE.2013.2280917}{IEEE Journal of Quantum Electronics \textbf{49} (2013) 910--918}.
\bibitem{Kra22}  W.O. Krawec, \textsl{Quantum random number generation with practical device imperfections}, \href{https://spie.org/Publications/Proceedings/Paper/10.1117/12.2618388?SSO=1}{Proc. SPIE 12093, Quantum Information Science, Sensing, and Computation XIV, 1209307 (2022)}.

\bibitem{DVJ08}  D.J. Daley, D. Vere-Jones, \textit{An Introduction to the Theory of Point Processes. Volume II: General Theory and Structure},
Second Edition,  (Springer, 2008).

\end{thebibliography}
\end{document}